\documentclass[twocolumn,  dvipsnames]{aastex63}

\received{\today}
\revised{\today}
\accepted{\today}
\submitjournal{ApJ}

\shorttitle{\textmyfont{Fe II} and \textmyfont{Ca II} emission in AGNs: Paper II}
\shortauthors{Mart\'inez-Aldama et al.}
\graphicspath{{./}{figures/}}

\newenvironment{myfont}{\fontfamily{lmtt}\selectfont}{\par}
\DeclareTextFontCommand{\textmyfont}{\myfont}
\usepackage{amsmath}
\usepackage{natbib}
\usepackage[caption=false]{subfig}
\usepackage[normalem]{ulem}
\usepackage[T1]{fontenc}
\usepackage{mathrsfs} 
\usepackage{txfonts}
\usepackage{multirow}
\usepackage{booktabs}
\usepackage{ulem}
\usepackage{xcolor,color}
\usepackage{comment}
\usepackage{rotating}
\usepackage{chngcntr}

\def\civ{C {\sc iv}$\lambda1549$}
\def\kms{\,km\,s$^{-1}$}
\def\ergs{erg\,s$^{-1}$}
\def\hb{{\sc{H}}$\beta$\/}
\def\ergscmA{erg\,s$^{-1}\,$cm$^{-2}\, \AA^{-1}$}

\def\nv{{N \sc{v}}$\lambda1240$}
\def\mgii{{Mg \sc{ii}}$\lambda2800$}
\def\feii{{Fe \sc{ii}}}
\def\caii{{Ca {\sc{ii}}}}
\def\heii{{He \sc{ii}}$\lambda1640$}
\def\rfe{R$_{\rm{FeII}}$}
\def\rcat{R$_{\rm{CaT}}$}
\def\cat{{CaT}}
\def\lbol{$L\mathrm{_{bol}}$}
\def\mbh{$M\mathrm{_{BH}}$}
\def\fblr{$f\mathrm{_{BLR}}$}

\def\eddr{$L\mathrm{_{bol}}/L\mathrm{_{Edd}}$}

\def\rblr{$R\mathrm{_{BLR}}$}
\def\oi{O {\sc i} $\lambda8446$}

\def\hb{H$\beta$}
\def\rca{R$_{\rm{CaT}}$}

\def\ewoi{EW$_{\rm OI}$}
\def\ewhb{EW$_{{\rm H}\beta}$}
\def\ewfe{EW${ _{\rm FeII}}$}
\def\ewca{EW$_{\rm CaT}$}

\def\fwhmhb{FWHM$_{\rm H\beta}$}
\def\fwhmoi{FWHM$_{\rm OI}$}
\def\fwhmca{FWHM$_{\rm CaT}$}
\def\n{$\rm{n_{H}}$}
\def\u{${U}$}
\def\hb{{\sc{H}}$\beta$\/}
\def\eddr{$L\mathrm{_{bol}}/L\mathrm{_{Edd}}$}
\def\lopt{$L_\mathrm{opt}$}
\def\ledd{$L\mathrm{_{Edd}}$}
\def\kbol{$k\mathrm{_{bol}}$}

\def\lnir{$L_{\rm NIR}$}
\def\femg{Fe {\sc ii}/Mg {\sc ii}}
\def\feca{Fe {\sc ii}/CaT}

\def\aliii{Al {\sc iii}$\lambda$1860\/}
\def\siiv{Si{\sc iv}$\lambda$1397\/}
\def\oiv{O{\sc iv]}$\lambda$1402\/}

\begin{document}

\title{The CaFe Project: Optical \textmyfont{Fe II} and Near-Infrared \textmyfont{Ca II} triplet emission in active galaxies. \\ II. The driver(s) of the \textmyfont{Ca II} and \textmyfont{Fe II} and its potential use as a chemical clock}

\correspondingauthor{Mary Loli Mart\'inez-Aldama}
\email{mmary@cft.edu.pl}

\author[0000-0002-7843-7689]{Mary Loli Mart\'inez-Aldama}
\affiliation{Center for Theoretical Physics, Polish Academy of Sciences, Al. Lotnik{\'o}w 32/46, 02-668 Warsaw, Poland}

\author[0000-0002-5854-7426]{Swayamtrupta Panda}
\affiliation{Center for Theoretical Physics, Polish Academy of Sciences, Al. Lotnik{\'o}w 32/46, 02-668 Warsaw, Poland}
\affiliation{Nicolaus Copernicus Astronomical Center, Polish Academy of Sciences, ul. Bartycka 18, 00-716 Warsaw, Poland}

\author[0000-0001-5848-4333]{Bo\.zena Czerny}
\affiliation{Center for Theoretical Physics, Polish Academy of Sciences, Al. Lotnik{\'o}w 32/46, 02-668 Warsaw, Poland}

\author[0000-0001-9719-4523]{Murilo Marinello}
\affiliation{Laborat\'orio Nacional de Astrof\'isica, R. dos Estados Unidos, 154 - Na\c{c}\~oes, Itajub\'a - MG, 37504-364, Brazil}

\author[0000-0002-6058-4912]{Paola Marziani}
\affiliation{INAF-Astronomical Observatory of Padova, Vicolo dell'Osservatorio, 5, 35122 Padova PD, Italy}

\author[0000-0001-5756-8842]{Deborah Dultzin}
\affiliation{Universidad Nacional Auton\'oma de M\'exico Instituto de Astronom\'ia: Ciudad de Mexico, Distrito Federal, MX 04510, Mexico}


\begin{abstract}

In this second paper in the series, we carefully analyze the observational properties of the optical \feii\ and NIR \caii\ triplet  {in Active Galactic Nuclei}, as well as the luminosity, black hole mass, and Eddington ratio in order to define the driving mechanism {behind the properties of our sample}. 
The \caii\  shows an inverse Baldwin effect, bringing out the particular behavior of {this ion} with respect to the other low--ionization lines such as \hb. We performed a Principal Component Analysis, where {$81.2\%$} of the variance can be explained by the first three principal components drawn from the FWHMs, luminosity, and equivalent widths.  {The first principal component {(PC1)} is primarily driven by the combination of black hole mass and luminosity with a significance over $99.9\%$, which in turn is reflected in the strong correlation of the {PC1} with the Eddington ratio. The observational correlations are better represented by the Eddington ratio,  thus it could be the primary mechanism behind the {strong correlations} observed in the \caii-\feii\ sample}. Since, calcium belongs to the $\alpha$-elements, the \feii/\caii\ flux ratio can be used as a chemical clock for determining the metal content in AGN and trace the evolution of the host galaxies. We confirm the { de-enhancement} of the ratio \feii/\caii\ by the Eddington ratio, suggesting {a metal} enrichment of the BLR in intermediate-$z$  with respect to low-$z$ objects. A larger sample, particularly at $z>2$, is needed to confirm the present results.

\end{abstract}

\keywords{galaxies: active, quasars: emission lines; quasars: supermassive black holes; galaxies: abundances}

\section{Introduction} 
\label{sec:intro}

The large diversity of the emission lines observed in the spectrum of the Active Galactic Nuclei (AGN) reveals a complex structure of the broad line region (BLR). The physical conditions of the BLR such as density, ionization parameter and metallicity can be estimated by the flux ratios of the emission lines and their profiles supply information of the dynamics in the BLR cloud { \citep{wandel1999, negrete2014, schnorr-muller, devereux2018}}. Emission lines can be divided considering their ionization potential (IP). Typically, high-ionization lines (HIL) show IP$>$40 eV, while low-ionization lines (LIL) have IP$<$20 eV \citep{,collin88, marziani_atoms2019}. Reverberation mapping studies have confirmed the stratification of the BLR \citep[e.g.][]{horne2020}, where HIL such as \civ\ or \heii\ are emitted closer to the central continuum source, and LIL such as \hb\ or \mgii\ are emitted, at least three times further. The presence of emission lines with very low-ionization potentials (IP$\sim$10 eV)  such as { the multiple {permitted} \feii{} transitions or the \caii{} triplet at $\lambda8498,\lambda8542,\lambda8662$} (hereafter \cat) suggests the existence of a zone shielded from the high energy photons { emanated by the central source and} { likely located in the outermost portion of the BLR \citep{joly1987,dul99,rodriguezardillaetal2002,rodriguez-ardilla2012, rissmann2012,murilo2016}} .

The physical conditions of the \feii\ have been widely explored in a broad wavelength range since it provides useful information about the energy-budget of the BLR \citep{osterbrock_ferland06,vestegaard01}. However, its complex electronic structure owing to the various ionization and excitation mechanisms complicates the model of the \feii\ \citep{col00,baldwin2004}. {This ionic species manifests as a pseudo-continuum due to the numerous blended multiplets ranging from the UV to the NIR. In our studies \citep[see e.g.][, hereafter Paper-1]{panda_cafe1}, we incorporate the \feii{} dataset from \citet{verner99} that includes a 371 level with ionization potential upto $\sim$11.6 eV, available in \textmyfont{CLOUDY} \citep{f17}. Newer \feii{} models are now available that have calculated more energy levels for this species, reaching upto 26.4 eV \citep[see][for a recent compilation]{sarkar2020}.} { This model reproduces well the UV and optical \feii\ contribution observed in I Zw 1, constraining in a better way the physical conditions of the \feii\ emitting clouds. For more details on the progress in understanding the  \feii{} emission in AGNs and its modelling we refer the readers to Paper-1.} 

{ The singly-ionized calcium emission can be approximately modeled by a five levels atom: (1) the optical H and K lines ($\lambda3933$, $\lambda3968$ \AA) are emitted from the 4p level to the 4s ground level, (2) the infrared multiplet ($\lambda8498$, $\lambda8542$ and $\lambda8662$ \AA, \cat) arises from the 4p level to the 3d metastable level, and (3) the forbidden multiplet ($\lambda7291$, $\lambda7324$ \AA) arises from the 3d metastable level to the ground level \citep{perssonferland89, marziani2014}. Due to  similarity between the ionization potentials of Ly$\alpha$ (10.2 ev) and the singly-ionized \caii\ (11.8 eV), the 3d metastable level is highly populated and the collisional excitation process leading to the infrared CaII triplet emission is efficient. Thus, the near-infrared \cat\ offers the possibility to study the properties of the very low-ionization lines {in the BLR}. \cat\ is {prominent} in Narrow-Line Seyfert 1 (NLS1) galaxies \citep{persson1988, murilo2016} and quasars \citep{martinez-aldamaetal15}. However, when the stellar continuum has a significant contribution, the emission profile shows a central dip or, in extreme cases, only an absorption profile is observed.  Therefore, a correct subtraction of the stellar component is needed, particularly in low-luminosity sources.
The \cat\ absorption is mainly observed in Seyfert 1 and Seyfert 2 galaxies, where it may be enhanced by a population of red supergiant stars associated with a starburst \citep{terlevich1990}. The velocity dispersion provided by the stellar \cat\ has been used to infer the stellar populations and determine the black hole mass throughout the relation $M_{\rm BH} - \sigma_\star$ \citep{garcia-rissmann2005}.} 

Some theoretical and observational studies have been devoted to look for the connections between the optical \feii\ and { \cat}. Both ions show a strong linear relation and similar widths, narrower than \hb\ or Pa$\beta$ \citep{persson1988, martinez-aldamaetal15, martinez-aldamaetal15b, murilo2016,panda_cafe1}, suggesting that both emission lines are emitted in the outer parts of the BLR. According to the  photoionization models, both emission lines share  { almost identical} physical  { conditions - large clouds (column densities $\sim$ 10$^{24}$ cm$^{-2}$) with high mean densities ($\sim 10^{12}$ cm$^{-3}$) and relatively low temperatures ($\lesssim$ 8000 K)} \citep{joly1987, joly1989, perssonferland89,panda_cafe1,panda_cafe2}. 

In the first paper of the presented analysis {(Paper-1)}, we updated on the observational correlation between the strengths of the two species { (i.e. the flux ratios \feii/\hb\ and \cat/\hb, hereafter \rfe\ and \rcat, respectively)} given by:
\begin{equation}
{    \log 
{\rm R}_{\rm CaT} \approx (0.974 \pm 0.119) \log {\rm R}_{\rm FeII}  - (0.657 \pm 0.041).}
    \label{equ:cafe}
\end{equation}

 We also looked extensively at the optical \feii{} and { \cat} emission from a theoretical standpoint, using the photoionization models, which are compared with an up-to-date sample of \feii\ and \cat. We tested various { photoionization} models in terms of ionization parameter, cloud density, metallicity, and column density, and found an overlapping range of physical conditions that are required to efficiently excite these two species. We also find the strong \feii{} emitters in order to be well modeled require { a range of metallicity from solar to super-solar} \citep{martinez-aldamaetal18, sniegowska+20}. { This result is obtained by comparing the observed UV flux ratios of emission lines such as \civ, \aliii, \siiv+\oiv\ or \nv\ over \heii\  with the ones predicted by \textmyfont{CLOUDY} simulations. The correlation between the stronger \feii{} emitters, metallicity, and Eddington ratio has been confirmed by several independent studies   \citep[e.g.][]{hamannferland92,shinetal13,panda19b}}.

{ In a subsequent paper, \citet{panda_cafe2}, we furthered the photoionization modelling to recover the EWs in the low-ionization emitting region in the BLR {and realize the anisotropy in the accretion disk emission leading to a better understanding of the photoionization of the low-ionization emitting regions of the BLR}.}

In this part of the series, we look at the observational properties and correlations from the up-to-date optical and near-infrared measurements centered around \feii{} and \cat{} emission, respectively. Usually, the stronger \feii\ and \cat\ emitters are associated with the Narrow Line Seyfert 1 (NLS1) AGN, but also AGN with higher luminosities and broader profiles show a strong emission { for these two species} \citep{martinez-aldamaetal15}. Since \feii\ strength {( or \rfe{})} is apparently driven by the Eddington ratio  \citep{borosongreen1992,marziani2003,dong2011, zamfiretal10,panda18b,panda19b}, it motivates us to explore the role of the Eddington ratio, black hole mass and luminosity in the \cat\ and \feii\ properties  to decipher the primary driver leading to this observed correlation between the two species. 

{ Additionally, since calcium belongs to the $\alpha-$elements { and iron is mainly produced by Type-1 supernovae on relatively longer timescales,}  the flux ratio  \feii/\caii\  can be used as a proxy for estimating the chemical {enrichment} \citep{martinez-aldamaetal15}, such as it has been tested with the UV \feii\ and \mgii\ \citep[][and references therein]{verner2009, dong2011, shin2019, onoue2020}. Therefore a deep observational analysis is required. }
 
  The paper is organized as follows: in Section~\ref{sec:data}, we include a short review of the sample. Section~\ref{sec:measurementes} describes the methods employed to estimate the black hole mass and Eddington ratio. In Section~\ref{sec:results}, we report the observational correlations of our sample, including the Eigenvector 1 sequence and the Baldwin effect. In order to confirm the correlations found, we performed a Principal Component Analysis (PCA), the results of which are shown in Section~\ref{sec:pca}. In Section~\ref{sec:discussions} we discuss the potential drivers of the \cat--\feii\ properties, the Baldwin effect, as well as the \feca\ ratio as a possible metal indicator. Conclusions are summarized in Section~\ref{sec:conclusions}. Throughout this work, we assume a standard cosmological model with $\Omega_{\Lambda}$ = 0.7, $\Omega_{m}$ = 0.3, and H$_0$ = 70 \kms{} Mpc$^{-1}$.

\section{Observational Data} \label{sec:data}

\begin{figure}
    \centering
    \includegraphics[width=0.5\textwidth]{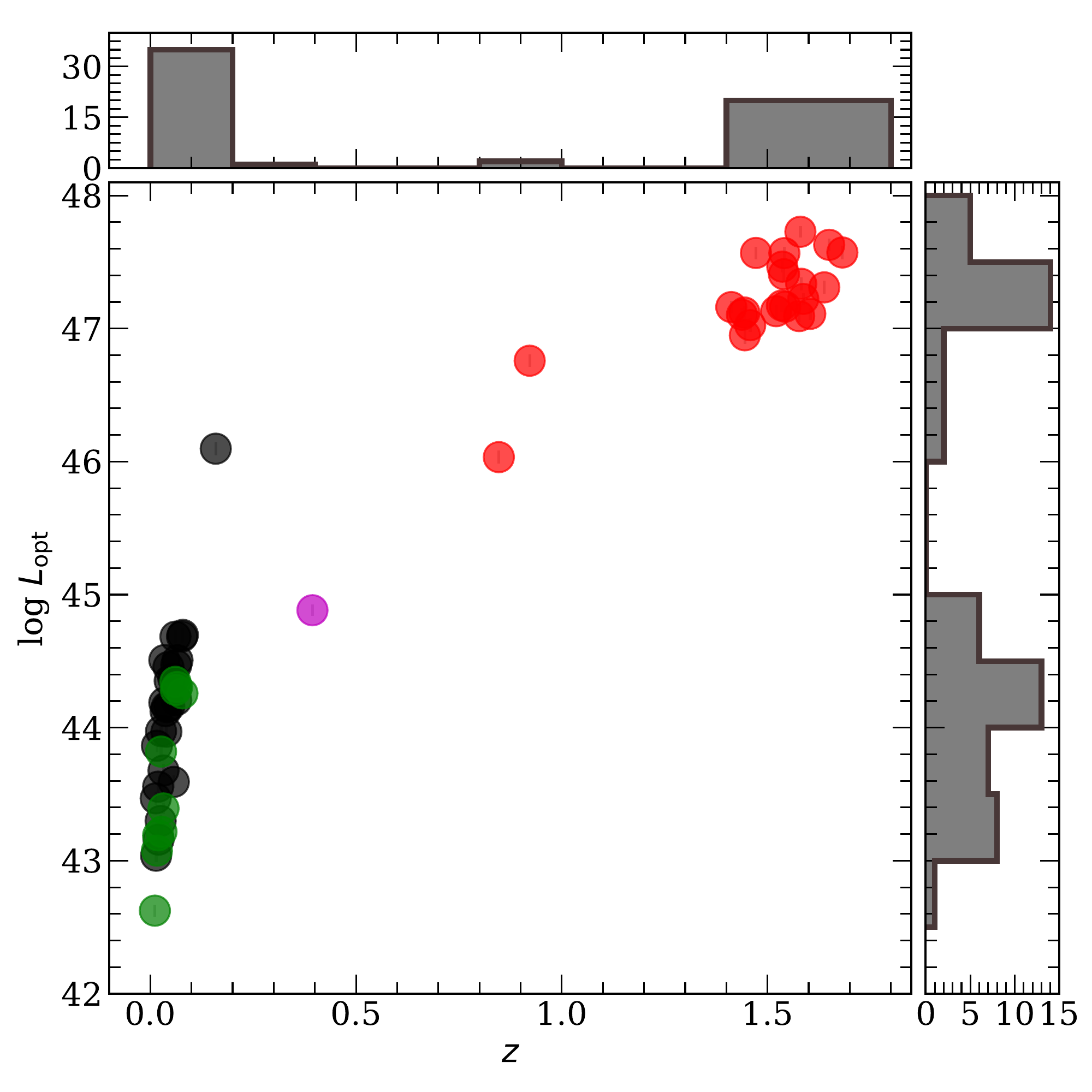}
    \caption{Redshift distribution of the sample as a function of optical luminosity { at 5100\AA} in units of \ergs. Black, red, green and magenta symbols correspond to  \citet{persson1988}, \citet{martinez-aldamaetal15, martinez-aldamaetal15b}, \citet{murilo2016} and \citet{2020arXiv200401811M} samples, respectively. }
    \label{fig:z_Lopt}
\end{figure}

Our analysis is based on the observational properties of \hb, optical FeII (4434--4684 \AA) and  NIR CaII triplet collected from \citet{persson1988}, \citet{ martinez-aldamaetal15, martinez-aldamaetal15b}, \citet{murilo2016} and \citet{2020arXiv200401811M}. { A detailed description of the full sample is discussed in Paper-1.} The full sample includes 58 objects with {$42.5<\mathrm{log}\,L_{\rm opt}\, { (5100\AA)}<47.7$} at $0.01<z<1.68$.  Due to the different selection criteria of the subsamples, the full sample shows a bimodal distribution in redshift and luminosity, where $58\%$ of the sample shows $z<0.1$ and {log~$L_{\rm opt}\sim 44$}, while the rest of the objects are located at $z\sim1.6$ with  log~$L_{\rm opt}\sim 47.4$ (see Figure~\ref{fig:z_Lopt}). Therefore, our sample is affected by  such biases, which could influence our results. These aspects are discussed in Sec.~\ref{sec:results} and Sec.~\ref{sec:discussions}.  

{ The optical measurements from \citet{persson1988} are originally reported by \citet{oster76}, \citet{oster1977}, \citet{koski1978}, \citet{oke1976}, \citet{kunth1979} and  \citet{debruyn}. However, the quality of the data  was not so high like in recent times, therefore, this sample should be treated with caution. There are five sources in common in Marinello's and Persson's samples. The variation in the different observational parameters are significant in three of them (Mrk~335, Mrk~493, and I~Zw~1, see Table~\ref{tab:table1}). This could be an indication of the quality of the measurements. However, Marinello and Persson's samples include typical NLS1y objects, thus a similar behavior is expected, such as Figure~\ref{fig:spectral_properties} shows. In order to disentangle this point, new observations of the Persson sample are needed. }

Table \ref{tab:table1} reports {the properties of the each source in the sample} such as redshift, optical (at 5100\AA) and NIR (at 8542\AA) luminosity, { {the flux ratios \rfe\ and \rcat}, as well as the equivalent width (EW) and Full-Width at Half Maximum (FWHM) of \hb, \cat\ and \oi. All the measurements were taken from the original papers \citep{persson1988,martinez-aldamaetal15, martinez-aldamaetal15b, murilo2016,2020arXiv200401811M}. Since \citep{persson1988} do not report the luminosities at 5100\AA, we have estimated them from their apparent V magnitudes  reported by \citet{veron-cetty2010} catalog. We have considered a zero point flux density of $3.55\times10^{-9}$ \ergscmA{} \citep{bessel1990} to estimate the flux at 5500\AA\ in the observed-frame. After correcting for the redshift, we assumed a slope of $\alpha_\lambda=-1.67$ \citep{vandenberk2001} to estimate the flux at 5100\AA. Finally, the distance to the source was obtained thought classical integration assuming the cosmological parameters specified at the end of Sec.\ref{sec:intro}. }


\section{Parameter Estimations}
\label{sec:measurementes}

\subsection{Black hole mass}

The black hole mass (\mbh) is estimated using the classical relation given by:
\begin{equation}
M\mathrm{_{BH}}=f\mathrm{_{BLR}}\frac{R\mathrm{_{BLR}} \, v^2}{G}
\label{equ:mass}
\end{equation}
where $G$ is the gravitational constant, \fblr\ is the virial factor, $R\mathrm{_{BLR}}$ is the broad line region size and $v$ is the velocity field in the BLR, which is represented by the FWHM of \hb. The virial factor includes information of geometry, kinematics, and inclination angle of the BLR. Typically, it is assumed as constant ($\sim 1$), however some results point out that this factor should vary along the AGN populations \citep[e.g.][]{collin2006,yu2019}. In this work, we assume the virial factor proposed by \citet{mejia-restrepoetal18a}, which is anti-correlated with the FWHM of the emission line: $f\mathrm{_{BLR}}=\left({{\mathrm{FWHM_{H\beta}}}}\,/\,{4550 {\pm 1000}}\right)^{ -1.17}$.

For single--epoch spectra, the \rblr\ is usually estimated through the Radius-Luminosity (RL) \citep{bentz13} given by:
 \begin{align}
\mathrm{log}\left(\frac{R\mathrm{_{BLR}}}{\mathrm{1 lt-day}}\right)=&(1.527\,\pm\,0.31)\, + \nonumber \\ &\,0.533{^{+0.035}_{-0.033}}\, \mathrm{log}\left(\frac{ L_{\rm opt}}{10^{44}L_\odot}\right).
\label{equ:mass_bentz}
\end{align}

where $L_{\rm opt}$ corresponds to the luminosity at 5100\AA. Black hole mass estimations are reported in Table~\ref{tab:table2}. { The sample  shows a clear distinction between low and high black hole masses (log \mbh $\sim7-10$ M$_{\odot}$). }

\subsection{Eddington ratio}

The {accretion rate} is estimated by the classical Eddington ratio defined by \eddr, where \lbol\ is the bolometric luminosity and \ledd\ is the Eddington luminosity defined by $L\mathrm{_{Edd}}=1.5\times10^{38}\left(\frac{M_{BH}}{M_\odot}\right)$. The bolometric luminosity formally can be estimated integrating the area under the broadband spectral energy distribution (SED) \citep[e.g.][and references therein]{richards2006}. However, since this process requires multi-wavelength data to constrain the SED fitting process, it is hard to get an estimation for individuals sources. Mean SEDs have been used to estimate average values called bolometric correction factors (\kbol), which scale the monochromatic luminosity ($\lambda L_\lambda$) to give a rough estimation of \lbol$=k_{\mathrm{bol}} \cdot \lambda L_\lambda$. Usually, \kbol\ is taken as a constant for a monochromatic luminosity; however, results like the well-known non-linear relationship between the UV and X-ray luminosities \citep[e.g.][and references therein]{lusso-risaliti2016} indicate that \kbol\ should be a function of luminosity \citep{marconi2004, krawczyk2013}. Along the same line, \citet{netzer2019} proposed new bolometric correction factors as a function of the luminosity assuming an optically thick and geometrically thin accretion disk, over a large range of black hole mass ($10^{7}$- $10^{10}$ M$_\odot$), Eddington ratios (0.007--0.5), spin (-1--0.998) and a disk inclination angle of $56^\circ$. For the optical range, the bolometric correction factor is given by:
\begin{align}
k_\mathrm{bol}=40\left(\frac{L_{\rm opt}}{10^{42}}\right)^{-0.2},
\label{equ:kbol}
\end{align}
where $L_{\rm opt}$ corresponds to the luminosity at 5100\AA. The wide option of parameters considered for the model process provide a better approximation corroborating previous results \citep{nemmen-brotherton2010, runnoe2012a, runnoe2012b}. In addition, it provides a better accuracy than the constant bolometric factor correction which led to errors as large as $50\%$ for individual measurements. Therefore, we explore the use of \kbol\ for estimating the Eddington ratio. 
Table \ref{tab:table2} reports the { Eddington ratios utilizing the BH masses obtained using the classical RL relation} (Eq.~\ref{equ:mass_bentz}). 


\section{The correlation analysis} \label{sec:results}

\subsection{Observational Pairwise Correlations }
\label{sec:pairwise_correlations}

Figure \ref{fig:spectral_properties} shows the  { correlation matrix of the observational parameters:} optical (\lopt\ at 5100\AA) and NIR (\lnir\ at 8542\AA) continuum luminosities, the flux ratios \rfe\ and \rca\ and the emission lines properties such as FWHM and the equivalent width (EW) of \hb, \oi\ and \cat, plus the EW of \feii{}. In order to stress the difference in luminosity and FWHM values in the subsamples, they are identified by different colors. Each panel also includes the Spearman's rank correlation coefficient ($\rho$) and the $p$-value, 
where significant correlations ($p<0.001$) are colored in red (otherwise shown in black). Optical and NIR luminosities {follow a linear relation (Fig.~\ref{fig:spectral_properties}), therefore} both luminosities show the same behavior with the rest of the observational properties. 

Top panel of Figure~\ref{fig:spectral_properties} shows the strong correlation between \rfe\ and \rcat, which is described by the Eq.~\ref{equ:cafe} { (dashed gray line; see also inset panel)}. { The anti-correlation between \rfe\ (or \rcat) and \ewhb\ is expected since the strength of \hb\ decreases as \rfe\ (or \rcat) increases. The linear correlation between \rfe\ and \ewfe\ is due to the fact that both parameters reflects the strength of the \feii\ emission, the first one is weighted by the \hb\ flux and the second by the luminosity. It is the same case for the correlation  \ewca-\rcat. Since \rcat\ is correlated with \rfe, we expect a positive linear relation between \ewca-\rfe\ and \ewfe-\rcat.} On the other hand, \rfe\ and \rcat\ show non-linear trends with the FWHM of the emission lines, which are { further discussed} in Section~\ref{sec:eigenvector}. The correlation between the EW and the continuum luminosity are extensively described in Sec.~\ref{sec:ew-L} and \ref{sec:beff}.

The correlations between the FWHM of \hb, \oi\ and \cat\ are strongest according to their Spearman coefficients and {their associated }$p-$values (Fig.~\ref{fig:spectral_properties}). 
In these panels, the 1:1 line is shown for reference. \hb\ shows broader profiles than the \oi, particularly for the sources with FWHM $>4000$ \kms. {We obtained the trend lines by least squares (OLS) fitting
 implemented in python packages \textmyfont{sklearn} and \textmyfont{statsmodels}}. The relation has a slope of {0.894$\pm$0.05 and a scatter of $\sigma_{\rm rms}\sim$0.115 dex. The deviation at 4000 \kms\ could be associated with the presence of a red-ward asymmetry in the broadest \hb\ profiles, i.e. with an emitting region closer to the continuum source \citep{marziani13, punsly2020}. The presence of this feature is hard to observe in the \oi\ profile, since it is blended with the \cat\ and the NIR \feii. On the other hand, \cat\ is also narrower than \hb, although the scatter is larger ($\sigma_{\rm rms}\sim$0.152 dex) and the relation is slightly shallower than the one given by \oi\ with a slope of 0.827$\pm$0.08. \oi\ and \caii\ show similar widths, the predicted relation gives a slope of 0.944$\pm$0.05 and the scatter is smaller ($\sigma_{\rm rms}\sim$ 0.103 dex)} than in the previous cases. This general behavior corroborates that \hb\ is emitted closer to the continuum source than \oi\ and \cat\ \citep{persson1988, martinez-aldamaetal15, murilo2016}. 


\subsection{Eigenvector 1 sequence}
\label{sec:eigenvector}

The correlation between \rfe\ and the FWHM of \hb\ is known as the Eigenvector 1 (EV1) sequence \citep{borosongreen1992}, which is also known as the quasar main sequence \citep{sulentic2000}. According to the EV1 scheme, the observational and physical properties of type 1 AGN change along the sequence \citep{marziani2018,panda18b,panda19b}. Based on the \rfe\ strength the accretion rate  can be inferred, where the sources with \rfe$>$1 are typically associated with the highest  Eddington ratios {  \citep[\eddr$>$0.2, ][]{marziani2003, panda19b, panda19c}}. The relation between these parameters is not linear \citep{wildy2019}, where orientation and luminosity are also involved \citep{shen-ho2014, negreteetal18}

\begin{figure*}
    \centering
    \includegraphics[width=\textwidth]{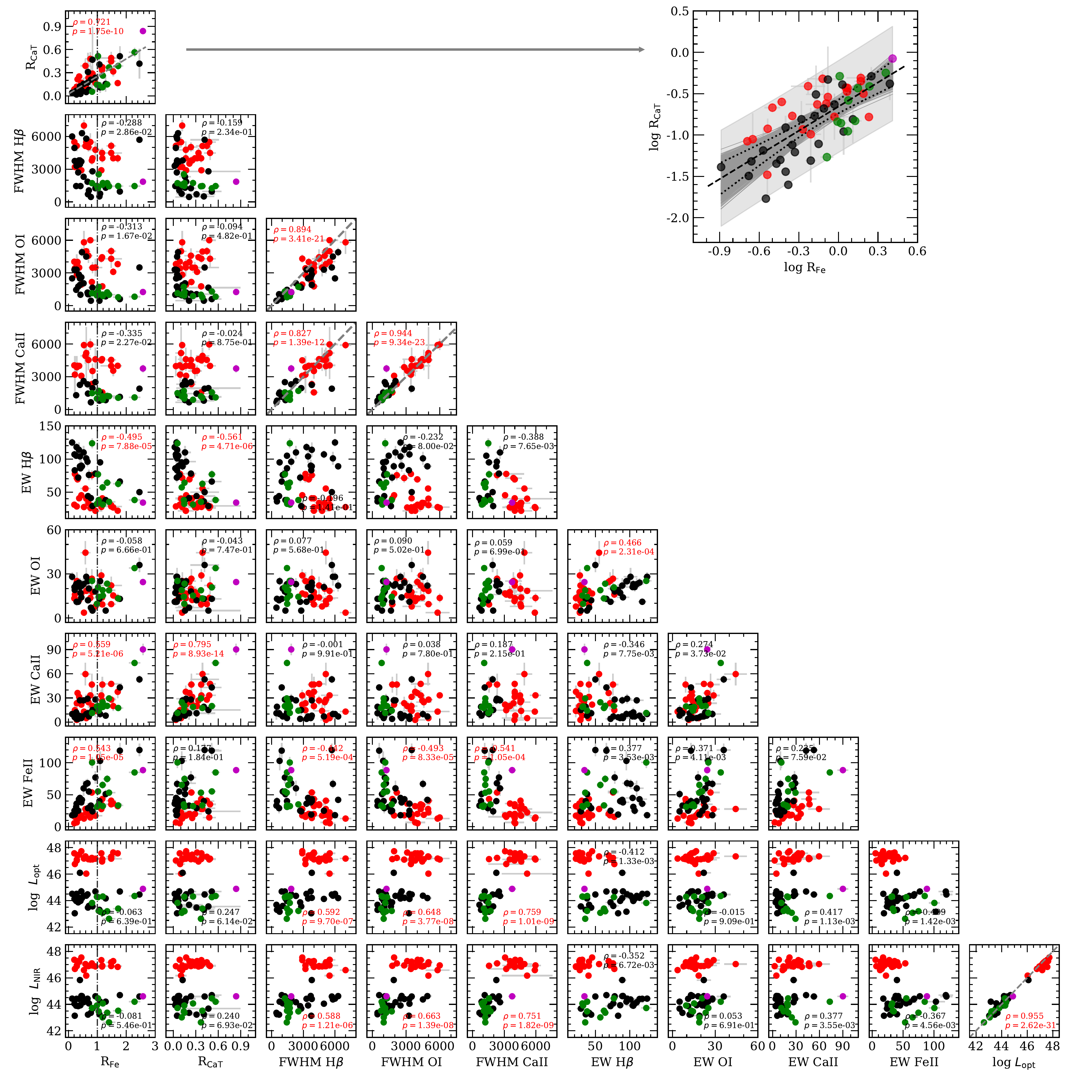}
    \caption{{ Correlation matrix} for emission lines and continuum properties. Black, red, green and magenta symbols correspond to  \citet{persson1988},  \citet{martinez-aldamaetal15, martinez-aldamaetal15b}, \citet{murilo2016} and \citet{2020arXiv200401811M} samples, respectively. Each panel  specifies the Spearman's rank correlation coefficient and the $p-$value, where the significant correlations are colored in red. Equivalent width (EW), FWHM, optical and NIR luminosities are given in units of \AA, \kms\ and \ergs, respectively. Gray vertical-lines in the first column mark the limit for super-Eddington sources at \rfe=1.0. In the diagrams where the FWHMs and luminosities are correlated, the gray dashed line marks the relation 1:1. { In the correlation \rcat-\rfe\, gray dashed line corresponds to Eq.~\eqref{equ:cafe}.  The inset right panel shows the relation \rfe-\rcat\ in log--scale with the results of the bootstrap analysis, see Sec.~\ref{sec:boots}. Black dotted lines mark the confidence intervals at 95\% for the 1000 realizations (dark gray lines) of the bootstrap analysis. Lightgray patch marks the corresponding prediction intervals bands for the sample. }}
    \label{fig:spectral_properties}
\end{figure*}

{ The EV1 sequence (\fwhmhb-\rfe\  relation) of our sample is shown in  Figure~\ref{fig:spectral_properties}.}  A displacement between the low-- and high--luminosity objects (HE sample) can be appreciated, however, both kinds of sources follow the same trend. This displacement is only a luminosity effect, where the HE sample is shifted to the larger FWHM values of the panel. An EV1-like sequence is also appreciated { in the relations  \rfe-\fwhmca\ and \rfe-\fwhmoi},
  which is expected due to the linear relation between the widths of the emission lines (Sec.~\ref{sec:pairwise_correlations}). 



{ The relation \fwhmhb-\rcat\ shows an EV1 sequence-like for the low-luminosity subsample, but  it is not appreciated in the high luminosity objects, the HE sample.} It seems that in some objects the \cat\ increases with increasing FWHM$_{\rm H\beta}$. The same effect is observed for the { relations \fwhmoi-\rcat\ and \fwhmca-\rcat}.  Surprisingly, the break occurs at \rcat$\sim0.2$ which corresponds to \rfe=1 (following the Eq.~\ref{equ:cafe}), the limit for the highly accreting sources according to \citet{marziani2003}. A similar decoupling is also observed in the relation between the \ewca\ and the FWHM of the emission lines, but the scatter is quite large. \citet{martinez-aldamaetal15} found a rough enhancement of \ewca\ for the HE sample at intermediate--$z$ with respect to the other objects at low-$z$ attributing this behavior to a burst of star formation and an enrichment at intermediate redshift sources. The new HE objects \citep{martinez-aldamaetal15b} added to the presented analysis seem to corroborate these results, however, some selection effect could also be involved. We discuss this result in Sec.~\ref{sec:star_formation}.


\subsection{Correlations with the equivalent width: the Baldwin effect}
\label{sec:ew-L}

The anti-correlation between the equivalent width and the luminosity is known as the {\it global} Baldwin effect (BEff) \citep[][]{osmer1999, green2001, baskin2004, bachev2004, dong2009, zamfiretal10}, which was first observed between the equivalent width of \civ\ and the continuum luminosity at 1450\AA\ \citep{baldwin1977}. {The BEff is clearly appreciated in the high--ionization lines, except NV~$\lambda1240$ due to a second enrichment \citep{osmer1994}. However, as the ionization potential decreases the slope of this anti-correlation gets shallower and it is hard to distinguish a strong correlation for low-ionization lines \citep{sulentic2000, dietrich02}} An {\it intrinsic} Baldwin effect \citep{poggeandpeterson1992} has been also detected in the multi-epoch observations for single variable AGNs. The BEff provides information about the ionizing continuum shape \citep{wandel1999}, structure, and metallicity \citep{korista1998} of the BLR. Also, it has been used for calibrating the luminosity in cosmological studies \citep{baldwin1978, korista1998}. 

\subsubsection{Luminosity}

Figure~\ref{fig:ew_ac} shows the equivalent widths of \hb, \oi, optical \feii\ and { \cat} as a function of the optical and NIR luminosities. 
{ Spearman rank correlation coefficients, $p-$values and the scatter of the correlations are reported in Table~\ref{tab:params_corr}. None of the trends between the EW and the optical and NIR luminosity satisfies the criteria for a significant relation and all of them show shallow relations with a slope of $\alpha<0.1$. The shallow slope confirms the weak relation between the luminosity and the EW for low ionization lines. This result is in agreement with larger samples at high redshift  \citep[][]{dietrich02}.}
{ The negative correlation between  \ewhb\ and $L_{\rm opt}$  ($\alpha=-0.064\pm0.032, \, \rho=-0.412$, $p=0.001$) is expected due to the behavior of individual variable sources \citep[e.g.][]{rakic2017}. This behavior is different {from the one of the} relation  \ewca-\lopt\  ($\alpha=0.068\pm0.053$, $\rho$=0.417 $p=0.001$), which suggests the presence of an  inverse Baldwin effect}. A positive correlation has also been observed between the continuum at 5100\AA\ and the optical \feii\ in the monitoring of the variable NLSy1 NGC~4051 \citep{wang2005}. The strong correlation between the \feii\ and \cat\ explain this behavior. However, in our sample the relation \ewfe-\lopt\ is negative { ($\alpha=-0.091\pm0.039$, $\rho=-0.409,\, p=0.001$)} and is just below the criteria assumed to consider a significant correlation. Other studies neither reported a BEff for optical nor for UV \feii\   \citep{dong2011}. Finally, the trend observed for \ewoi-\lopt\ is not significant and show a slope consistent with zero ($\alpha = -0.007 \pm0.034$), also confirmed by  previous studies \citep{dietrich02}.

\subsubsection{Black hole mass and Eddington ratio}

\begin{figure*}
    \centering
    \includegraphics[width=1\textwidth]{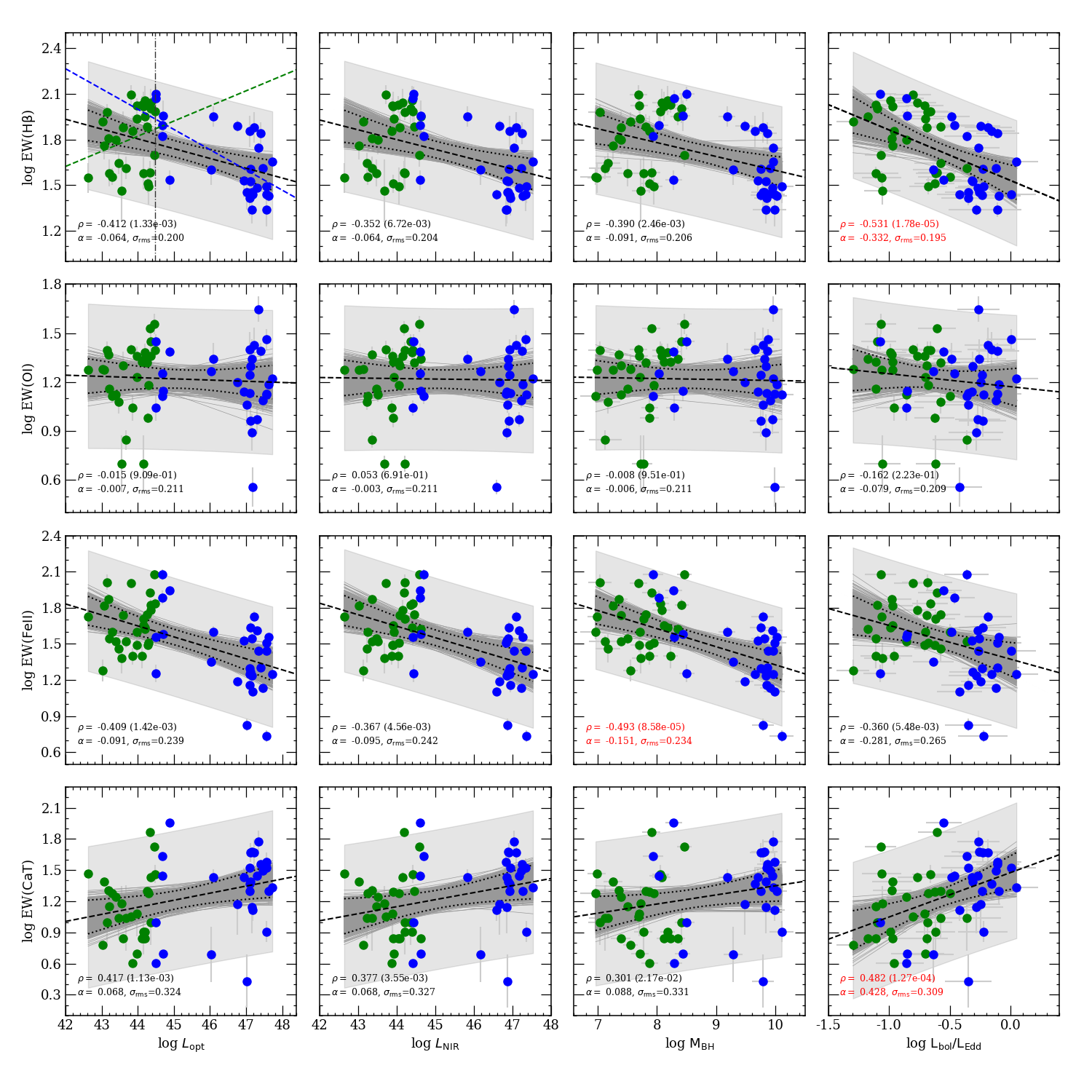}
    \caption{{ Correlation matrix for the optical and NIR luminosities, black hole and the equivalent width of \hb, optical \feii, \oi\ and { \cat}. Green and blue symbols indicate the low-- and high--luminosity subsamples, respectively (Sec~\ref{sec:division_sample}). Black dashed line represents the best OLS fitting for the full sample. The Spearman's rank correlation coefficient, $p-$values, slope ($\alpha$), and scatter ($\sigma_{\rm rms}$) are also shown, where the significant correlations are colored in red.  Black dotted lines mark the confidence intervals at 95\% for the 1000 realizations (dark gray lines) of the bootstrap analysis. Lightgray patch marks the corresponding prediction intervals bands for the sample.}}
    \label{fig:ew_ac}
\end{figure*}

Since the black hole mass and the Eddington ratio have been considered as the main drivers of the BEff \citep{wandel1999, dong2011}, we also present the correlations { EW-\mbh\ and EW-\eddr\ in Fig.~\ref{fig:ew_ac}. { The parameters of the correlations are reported in Table~\ref{tab:params_corr}}. The only significant relation involving the black hole mass is  \ewfe-\mbh\  ($\rho=-0.493$, $\alpha=-0.151\pm0.060$ $\sigma_{\rm rms}\sim0.234$ dex).} 

In the correlations between the equivalent width and the Eddington ratio, the significant relations are the ones involving \ewhb\ and {\ewca}. In both cases the correlations are sharper ($\alpha_{\rm H\beta}=-0.332\pm0.149$, $\alpha_{\rm  CaT}=0.428\pm0.237$) and stronger ($\rho_{\rm H\beta}=0.531$, $\rho_{\rm CaT}=0.482$) than the luminosity case. Although the correlations for \feii\ and \oi\ are below the significance level, their slopes are steeper than the correlations with respect to the luminosities and the black hole mass. Hence, the Eddington ratio highlights the correlations  with the equivalent width, as  \citet{baskin2004} and \citet{dong2011}  previously reported.

\subsubsection{Division of the sample}
\label{sec:division_sample}

According to \citet{dietrich02} to avoid selection effects in the global BEff a sample with a wide luminosity range is needed, $42<{\rm log }\,L<48$. Our sample covers this range, however at high redshift only high luminosity sources are available (${\rm log }\, L_{\rm opt} > 47.5$ \ergs). In order to clarify the results of Sec.~\ref{sec:ew-L} and the presence of possible bias, we divided the sample into two subsamples considering the median luminosity, log $L_{\rm opt}=44.49$ \ergs. In Fig.~\ref{fig:ew_ac} the low-- and high--luminosity subsamples are represented by green and blue symbols, respectively. 
The division of the sample directly affects the relations EW$-L_{\rm opt}$, EW$-L_{\rm NIR}$ and EW$-{\rm M}_{\rm BH}$ where no significant correlations are observed, which also reflects the bias involved in this consideration. For example, the relation \ewhb$-L_{\rm opt}$ is positive for the low-luminosity subsample ($\alpha=0.1\pm0.073$, $\rho=0.235$, $p=0.281$),  while for the the high luminosity case the relation has a different direction ($\alpha=-0.13\pm0.031$, $\rho=-0.53$,  $p=0.003$), similar to the behavior of the full sample. {The difference in the subsamples is also pointed out by the PCA (Appendix~\ref{appendix:pca_sub}).} Therefore, the correlations have some relevance only when the full sample is considered . 

{ However, the correlations EW--\eddr\ are less affected by the division of the sample, at least, in the significant correlations provided by the full sample analysis. In the relation \ewhb--\eddr, the direction of the best fits in the subsamples is still negative ($\alpha_{\rm low}=-0.215\pm0.162, \alpha_{\rm high}=-0.472\pm0.144$), although none of the relations are significant ({$\rho_{\rm low}=-0.26, \, p_{\rm low}=0.181, \, \rho_{\rm high}=-0.35, \, p_{\rm high}=0.07$}). While in the \ewca--\eddr\ relation, the slope for the subsamples is positive ($\alpha_{\rm low}=0.240\pm0.158, \, \alpha_{\rm high}=0.694\pm0.242$) such as in the full sample, but without any significance ($\rho_{\rm low}=0.222, \, p_{\rm low}=0.245, \, \rho_{\rm high}=0.313, \, p_{\rm high}=0.098 $)}. This result suggests that \eddr\ is less influenced by a bias and it then regulates the correlation between the equivalent width and the luminosity, {as originally suggested by \citet{baskin2004} and \citet{bachev2004}}.



\subsection{The behavior of \rfe, \rcat\ and the ratio \feii/\cat}


Fig.~\ref{fig:rfe-rcat_corr} shows the behavior of \rfe, \rcat\ and the ratio \feca\ as a function of optical and NIR luminosity, black hole mass and Eddington ratio. 
{ \rfe\ and \rcat\ do not show any significant correlation with the luminosity and black hole mass for the full sample. Only the \feii/\cat\  shows a significant anti-correlation with the optical ($\rho=-0.441, \, p=5.3\times10^{-4}$) and NIR luminosity ($\rho=-0.456, \, p=3.2\times10^{-4}$). If the subsamples are considered, all the best fits are below the statistical significance limit.  }

In the panels where \eddr\ is involved, the strongest correlation is the one corresponding to the ratio \feca\ ($\rho=0.554$, $p=6.4\times 10^{-6}$), followed by the one with \rcat\ ($\rho=$0.425, $p=8.9\times 10^{-4}$). In both cases the trend lines for the full samples and subsamples have the same direction such as in Fig.~\ref{fig:ew_ac}, although the significant correlation arises only for the full sample. 

{ The positive correlation between \rcat\ and \eddr\ confirm that the strength of \cat\ is driven by the accretion rate, and it remains even after the division of the sample. Although the same behavior is expected for \rfe, we cannot confirm this result in our sample.}
The positive correlation between the optical and UV \rfe\ and \eddr\ is robust  \citep[e.g][]{zamfiretal10, dong2011, martinezaldama2020}. Besides, the \rfe\ has been used as a proxy for the \eddr\ to correct the time delay by the accretion effect and decrease the scatter in the optical and UV Radius-Luminosity relation \citep{du2019, martinezaldama2020}. This indicates that the \feii\ (and \rfe) in our sample is affected by several factors: the sample size, the quality of the observations and the \feii\ templates employed, which could decrease the accuracy of its estimation. For instance, 10 objects from the \citet{persson1988} sample have only upper limits, such as Mrk~335. It is one of the five common sources observed by \citet{murilo2016}, where the \rfe\ value is $\sim50\%$ higher than the value estimated by \citet{persson1988}. It confirms that the objects with upper values are highly inaccurate and thus it could be reflected in the loss of the correlation with other parameters. A homogeneous fitting process considering the same analysis spectral procedure could help to decrease the scatter and clarify the trends which we aim to pursue in a future work.


\subsection{Bootstrap analysis}
\label{sec:boots}

\subsubsection{Random distributions}
\label{sec:random}

{ Which is the probability that an uncorrelated data set gives a correlation with Spearman rank coefficient as high as the one we observe? In order to answer this question, we modeled the distributions of each of the parameters in Fig.~\ref{fig:ew_ac} and  \ref{fig:rfe-rcat_corr} considering a random sample of 1000 elements and the probability distributions implemented in the module \textmyfont{stats} in python. To determine how good of a fit this distribution is, we used the Kolmogorov-Smirnov test which compare a sample with a reference probability distribution and we chose the distribution with the highest $p-$value$_{\rm ran}$. Since the luminosity and the black hole mass show bimodal distributions, we used two distributions to reproduce the observational one. In the rest of the cases a good fit was obtained with only one  distribution. The probability distributions considered  and the $p-$value are reported in column (11) and (12) of Table~\ref{tab:params_corr}. The distribution fitting of the correlation \eddr-\feca\ is shown in Fig.~\ref{fig:dist_ran} as an example of this analysis. }

{ Later, we randomly selected 58 realizations from the original 1000,  which later we correlated following the correlations in Figures \ref{fig:ew_ac} and \ref{fig:rfe-rcat_corr}, and in the correlation \rfe-\rcat.  Finally, we repeated the process 2000 times and estimated the Spearman rank correlation coefficient ($\rho_{\rm ran}$), the corresponding $p-$value and estimated the fractions of significant correlations ($f_{\rm ran}$). Results are reported in column (13) of Table~\ref{tab:params_corr}. In all the cases $f_{\rm ran}<1\times10^{-3}$, it means that is very unlikely at $3\sigma$ confidence level that two independent correlations provide high correlations coefficients such as the observational sample does. Therefore, our analysis supports the reliability of the observed correlations.}

\subsubsection{Linear regression fitting}

Due to the small size of our sample and the gaps in luminosity and redshift, we proved the statistical reliability of the correlations in Fig.~\ref{fig:ew_ac} and \ref{fig:rfe-rcat_corr} via a bootstrap analysis \citep{efron93}. The bootstrap sample is formed by the selection of a subset of pairs from each one of the correlations by random resampling with replacement. We created 1000 realizations and then performed a linear regression fitting. The gray lines in Fig.~\ref{fig:spectral_properties} and \ref{fig:ew_ac}   correspond to the 1000 realizations, which are in agreement with the best fit of each correlation at 2$\sigma$ confidence level (dotted black lines). As a reference, the figures also shows the prediction intervals bands  (lightgray patch), which indicate the variation of the individuals measurements and predict that 95$\%$ of the individual point lies within the patch. As a reference, we also analyzed the relation \rfe-\rfe\ (inset panel, Fig.~\ref{fig:spectral_properties}) to compare the behavior of the bootstrap analysis in a very well-know correlation, obtaining an agreement within $2\sigma$ confidence level.

In order to quantify the bootstrap results, we considered the percentiles at $2\sigma$ confidence level and estimated the errors of the slope ($\alpha_{\rm BS}$) and ordinate ($\beta_{\rm BS}$) of the normal distribution drawn from the 1000 realizations for each correlation. Results are reported in  Table~\ref{tab:params_corr}. As it is expected the distributions are centered in the slope and ordinates values of each correlations, which are completely equivalent to the ones from the observational correlations. The magnitude of the errors indicates the reliability of the correlation. The larger errors are associated with the correlations below the significance criteria ($-0.4<\rho_{\rm BS}<0.4$, $p>0.001$). A clear example are the errors in the slope of the relations involving \oi\ (Fig.~\ref{fig:ew_ac}) or \rfe\ (Fig.~\ref{fig:rfe-rcat_corr}), which indicates the inaccuracy of the results, such as the Spearman correlation coefficient shows. Meanwhile, good correlations, such as \rfe-\rcat, will show errors $<20\%$. As the correlation coefficients indicate, the errors decrease considerably in the correlations where \eddr\ is involved. This result points out the relevance of \eddr\ in the behavior of our sample and its role in the Baldwin effect.

On the other hand, we also estimated the Spearman correlation coefficient (${\rho_{\rm BS}}$) for the 1000 realizations and estimated the fraction of significant realizations respect to the total number ($f_{\rm sig}$), which satisfy the significance criteria ($|\rho|>0.4$ and $p<0.001$). We also modeled the distribution of ${\rho_{\rm BS}}$ using a skewnorm distribution and estimated the error at $2\sigma$ confidence level. The maximum of the ${\rho_{\rm BS}}$ distribution and  $f_{\rm sig}$ are reported in columns (9) and (10) of Table \ref{tab:params_corr}. In the strongest correlation of the sample, \rfe-\rcat, we obtained $f_{\rm sig}$=1. It means that the 1000 bootstrap realizations satisfy the significant criteria and confirm the reliability of the correlation. This is also highlighted by the errors of ${\rho_{\rm BS}}$, where the correlation remains significant within the uncertainty range. In the correlations with $|\rho_{\rm BS}|>0.5$, $f_{\rm sig}>0.75$ indicating a reliable correlations. However, if the errors of ${\rho_{\rm BS}}$ are considered, there is an small possibility to dismiss the significance of the correlation. It can be expressed by the parameter 1-$f_{\rm sig}$, which expresses the probability of failing to detect a correlation.  Thus,  there is probability of $<25\%$ to detect a false positive correlation in  \ewhb--\eddr\ and \feca--\eddr. If $|\rho_{\rm BS}|= 0.4 - 0.5$, great care should be considered because the probability to detect a false positive correlation increases to $(1-f_{\rm sig})\sim50\%$. It is the case of the correlations such as  \ewca-\eddr\ or \feca-$L_{\rm opt}$. The same interpretation of false positive probability applies in the case of no detected correlation in the observed or bootstrap samples ($\rho < 0.4$), when the probabilities are always low, particularly for the weakest correlations $(1-f_{\rm sig}>80\%)$.




\subsection{Residuals behavior}
\label{sec:residuals}

\begin{figure*}
    \centering
    \includegraphics[width=1\textwidth]{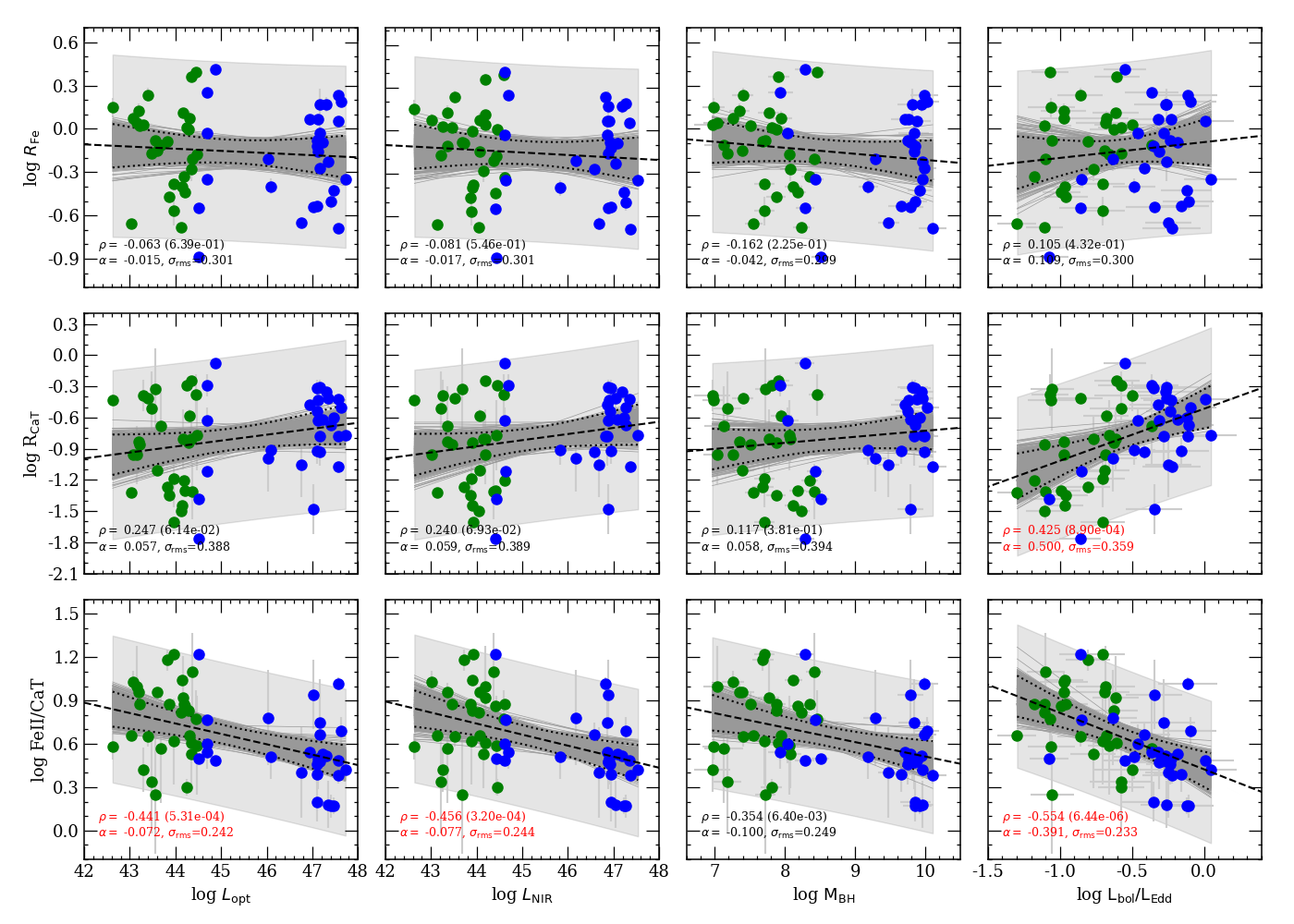}
    \caption{{ Correlation matrix for optical and NIR luminosity, black hole mass, Eddington ratio (\eddr), \rfe, \rcat\ and \feca{} ratio. Colors and symbols are the same as Fig.\ref{fig:ew_ac}.}}
    \label{fig:rfe-rcat_corr}
\end{figure*}

{ In order to assess a possible redshift effect in our results, we estimated the residuals with respect to the best fit for the correlations in Fig.~\ref{fig:ew_ac} and Fig.~\ref{fig:rfe-rcat_corr}. We divided the sample into low-- and high--$L$ subsamples, which is equivalent to a division into low-- and high--redshift. The behavior of the distributions is shown in Fig.~\ref{fig:residual_fig3} and \ref{fig:residual_fig4}. If any significant difference of the median of the distribution with respect to the zero residual level is observed, it could indicated a redshift effect. In all the correlations of Fig.~\ref{fig:residual_fig3}, we observed a difference within  $2\sigma$ confidence level. On the other hand, the relations of Fig.~\ref{fig:rfe-rcat_corr} show the same behavior, however the width of the distribution increases significantly as well as the median values,  particularly for the correlations involving \rfe. Since this behavior is only observed in these correlations and they still show a dependency within $2\sigma$ level, we cannot claim a redshift effect. As we mentioned previously, the \rfe\ is not well behaved in our sample compared to previous results. Therefore any trend involving \rfe\ must be taken with caution.}



\section{Principal Component analysis}
\label{sec:pca}

Principal component analysis (hereafter PCA) allows to get a \textit{better view} of the data where it can be separated in a quantitative manner, such that the relevant properties explain the maximum amount of variability in the dataset. PCA works by initially finding the principal axis along which the variance in the multidimensional space (corresponding to all recorded properties) is maximized. This axis is known as eigenvector 1. Subsequent orthogonal eigenvectors, in order of decreasing variance along their respective directions, are found, until the entire parameter space is spanned \citep[see, for example,][]{borosongreen1992, kur09,wildy2019,tripathietal2020}. The PCA method is particularly useful when the variables within the data set are highly correlated. Correlation indicates that there is redundancy in the data. Due to this redundancy, PCA can be used to reduce the original variables into a smaller number of new variables (principal components) explaining most of the variance in the original variables. This allows to determine correlated parameters and, in the context of our work, we utilize this technique to determine the physical parameter(s) that lead to the correlations illustrated in Fig.~\ref{fig:ew_ac} and \ref{fig:rfe-rcat_corr}.

Eigenvalues (or loadings) can be used to determine the {numbered  principal components} to retain after PCA \citep{kaiser1961}: (a) An eigenvalue $>$ 1 indicates that principal components (PCs) account for more variance than accounted by one of the original variables in standardized data. This is commonly used as a cutoff point for which PCs are retained. This holds true only when the data are standardized, i.e., they are scaled to have 
standard deviation one and mean zero; and (b) one can also limit the number of component that accounts for a certain fraction of the total variance. Since there is no well-accepted way to decide how many principal components are enough, in this work we evaluate this using the \textit{scree plot} (see, for example, Figure \ref{fig:pca2}), which is the plot of eigenvalues ordered from largest to the smallest. A scree plot shows the variances (in percentages) against the {numbered} principal component, and allows visualizing the number of significant principal components in the data. The number of components is determined at the point beyond which the remaining eigenvalues are all relatively small and of comparable size \citep{PERESNETO2005974,Jolliffe2011}. This helps to realize whether the given data set is governed by a single or more-dimensions, where a dimension refers to a variable. Subsequently, these principal components are investigated against the original variables of the dataset to extract information of the importance of these variables in each principal component.

We consider the 58 sources in our sample and collect the properties that are common among them. Due to the diversity in the studied subsamples, we only have 12 parameters that are obtained/estimated directly from the observation: $z$, optical and NIR luminosity, \rfe, \rcat, FWHM and EW of \hb, \oi\ and \cat, as well as the EW of \feii. Among these 12 parameters, the redshift and the optical luminosities are correlated by a bias effect as shown in Figure \ref{fig:z_Lopt} and hence we drop the redshift and only choose the optical luminosity. Thus we have a 11 parameters that are considered in the initial PCA.  Later, we only adopt the NIR luminosity which is equivalent to the optical (Fig.~\ref{fig:spectral_properties}). We refer the readers to Appendix \ref{appendix-pca} for more details on the initial PCA tests, a note on the effect redundant parameters play in this analysis and the final set of parameters used to infer the observed correlations.

Next, we have the derived parameters - bolometric luminosity (\lbol{}), black hole mass (\mbh{}) and Eddington ratio (\eddr{}),
which are predicted using one or more of the observed parameters that are already taken into account for the PCA run. PCA is an orthogonal linear transformation technique applied to the data and assumes that the input data contains linearly independent variables such that the eigenvectors can be represented as a sum of linear combination of variables with associated weights (eigenvalues or loadings) corresponding to each variable. Thus, in order to remove any redundancy in the result obtained via the PCA, one needs to make sure that the parameters that go in as input are not scaled up version of another parameter, thereby saving us the trouble of unwanted bias that comes out of it. The effect of the inclusion of derived variables in the PCA is illustrated in Appendix \ref{appendix-pca}.

Similar to \citet{wildy2019}, we use the \textmyfont{prcomp} module in the \textit{R} statistical programming software. In addition to \textmyfont{prcomp}, we use the \textmyfont{factoextra\footnote{\href{https://cloud.r-project.org/web/packages/factoextra/index.html}{https://cloud.r-project.org/web/packages/factoextra/index.html}}} package for visualizing the multivariate data at hand, especially to extract and visualize the eigenvalues/variances of dimensions.

\subsection{PCA on the full sample}

\begin{figure*}
    \centering
    \includegraphics[width=2\columnwidth]{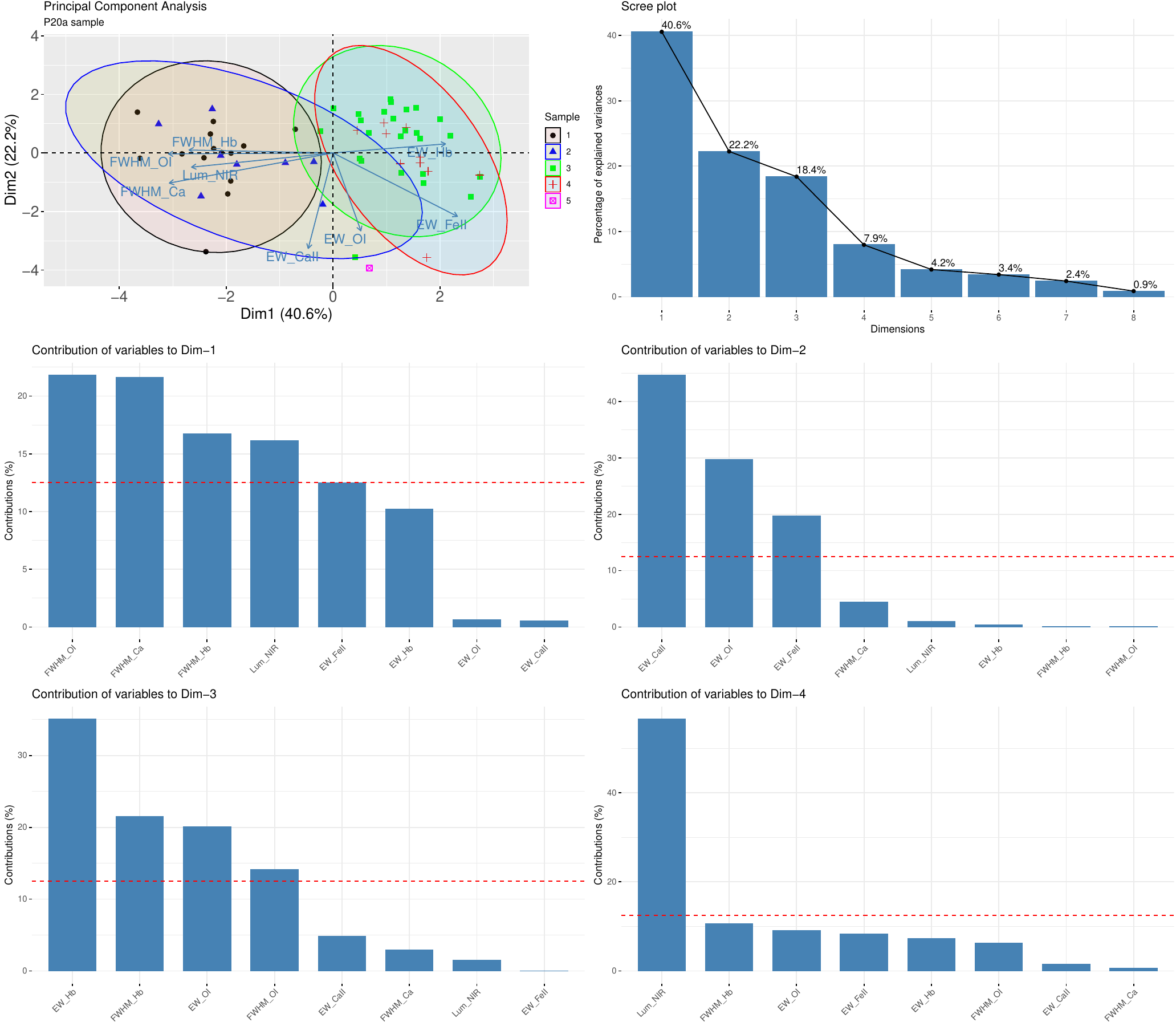}
    \caption{Graphical representation (factor-map) of the principal component analysis (PCA) decomposition of our sample (58 sources) is shown in the first panel. The dots represent individual objects on the standardized PC1 - PC2 plane that have variables indicated by the axis labels. The arrows represent the prominence of the variables in the PC1 - PC2 plane. The dashed lines mark the coordinate axes in the PC1 - PC2 plane and the ellipses depict the 95\% occupancy of the sources in their respective subsamples. The sample is categorized based on their original source catalogues \citep[see][for details on the observational sample]{panda_cafe1} - (1) \citet{martinez-aldamaetal15}; (2) \citet{martinez-aldamaetal15b}; (3) \citet{persson1988}; (4) \citet{murilo2016}; and (5) PHL1092 \citep{2020arXiv200401811M}. 
    The other panels illustrate the precedence of the first 10 principal components in the form of scree plots, followed by the contributions of the original variables to the first four principal components.}
    \label{fig:pca2}
\end{figure*}

The tests aimed to reduce the redundancy of the variables were applied as described in (see appendix \ref{appendix-pca}) and now allow us to perform a final PCA run with the dataset that contain variables which are obtained directly from the observations and have as little redundancy as possible. The final input contains 8 variables, namely, the NIR luminosity (at 8542~\AA), the EWs of \feii{}, \hb{}, \oi{} and \cat{}, and, the FWHMs of \hb{}, \oi{} and \cat{}. The result of the PCA is illustrated in Figure \ref{fig:pca2}.

In this section, we present the results of the PCA on the full sample and infer the relative importance of the \textit{eigenvectors} as a function of the input variables. {In Appendix \ref{appendix:pca_sub} is described the PCA analysis for the low- and high-luminosities samples described in Sec.~\ref{sec:division_sample}. Figure \ref{fig:pca2}} shows the two-dimensional factor-map between the first two principal components, scree plot and the contributions of the input variables to the first four principal components for the full sample.  The factor-map shows the distribution of the full sample categorically colored to highlight the different sources (see Sec. \ref{sec:data}) in the eigen-space represented by the two principal components, Dim-1 and Dim-2 (i.e., the PC1 and PC2). The first and the second PC contribute 40.6\% and 22.2\% to the overall dispersion in the dataset. Combining the contributions from the two subsequent PCs (PC3 and PC4), one can explain 89.1\% of the variation present in the data. We demonstrate the contributions of the original variables on these four principal components to draw conclusions on the primary driver(s) of the dataset.

\textit{First principal component}: From the factor-map we find that the vectors corresponding to the FWHMs of \hb{}, \oi{}, \cat{} and the NIR luminosity are nearly co-aligned, with the FWHM vectors of \hb{} and \oi{} having almost similar orientation. These four vectors are also the major contributors to the variance along the primary principal component (see third panel on the left of Figure \ref{fig:pca2}). The red dashed line on the graph above indicates the expected average contribution. If the contribution of the variables were uniform, the expected value would be 1/length(variables) = 1/8 $\approx$ 12.5\%. For a given component, a variable with a contribution larger than this cutoff could be considered as important in contributing to the component. The \ewfe\ is barely over this cutoff limit.

\textit{Second principal component}: The factor-map highlights the prevalence of the \ewca\ which contributes $\sim$45\% to this component, followed by the \ewoi\ and \ewfe. The overall contribution accounts for $\sim$95\% from these three variables.

\textit{Third and fourth principal components}: The third PC is dominated by the \ewhb\ with a minor contribution from \fwhmhb, \ewoi\ and FWHM$_{\rm OI}$. Whereas, the fourth PC is singularly governed by NIR luminosity. The other variables are below the expected average contribution limit.



\subsection{Correlations between the principal eigenvectors and observed/derived parameters}

\begin{figure*}
    \centering
    \includegraphics[width=2.1\columnwidth]{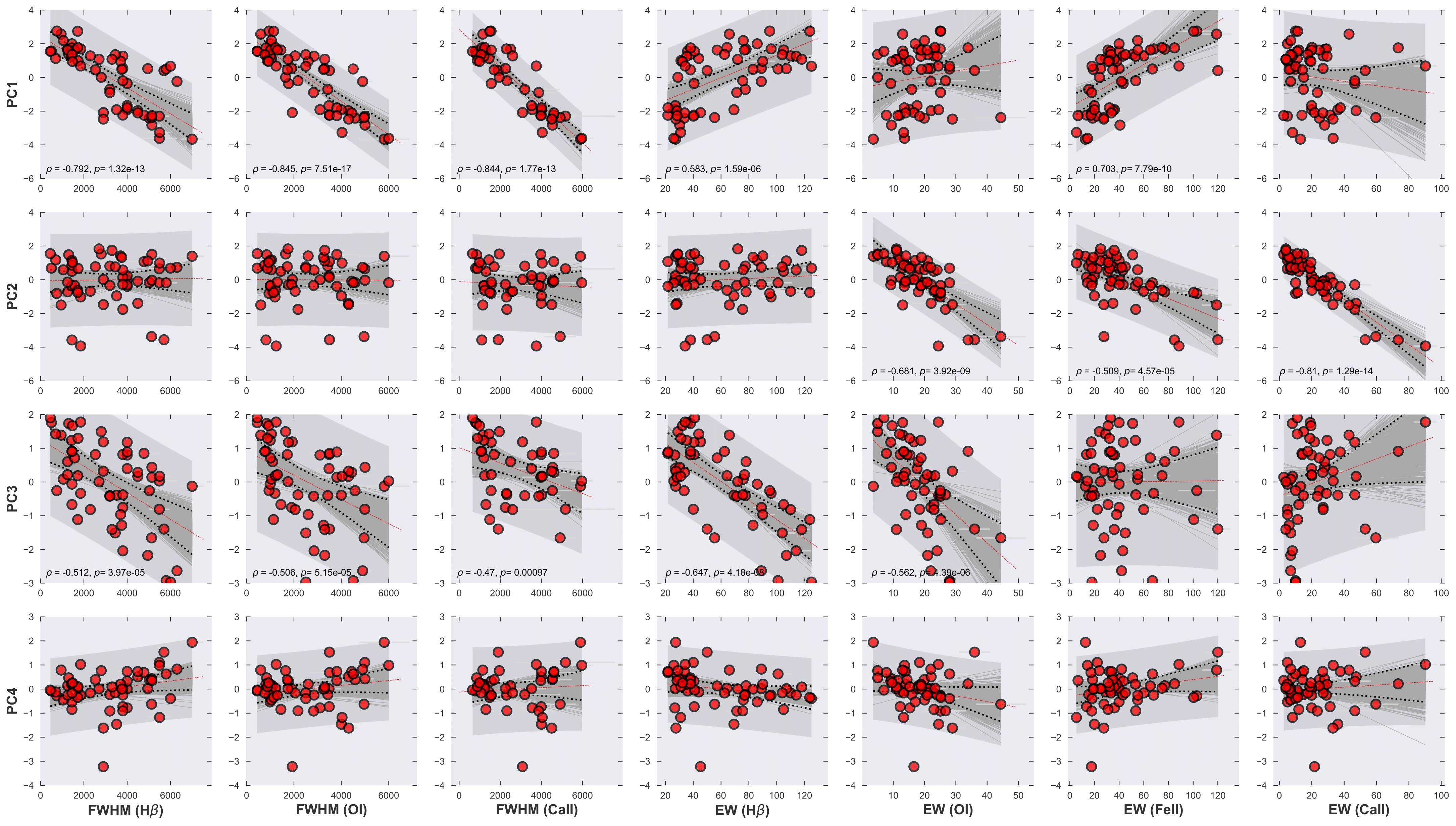}
    \caption{Correlation matrix showing dependence of the first four PCA vectors' loadings versus the physical parameters (\textit{observed}) for our full sample. The Spearman's rank correlation coefficient ($\rho$) and the $p-$value are reported for the correlations whenever $p$-value $<$ 0.001. The OLS fits in each panel are shown with red dashed lines. { Black dotted lines mark the confidence intervals at 95\% for the 1000 realizations (dark gray lines) of the bootstrap analysis. Lightgray patch marks the corresponding prediction intervals bands for the sample.}}
    \label{fig:pca_corr2_full}
\end{figure*}

\begin{figure*}
    \centering
    \includegraphics[width=2.1\columnwidth]{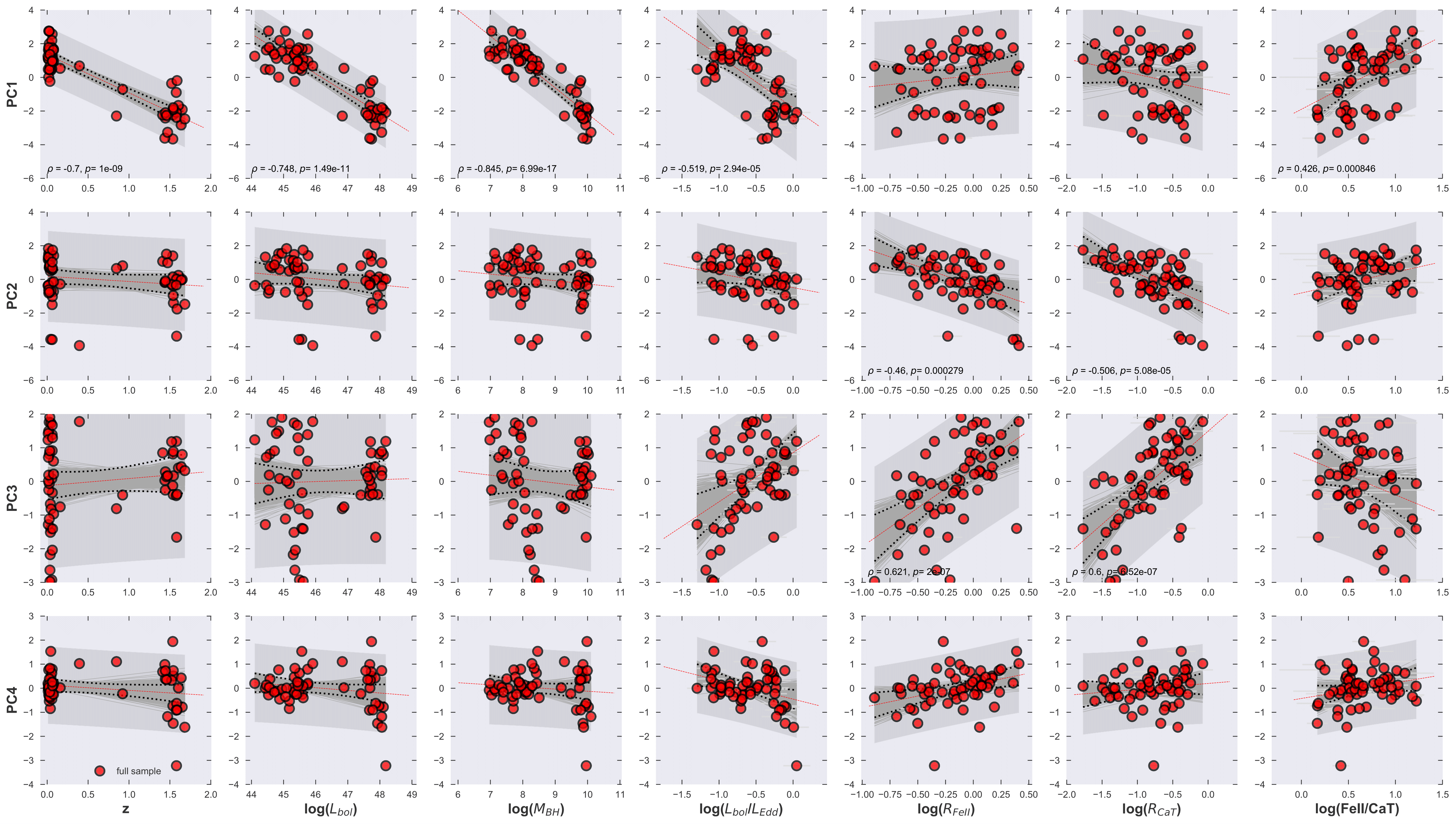}
    \caption{Correlation matrix showing dependence of the first four PCA vectors' loadings versus the physical parameters (\textit{derived}) for our full sample. The Spearman's rank correlation coefficients ($\rho$) and the $p-$values are reported for the correlations whenever $p-$value $<$ 0.001. The OLS fits in each panel are shown with red dashed lines. { Black dotted lines mark the confidence intervals at 95\% for the 1000 realizations (dark gray lines) of the bootstrap analysis. Lightgray patch marks the corresponding prediction intervals bands for the sample.}}
    \label{fig:pca_corr1_full}
\end{figure*}

To quantitatively assess the relevance of the observational variables and the derived physical parameters we show their correlation against the contributions (loadings) of the the first four principal components (PC1, PC2, PC3 and PC4) {for the full sample in Figures \ref{fig:pca_corr2_full} and \ref{fig:pca_corr1_full}. The full-sample is then separated into two subsamples based on the median optical luminosity of the full-sample (i.e. log \lopt{} = 44.49 erg s$^{-1}$). A comparative analysis {of the full-sample against the two subsamples is carried out} in Appendix \ref{appendix:pca_sub} and in Figures \ref{fig:pca_corr2} and \ref{fig:pca_corr1}.}


Figure \ref{fig:pca_corr2_full} is basically the correlation matrix representation for the Figure \ref{fig:pca_corr2_full} that includes all the intrinsic variables (except for the NIR luminosity). We only state the values of the Spearman's correlation coefficient and the corresponding $p$-value when the correlation is significant ($p<$ 0.001). In the full sample, the strongest correlation with respect to PC1 (in decreasing order) are exhibited by \fwhmoi\ ($\rho$=-0.845, $p$=7.51$\times 10^{-17}$), \fwhmca\ ($\rho$=-0.844, $p$=1.77$\times 10^{-13}$), \fwhmhb\ ($\rho$=-0.792, $p$=1.32$\times 10^{-13}$), \ewfe\ ($\rho$=0.703, $p$=7.79$\times 10^{-10}$), and by \ewhb\ ($\rho$=0.583, $p$=1.59$\times 10^{-6}$). 
In case of PC2, significant correlations are obtained only for the EWs of \cat{} ($\rho$=-0.81, $p$=1.29$\times 10^{-14}$), \oi{} ($\rho$=-0.681, $p$=3.92$\times 10^{-9}$) and \feii{} ($\rho$=-0.509, $p$=4.57$\times 10^{-5}$). For the PC3, there are negative correlations right above the significance limit for the three FWHMs and the EWs for \hb{} and \oi{}.  The correlations for the subsamples are described in Appendix \ref{appendix:pca_sub}. 

Figure \ref{fig:pca_corr1} corresponds to the derived parameters - bolometric luminosity, black hole mass, Eddington ratio, \rfe{}, \rcat{} and the ratio of the two species, \feca{}. 
For the full sample, we see clear, strong anti-correlations with PC1 for the {black hole mass ($\rho$=-0.845, $p$=6.99$\times 10^{-17}$), the bolometric luminosity ($\rho$=-0.748, $p$=1.49$\times 10^{-11}$)}, followed by the redshift ($\rho$=-0.7, $p$=1$\times 10^{-9}$) and Eddington ratio {($\rho$=-0.519, $p$=2.94$\times 10^{-5}$)}. Although, in the case of the correlation with respect to the redshift, this is biased due to the gaps in the sample distribution which is highlighted in the panels in the left column (this is also illustrated in Figure \ref{fig:z_Lopt}). For the remaining trends, they corroborate with the correlations that were obtained with the FWHMs in the previous figure. We also see a mild positive correlation of the PC1 with the ratio, \feii{}/\cat{} ($\rho$=0.426, $p$=8.46$\times 10^{-4}$). There are only two significant correlations obtained with respect to PC2, i.e. for the two line ratios - \rcat{} ($\rho$=-0.506, $p$=5.08$\times 10^{-5}$) and \rfe{} ($\rho$=-0.46, $p$=2.79$\times 10^{-4}$). This checks the observed correlation that is obtained and studied in this work and in our previous studies. The correlations with PC3 and the two ratios - \rfe{} ($\rho$=0.621, $p$=2$\times 10^{-7}$) and \rcat{} ($\rho$=0.6, $p$=1.52$\times 10^{-7}$) indicates that the \rfe{}-\rcat{} correlation in the full dataset is at least two dimensional and may have multiple drivers for this observed correlation. { Following the same analysis carried out in Sec.~\ref{sec:boots}, we also performed a bootstrap analysis for the correlations in Fig.~\ref{fig:pca_corr2_full} and  \ref{fig:pca_corr1_full} which reflects the behavior observed in Sec.~\ref{sec:boots}.} The correlations for the subsamples are described in  Appendix \ref{appendix:pca_sub}.

\section{Discussion} \label{sec:discussions}

In this paper, we carefully look at the observational correlations from the up-to-date optical and near-infrared measurements centered around \feii{} and { \cat} emission, respectively. Since our sample is affected by bias, we explored its influence in our results, showing that the correlations with the Eddington ratio and the independent observational parameters are less biased than the ones involving luminosity or black hole mass. { These results are supported by a bootstrap analysis, which corroborated the statistical meaning of the correlations. {We did not detect any redshift dependence in the correlations with luminosity and Eddington above 2$\sigma$ confidence level.}} We also performed a Principal Component Analysis in order to define the main driver(s) of our sample where the black hole mass, luminosity, Eddington ratio and the \feca\  show the main correlation with the PC1. 

\subsection{The primary driver(s) of our sample}

The PCA is a powerful tool, however, the principal eigenvectors are just mathematical entities and it not easy to  { connect them with a direct physical} meaning. As is shown in 
Sec.~\ref{sec:eigenvector} and {Appendix~\ref{appendix:pca_sub}},
the PCA gives different results if the full, low- or high-luminosities samples are considered. The correlations between the PCi values and the observational parameters such as \fwhmhb, \rfe\ or \ewhb\ for the low-luminosity subsample resembles the \citet{borosongreen1992} PCA results. However, the trends are different for the high-luminosity subsample. { This difference cannot be associated with luminosity or redshifts effects, since PCA results are based on the space parameter considered.} Hence, since the objective is to understand the general drivers of the sample, we only discuss the PCA  for the full sample.

Figures \ref{fig:pca_corr2_full} and \ref{fig:pca_corr1_full} describe the relation between the observational and independent parameters with the principal eigenvectors, where the FWHM shows the primary correlations with a significance over 99.9$\%$, followed by the relations with the equivalent widths again, with a high significance ($62.8\%$ of the variation). Hence, due to the relevance of the FWHM, a high correlation between PC1 and the black hole mass is expected (Fig.~\ref{fig:pca_corr1_full}). The luminosity and black hole mass show the strongest correlations with the PC1, followed by the Eddington ratio and the ratio \feca\ {  (see Sec.~\ref{sec:star_formation}).} The main correlations with the PC2 are with \ewca, \ewoi, \ewfe, \rcat, \ewfe\ and \rfe\ ordered in decreasing order of significance. Similar to the observational trends (Figures \ref{fig:ew_ac} and \ref{fig:rfe-rcat_corr}), all the correlations are stronger for the \cat\ than for \feii, indicating the relevance of the \cat\ in our sample.

On the other hand, in Figures \ref{fig:ew_ac} and \ref{fig:rfe-rcat_corr} the main correlations are the ones involving the \eddr,  { which is also supported by the lowest errors provided by the bootstrap results (Fig.~\ref{fig:residual_fig3} and \ref{fig:residual_fig4})}. 
These facts point towards the Eddington ratio as the main driver. However, from the PCA black, hole mass and luminosity have similar relevance ($\rho_{M_{\rm BH}}=0.845, \, \rho_{L}=0.748$) followed by the Eddington ratio ($\rho \sim -0.519$), all of them with a significance over 99.9$\%$.  Since the PCA reduces the dimensionality of the object to object variations, it is expected that the main correlations are associated with luminosity, black hole mass and Eddington ratio, since the third one can be expressed by: \eddr
$\propto L_{\rm opt}^{1/2} {\rm FWHM_{H\beta}^{-2}}$.
{ In order to test the self-dependence of the Eddington ratio and hence its role in our sample, we performed a multivariate linear regression fitting in the form {$\log$} \eddr\ $ \propto {\rm log} L_{\rm opt} + a {\rm log FWHM_{\rm H\beta}}$. We got $a=3.1$  ($\sigma_{\rm rms}\sim7.71\times 10^{-5}$ dex), a variation of $25\%$ respect to the expected value ($a=-4$) from definition of \eddr.  
Therefore, this highlights the  Eddington ratio as the driver of the correlations in the sample.}

However, this result must be tested with the inclusion of more objects. In our sample, the highest \eddr\ values are always associated with the highest luminosities, largest black hole masses and highest redshifts, which is an artifact of the flux-limited sample. 
{To verify the Eddington ratio as the main driver, one should consider samples reducing flux-limit and small-number biases, for example including low accretors at high redshift, or enlarging the sample at low--z.}
In addition, our sample does not include sources with \fwhmhb\ $>$ 7000 \kms, which usually show weak or negligible \feii\ contribution. Hence newer sources with { \cat}-\feii\ estimates are required to confirm the current results and to certify the Eddington ratio as the driving mechanism.

\subsection{Is the Eddington ratio the driver of the Baldwin effect? }
\label{sec:beff}


The driver of the Baldwin effect is still under discussion. The most accepted explanation for this effect is that high luminosity AGNs have a soft ionizing continuum, so the number of ionizing photons available for the emission line excitation decrease. It is supported by the fact that the spectral index between the UV (2500\AA) and X-ray continuum (2 keV) increases with luminosity \citep{baskin2004,shields2007}. Thus the UV bump will be shifted to longer wavelengths provoking a steeper EW--$L$ relation as a function of the ionization potential. { Metallicity also has an important role \citep{korista97}, due to the correlation with the black hole mass and luminosity \citep{hamann-ferland1993, hamann99}. An increment in the metallicity reduces the equivalent width of the emission lines. 

{ In our analysis only low-ionization lines are considered, therefore we expect weak relation between the EW and the luminosity.} The values of the slopes are around zero, $-0.1<\alpha<0.1$ in all the correlations, {as predicted by \citet{dietrich02}}. And the correlation coefficients are below the significance level considered, except in the correlations \feca-luminosity, although the bootstrap results predict a $\sim50\%$ to detect false positive in this case. Therefore,  the statistical significance of the EW--$L$ relations {is called into question.}}


{ In the correlations where the Eddington ratio is involved, the slope is stronger than in the luminosity case, $\alpha>0.3$. The same effect is found for \civ, a high--ionization emission line. Considering the equivalent width of \civ, luminosities and Eddington ratios reported by \citet{baskin2004}, we found a stronger correlation and higher slope ($\rho=-0.5, \, \alpha=-0.3\pm0.06$) in the relation CIV-\eddr\ than for the correlation with the luminosity ($\rho=-0.04\pm0.05, \, \alpha=-0.05\pm0.09$).  Additionally,  the bootstrap results predicts the smallest errors and a low probability to detect false positive in the correlations EW-\eddr.  These results suggest that the Eddington ratio has more relevance than the luminosity \citep{baskin2004, bachev2004, dong2009} and thus the behavior of the equivalent width of low-- and high--ionization lines is driven by the Eddington ratio. }


{ We can probe the role of the Eddington ratio in an independent way throughout} a  multivariate linear regression fitting in the form 
EW $\propto L_{\rm opt}+a{\rm FWHM}$. { For \hb\ we obtain $a=-2.5 \pm1.4$ ($\sigma_{\rm rms}\sim0.194$ dex), while for \ewca\ $a=-3.8 \pm1.9$ ($\sigma_{\rm rms}\sim0.308$ dex).} In the last case, the slope is almost similar to the expected value ($a=-4$), which again highlights the strong correlation between \cat\ and \eddr. At least in our sample, the \cat\ is better proxy for the Eddington ratio than \feii, { although there is a $50\%$ of probability to detect a false positive correlation as the bootstrap results pointed out. } 

A novel results from the PCA is the relevance of the metallicity expressed as the ratio \feca\ (Sec.~\ref{sec:star_formation}). According to \citet{korista1998}, the metallicity has a secondary role in the Baldwin effect, so we included this parameter in the multivariate linear regression fitting: EW $\propto L_{\rm opt}+a{\rm FWHM_{H\beta}}+b{\rm {FeII/CaII}}$. There is no improvement for the \hb\ correlation ($a=-2.4 \pm 1.6 ,\, b=-0.786\pm1.4, \sigma_{\rm rms}\sim0.193$), while for \cat\ the uncertainties are high ($a=-4.8\pm4.1,\, b=-9.705\pm6.35, \sigma_{\rm rms}\sim0.279$). { Therefore, the Eddington ratio have the main role in the Baldwin effect than the metallicity if it is expressed as the \feca\ ratio.}



\subsection{Implication for the chemical evolution}
\label{sec:star_formation}

The relative abundance of iron with respect to the $\alpha$--elements has been used as a proxy for the chemical abundance in AGN \citep[see][for a review]{hamannferland92}. The $\alpha$--elements (O, Mg, Si, Ca and Ti) are predominantly produced by Type II supernovae (SNe) after the explosion of massive stars (7 M$_\odot$ $< M_\star <100$ M$_\odot$) on timescales of $10^7$ yr, while  Fe is mostly produced  by Type Ia SNe from white dwarf progenitors on longer timescales $\sim1$ Gyr \citep{hamann-ferland1993}.  {Depending on the host galaxy type, the time delay between the manufacturing timescales varies from 0.3 to 1 Gyr for massive elliptical and Milky Way-like galaxies, respectively \citep{matteucci2001}. Thus, the ratio Fe/$\alpha$ can be used as a clock for constraining the star formation, the metal content and the age of the AGN \citep{mateucci1994, hamannferland92}. The UV \feii\ and \mgii\   have been widely used for this purpose since the UV spectrum is accessible in a wide redshift range, up to $z \sim 7.5$ \citep[e.g.][and references therein]{dietrich2003, verner2009, dong2011, derosa2014, sameshima2017, shin2019, onoue2020,  sarkar2020}. However, the  \femg\ flux ratio does not show a significant redshift evolution suggesting a large number of Type Ia SNe \citep{onoue2020} or AGN being chemically mature {also at high--$z$} \citep{shin2019}}.

\begin{figure*}
    \centering
    \includegraphics[width=15cm,keepaspectratio=true]{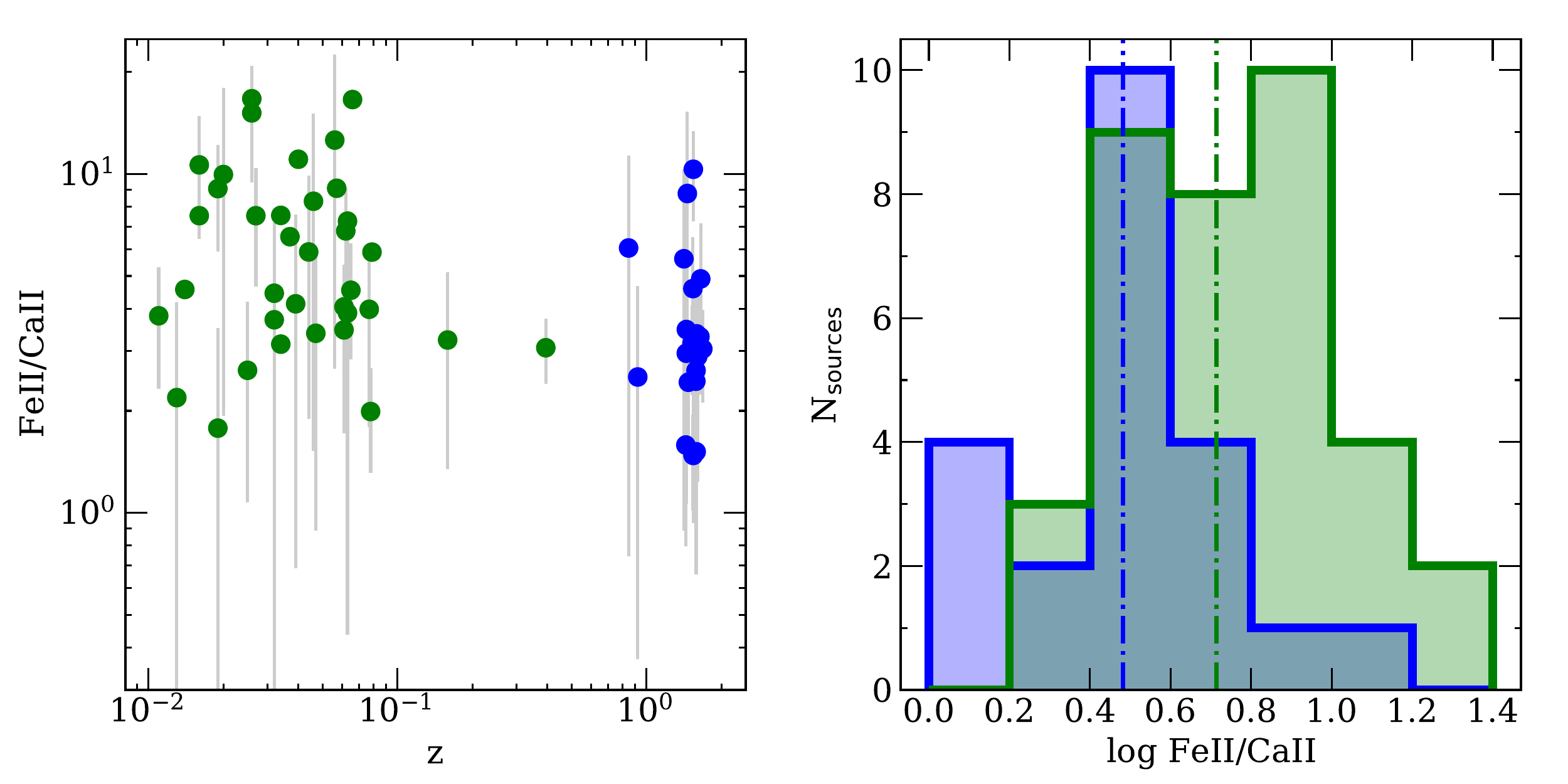}
    \caption{Left panel: \feca\ distribution as a function of redshift in log-scale. Green symbols correspond to sources with $z<0.8$, the rest of the sources are marked with blue symbols. Right panel: \feca\ distribution, colors are the same as the left panel. Vertical lines mark the median redshift for the low- and high-redshift subsamples.}
    \label{fig:hist_feca}
\end{figure*}

{ The optical \feii\ and { \cat} have similar ionization potentials and both are emitted by the same region in the BLR, although the { \cat} region is seemingly more extended \citep{panda_cafe2}. Assuming that { \cat} scales with the rest of the $\alpha$-elements and the ratio \feca\ traces the abundance iron over calcium, we can use the ratio \feca\ as a metal estimator.} Figure \ref{fig:hist_feca} shows the distribution of the ratio \feca\ as a function of the redshift. Dividing the sample at $z=0.8$, which basically separates the Persson and Marinello et al. samples from the HE-sample (Martínez-Aldama et al. sample), we get that low-redshift sample has a median \feca\ ratio of $\sim5.8$, while the intermediate-redshift show a \feca $\sim3.0$. A two-sample Kolmogorov-Smirnov test provides a value of 0.489 with a probability of $p_{KS}\sim0.001$. It means that both samples are not drawn from the same distribution. 
{The higher ratio \feca\ at low redshift  suggests some form of chemical evolution.} {Our sample reaches a maximum redshift at $z\sim1.7$, just after the maximum star formation peak  \citep{madau2014}. 
{Thus,  the \feca\ ratio will be lowered  by the effect of a recent starburst enhancing the $\alpha$--elements with respect to iron at intermediate-redshifts \citep{martinez-aldamaetal15}}
. Surprisingly, the \feca\ ratio also has a mild correlation with the PC1 (Fig.~\ref{fig:pca_corr1_full}), suggesting that the metal content has a relevant role in governing the properties of our sample.}

{\citet{hamann-ferland1993}  found a positive relation between the metallicity, black hole mass and luminosity, therefore the highest metallicity AGN might also be the most massive, such as the last row of Fig.~\ref{fig:rfe-rcat_corr} shows.
}
{An exception to the \citet{hamann-ferland1993} results are the Narrow-Line Seyfert 1 (NLSy1) galaxies which exhibit a high NV/CIV flux ratio (an alternative proxy of the metal content) despite their low-luminosity \citep{shemmer2004}. In our sample a clear NLSy1 is the object PHL~1092 \citep{2020arXiv200401811M}, that shows \feca$\sim3.1$, which is close to the mean \feca\ value of the high-redshift sample, putting this source within the regime of high metal content. Previous studies indicate that NLSy1 show a deviation from the relations NV/CIV-$L$ and -\mbh\ \citep{shemmer-netzer2002}, in our sample we cannot confirm these results using the ratio \feca, since the scatter for a fixed $L$ or \mbh\ is too large.} 

{Based on the flux ratio NV/CIV, \citet{shemmer2004} found a positive correlation between the abundance ($Z$) and the Eddington ratio. {Our sample shows an anti-correlation between \eddr\ and the ratio \feca\ with a Spearman rank coefficient of $\rho\sim 0.554$ and a significance over $99.9\%$. { The bootstrap results proved the reliability of this correlation, with a probability less than $11\%$ to detect a false positive trend.} Considering that it is very likely that a recent starburst concomitantly increases $Z$ and lowers \feca\ \citep[see for example][]{bizyaev2019}, we expect that the high Eddington ratio sources in our sample are associated with high metal content and/or low Fe/$\alpha$ values.} 
Similar to the Baldwin Effect, the relation between the metal content estimator and the Eddington ratio is stronger than with luminosity, black hole mass or redshift \citep{shemmer2004, dong2011, sameshima2017, shin2019}. Conversely, the correlation between the Eddington ratio and \femg\ remains unclear; depending on the sample considered there is a positive \citep{shin2019,  sameshima2017, onoue2020} or null correlation \citep{sarkar2020}. Since \feii\ and \mgii\ are affected by non-abundance parameters such as the density or the microturbulence, the ratio \femg\ might be affected \citep{sameshima2017, shin2019}. After a correction by these factors, the correlation \femg-\eddr\ roughly remains positive or disappears.}


{Under the assumption that \feca\ flux ratio is a first order proxy of the [Fe/Ca] abundances, we found that the behavior shown by the \feca\ flux ratio is in agreement with the normal chemical evolution \citep{hamann1993, hamann99}, {where the main enrichment occurs in the early epochs. Our results also support that main \feii\ enrichment occurs 1-2 Gyr after the first starburst \citep{hamann1993, hamann99} and also suggest that the strong \feii\ could be associated with a second starburst.} However, the overabundance of \feii\ depends on the SNe Ia lifetime and the star formation epoch \citep{sameshima2020}. In order to confirm these results models incorporating chemical evolution and [Fe/Ca] abundances are required. In addition, we must explore the dependence on non-abundance parameters, which could modify the relation of Figs.~\ref{fig:ew_ac} and \ref{fig:rfe-rcat_corr}, or the effect of the Baldwin effect in the abundance determination as \citet{sameshima2020} have tested for \femg. Enlarging the sample is one of the main challenges.} For observing the \cat\ in sources with $z>2$, { { near-infrared spectrometers} with higher sensitivity are required. However, due to the fact that  the near- and mid-infrared spectral regions are strongly affected by atmospheric telluric bands, some redshift ranges will remain inaccessible from ground- based telescopes} 
The most attractive possibility to study the \feca\ ratio at larger redshifts is offered by upcoming space observatories, such as Near InfraRed Spectrograph (NIRSpec) from the James Webb Space Telescope (JWST).

\section{Conclusions} \label{sec:conclusions}

We performed a detailed analysis of the observational correlations present in our \cat-\feii\ sample together with a { bootstrap analysis to asses the statistical reliability  of  the  correlations. Throughout a Principal Component Analysis, we identify the primary driver of our sample.  We could not find any redshift dependence above $2\sigma$ confidence level.} Since our sample is flux-limited, the presented analysis must be confirmed by larger samples in the future. The presented analysis shows the following:

\begin{itemize}


\item {The correlation with luminosity, black hole mass and Eddington ratio are stronger for \cat\ than \feii. { It suggest that CaT is a better proxy for the Eddington ratio than \feii. A potential application of this results could be, for example, the correction of the the Radius-Luminosity relation for effects dependent on the accretion rate.  However, the bootstrap analysis provides a probability of $50\%$ to detect a false positive relation. Therefore a more complete a sample is needed to confirm this result. }}


\item The \ewhb\ correlates negatively with luminosity  (Baldwin effect), while the \ewca\ shows a positive correlation. It stresses the different nature of both low ionization emission lines.  { In general, the correlations with Eddington ratio are stronger, more reliable and show the smallest errors according to the bootstrap analysis.}  This supports the Eddington ratio as the driver of the Baldwin effect. 

\item We performed a principal component analysis (PCA) to deduce the driver for the \rcat{}-\rfe{} relation observed in our sample. We confirm that the results of the PCA are dependent on the selection of the sample and the chosen
quantities. We consider only the directly observable variables (FWHMs, luminosity and EWs) in the final PCA and later correlate the first four eigenvectors to the derived variables (bolometric luminosity, black hole mass, Eddington ratio, the ratios - \rfe{} and \rcat{}, and the metallicity indicator, \feca). The dominant eigenvector is primarily driven by the combination of black hole mass and luminosity which in turn is reflected in the strong correlation of the first eigenvector with the Eddington ratio. We also notice a noticeable correlation of the primary eigenvector with the metallicity tracer, \feca{}.

\item Combining the PCA results and the observational correlation, we conclude that luminosity, black hole mass, and Eddington ratio are the main drivers of our sample, however, the correlations are better described by the Eddington ratio, setting this parameter one step ahead in comparison to the other parameters. A larger, more complete sample spanning a larger redshift range including a wide variety of AGN is needed to assert the driver(s) of the \cat-\feii\ correlations.

\item {We found a significant negative increment of the \feca\ ratio as a function of redshift, suggesting the effect of a recent starburst which enhanced the $\alpha$-elements with respect to iron at intermediate-redshift. So, the \feca\ ratio  could be used to map the metal and the star formation in AGN}. The \feca\ ratio is also highlighted by the PCA, pointing out the relevance of the metal content in our sample. The negative correlation with the Eddington ratio {corroborated by the bootstrap analysis } supports the \feca\ as a metal indicator instead of the typical \feii/\mgii\ ratio used for this purpose. { However, a more complete sample and sources at large redshift are needed to verify these results.}

\end{itemize}

\section*{Acknowledgements}
We are grateful to the anonymous referee for her/his comments and suggestions that helped to improve the manuscript. The project was partially supported by the Polish Funding Agency National Science Centre, project 2017/26/\-A/ST9/\-00756 (MAESTRO  9) and MNiSW grant DIR/WK/2018/12. DD acknowledges support from grant PAPIIT UNAM, 113719.
\software {\textmyfont{CLOUDY} v17.01 (\citealt{f17}); \textmyfont{MATPLOTLIB}  (\citealt{hunter07}); \textmyfont{NUMPY} (\citealt{numpy}); \textmyfont{SKLEARN} (\citealt{scikit-learn}); \textmyfont{STATSMODELS} (\citealt{seabold2010statsmodels}); \textmyfont{TOPCAT} (\citealt{taylor2005})}


\bibliography{main}


\appendix

\setcounter{table}{0}
\renewcommand\thetable{\Alph{section}.\arabic{table}}

\counterwithin{figure}{section}


\section{Tables}
\label{appendix:tables_lum}

In this section, we summarize the observational properties
of the full sample employed in the analysis. The description of the columns is included in the notes of each table.

\begin{table*}[hbt!]
\caption{Observational parameters }
\resizebox{\columnwidth}{!}{
\begin{tabular}[c]{c c c c c c c c c c c c c}
\label{tab:table1} \\
\hline\hline\noalign{\vskip 0.1cm}         
\multirow{2}{*}{Object}  & \multirow{2}{*}{$z$} & log {$L_{\rm opt}$} & log {$L_{\rm NIR}$} & \multirow{2}{*}{\rfe} & \multirow{2}{*}{\rca} & FWHM \hb\ & FWHM OI & FWMH CaII & EW \hb\ & EW OI & EW CaII & EW FeII \\

 &  & [\ergs] & [\ergs] & & & [\kms] & [\kms] & [\kms] & [\AA] & [\AA] & [\AA] & [\AA]  \\

(1) & (2) & (3) & (4) & (5) & (6) & (7) & (8) & (9) & (10) & (11) & (12) &  (13) \\

\hline\hline\noalign{\vskip 0.1cm}
\noalign{\vskip 0.1cm} 																
\multicolumn{13}{c}{\citet{persson1988} Sample} \\ 
				
\hline \noalign{\vskip 0.1cm}													
IZw1$\dagger$	&	0.061	&	44.684	$\pm$	0.048	&	44.694	$\pm$	0.049	&	1.78	$\pm$	0.05	&	0.51	$\pm$	0.13	&	950	$\pm$	100	&	600	$\pm$	150	&	1150	$\pm$	150	&	66	$\pm$	5	&	13	$\pm$	1	&	43	$\pm$	4	&	119	$\pm$	12	\\
Mrk~42	&	0.025	&	43.299	$\pm$	0.048	&	43.247	$\pm$	0.049	&	1.07	$\pm$	0.07	&	0.41	$\pm$	0.14	&	500	$\pm$	100	&	450	$\pm$	150	&	800	$\pm$	150	&	36	$\pm$	2	&	13	$\pm$	1	&	19	$\pm$	2	&	40	$\pm$	6	\\
Mrk~478$\dagger$	&	0.077	&	44.685	$\pm$	0.048	&	44.607	$\pm$	0.049	&	0.93	$\pm$	0.06	&	0.23	$\pm$	0.07	&	1250	$\pm$	100	&	1100	$\pm$	150	&	2500	$\pm$	200	&	78	$\pm$	5	&	18	$\pm$	2	&	28	$\pm$	3	&	77	$\pm$	10	\\
II~Zw~136	&	0.063	&	44.47	$\pm$	0.048	&	44.424	$\pm$	0.049	&	0.66	$\pm$	0.05	&	0.17	$\pm$	0.13	&	1850	$\pm$	100	&	1200	$\pm$	150	&	2300	$\pm$	250	&	96	$\pm$	7	&	25	$\pm$	3	&	29	$\pm$	6	&	68	$\pm$	7	\\
Mrk~231	&	0.044	&	44.456	$\pm$	0.048	&	44.59	$\pm$	0.049	&	2.46	$\pm$	0.05	&	0.42	$\pm$	0.19	&	5700	$\pm$	100	&	3500	$\pm$	150	&	1900	$\pm$	250	&	50	$\pm$	4	&	36	$\pm$	5	&	53	$\pm$	5	&	120	$\pm$	10	\\
3C~273	&	0.159	&	46.097	$\pm$	0.048	&	45.833	$\pm$	0.049	&	0.40	$\pm$	0.05	&	0.12	$\pm$	0.04	&	3800	$\pm$	100	&	1900	$\pm$	150	&	2250	$\pm$	250	&	89	$\pm$	6	&	22	$\pm$	5	&	27	$\pm$	5	&	40	$\pm$	4	\\
Mrk~6	&	0.019	&	43.556	$\pm$	0.048	&	43.675	$\pm$	0.049	&	0.83	$\pm$	0.16	&	0.47	$\pm$	0.43	&	2800	$\pm$	200	&	1650	$\pm$	200	&	1950	$\pm$	200	&	29	$\pm$	6	&	5	$\pm$	2	&	15	$\pm$	6	&	24	$\pm$	7	\\
Mrk~486	&	0.039	&	43.969	$\pm$	0.048	&	43.892	$\pm$	0.049	&	0.27	$\pm$	0.06	&	0.07	$\pm$	0.04	&	1450	$\pm$	100	&	1450	$\pm$	150	&	1300	$\pm$	250	&	105	$\pm$	5	&	23	$\pm$	3	&	12	$\pm$	6	&	31	$\pm$	4	\\
Mrk~1239$\dagger$	&	0.02	&	43.158	$\pm$	0.048	&	44.2	$\pm$	0.049	&	1.10	$\pm$	0.07	&	0.11	$\pm$	0.07	&	750	$\pm$	100	&	950	$\pm$	150	&	1550	$\pm$	400	&	95	$\pm$	5	&	25	$\pm$	3	&	10	$\pm$	5	&	103	$\pm$	15	\\
Mrk~766	&	0.013	&	43.468	$\pm$	0.048	&	43.221	$\pm$	0.049	&	0.68	$\pm$	0.12	&	0.31	$\pm$	0.25	&	700	$\pm$	100	&	1050	$\pm$	150	&	1450	$\pm$	300	&	44	$\pm$	3	&	12	$\pm$	2	&	11	$\pm$	5	&	29	$\pm$	8	\\
Zw~0033+45	&	0.047	&	44.352	$\pm$	0.048	&	44.159	$\pm$	0.049	&	0.53	$\pm$	0.10	&	0.16	$\pm$	0.08	&	2300	$\pm$	200	&	2050	$\pm$	200	&	2600	$\pm$	400	&	110	$\pm$	10	&	23	$\pm$	4	&	27	$\pm$	3	&	60	$\pm$	12	\\
Mrk~684	&	0.046	&	44.161	$\pm$	0.048	&	44.2	$\pm$	0.049	&	1.29	$\pm$	0.07	&	0.16	$\pm$	0.10	&	1300	$\pm$	100	&	850	$\pm$	150	&	850	$\pm$	250	&	38	$\pm$	4	&	5	$\pm$	2	&	8	$\pm$	4	&	51	$\pm$	7	\\
Mrk~335$\dagger$	&	0.026	&	43.974	$\pm$	0.048	&	43.932	$\pm$	0.049	&	$\leq$0.417			&	$\leq$0.025			&	1450	$\pm$	100	&	1000	$\pm$	150	&	--			&	86	$\pm$	5	&	17	$\pm$	2	&	$\leq$5			&	40	$\pm$	6	\\
Mrk~376	&	0.056	&	44.367	$\pm$	0.048	&	44.367	$\pm$	0.049	&	0.62	$\pm$	0.06	&	0.05	$\pm$	0.03	&	5800	$\pm$	100	&	4500	$\pm$	200	&	--			&	101	$\pm$	10	&	28	$\pm$	6	&	10	$\pm$	5	&	67	$\pm$	7	\\
Mrk~493$\dagger$	&	0.032	&	43.678	$\pm$	0.048	&	43.356	$\pm$	0.049	&	0.78	$\pm$	0.06	&	0.21	$\pm$	0.18	&	450	$\pm$	100	&	450	$\pm$	150	&	650	$\pm$	300	&	41	$\pm$	3	&	7	$\pm$	1	&	11	$\pm$	8	&	33	$\pm$	4	\\
Mrk~841	&	0.037	&	44.124	$\pm$	0.048	&	44.045	$\pm$	0.049	&	$\leq$0.209			&	$\leq$0.032			&	4950	$\pm$	100	&	3300	$\pm$	150	&	--			&	107	$\pm$	6	&	21	$\pm$	3	&	7	$\pm$	0	&	25	$\pm$	6	\\
Ton~1542	&	0.063	&	44.207	$\pm$	0.048	&	44.404	$\pm$	0.049	&	$\leq$0.363			&	$\leq$0.05			&	3800	$\pm$	100	&	2850	$\pm$	250	&	--			&	114	$\pm$	11	&	24	$\pm$	3	&	8	$\pm$	0	&	43	$\pm$	4	\\
VII~Zw~118	&	0.079	&	44.697	$\pm$	0.048	&	44.63	$\pm$	0.049	&	$\leq$0.447			&	$\leq$0.076			&	3700	$\pm$	100	&	2600	$\pm$	300	&	--			&	90	$\pm$	9	&	14	$\pm$	2	&	5	$\pm$	0	&	38	$\pm$	4	\\
Mrk~124	&	0.057	&	43.592	$\pm$	0.048	&	44.073	$\pm$	0.049	&	$\leq$0.708			&	$\leq$0.078			&	1050	$\pm$	100	&	1050	$\pm$	250	&	--			&	76	$\pm$	8	&	20	$\pm$	4	&	7	$\pm$	0	&	55	$\pm$	6	\\
Mrk~9	&	0.04	&	44.155	$\pm$	0.048	&	43.915	$\pm$	0.049	&	$\leq$0.398			&	$\leq$0.036			&	3450	$\pm$	100	&	2500	$\pm$	250	&	--			&	104	$\pm$	10	&	21	$\pm$	2	&	7	$\pm$	0	&	45	$\pm$	5	\\
NGC~7469	&	0.016	&	43.864	$\pm$	0.048	&	43.863	$\pm$	0.049	&	$\leq$0.339			&	$\leq$0.045			&	2700	$\pm$	200	&	1750	$\pm$	250	&	--			&	72	$\pm$	7	&	11	$\pm$	2	&	4	$\pm$	0	&	25	$\pm$	4	\\
Akn~120	&	0.034	&	44.188	$\pm$	0.048	&	44.62	$\pm$	0.049	&	$\leq$0.468			&	$\leq$0.062			&	6300	$\pm$	100	&	4900	$\pm$	400	&	--			&	89	$\pm$	5	&	22	$\pm$	4	&	7	$\pm$	0	&	42	$\pm$	4	\\
Mrk~352	&	0.014	&	43.037	$\pm$	0.048	&	43.136	$\pm$	0.049	&	$\leq$0.219			&	$\leq$0.048			&	3750	$\pm$	100	&	3250	$\pm$	300	&	--			&	83	$\pm$	5	&	19	$\pm$	4	&	6	$\pm$	0	&	19	$\pm$	4	\\
Mrk~304	&	0.066	&	44.507	$\pm$	0.048	&	44.416	$\pm$	0.049	&	$\leq$0.282			&	$\leq$0.017			&	3300	$\pm$	200	&	3300	$\pm$	400	&	--			&	118	$\pm$	6	&	11	$\pm$	2	&	4	$\pm$	0	&	36	$\pm$	6	\\
Mrk~509	&	0.034	&	44.508	$\pm$	0.048	&	44.43	$\pm$	0.049	&	$\leq$0.129			&	$\leq$0.041			&	6000	$\pm$	200	&	2500	$\pm$	200	&	--			&	125	$\pm$	6	&	28	$\pm$	3	&	10	$\pm$	0	&	18	$\pm$	3	\\
														
\hline\hline\noalign{\vskip 0.1cm}																		
\noalign{\vskip 0.1cm} 												
\multicolumn{12}{c}{\citet{martinez-aldamaetal15, martinez-aldamaetal15b} Sample} \\    												
\hline \noalign{\vskip 0.1cm}																					
HE~1349+0007	&	1.444	&	47.119	$\pm$	0.047	&	46.887	$\pm$	0.045	&	0.692	$\pm$	0.143	&	0.234	$\pm$	0.086	&	5027	$\pm$	430	&	4580	$\pm$	680	&	4530	$\pm$	940	&	33.5	$\pm$	4	&	19.9	$\pm$	4	&	24.7	$\pm$	7	&	18.5	$\pm$	1	\\
HE~1409+0101	&	1.65	&	47.629	$\pm$	0.044	&	47.264	$\pm$	0.048	&	1.549	$\pm$	0.107	&	0.316	$\pm$	0.066	&	4000	$\pm$	160	&	3100	$\pm$	310	&	3550	$\pm$	500	&	26.8	$\pm$	1	&	15.3	$\pm$	2	&	19.9	$\pm$	6	&	36.2	$\pm$	2	\\
HE~2349-3800	&	1.604	&	47.11	$\pm$	0.044	&	46.933	$\pm$	0.045	&	0.832	$\pm$	0.057	&	0.288	$\pm$	0.093	&	4000	$\pm$	160	&	3480	$\pm$	520	&	3520	$\pm$	700	&	27.1	$\pm$	1	&	17.6	$\pm$	4	&	19.9	$\pm$	6	&	19.9	$\pm$	1	\\
HE~2147-3212	&	1.543	&	47.163	$\pm$	0.049	&	46.891	$\pm$	0.045	&	1.479	$\pm$	0.375	&	0.49	$\pm$	0.079	&	4491	$\pm$	660	&	4300	$\pm$	860	&	3990	$\pm$	150	&	28.4	$\pm$	3	&	22.1	$\pm$	6	&	47.4	$\pm$	4	&	35	$\pm$	4	\\
HE~2202-2557	&	1.535	&	47.172	$\pm$	0.048	&	46.585	$\pm$	0.043	&	0.537	$\pm$	0.062	&	0.117	$\pm$	0.019	&	7000	$\pm$	540	&	5810	$\pm$	1060	&	5900	$\pm$	330	&	27.4	$\pm$	1	&	3.6	$\pm$	1	&	13	$\pm$	3	&	12.7	$\pm$	1	\\
HE~2340-4443	&	0.922	&	46.758	$\pm$	0.046	&	46.674	$\pm$	0.049	&	0.224	$\pm$	0.021	&	0.089	$\pm$	0.064	&	3200	$\pm$	100	&	3430	$\pm$	220	&	3190	$\pm$	1700	&	77.5	$\pm$	3	&	15.9	$\pm$	1	&	14.8	$\pm$	10	&	15.4	$\pm$	1	\\
HE~0248-3628	&	1.536	&	47.465	$\pm$	0.045	&	47.222	$\pm$	0.044	&	0.372	$\pm$	0.034	&	0.251	$\pm$	0.023	&	3800	$\pm$	150	&	3490	$\pm$	260	&	3990	$\pm$	150	&	40.7	$\pm$	2	&	12.2	$\pm$	2	&	31.3	$\pm$	2	&	13.6	$\pm$	0.5	\\
HE~2352-4010	&	1.58	&	47.727	$\pm$	0.044	&	47.537	$\pm$	0.05	&	0.447	$\pm$	0.031	&	0.17	$\pm$	0.023	&	2900	$\pm$	90	&	1930	$\pm$	110	&	3080	$\pm$	180	&	45.3	$\pm$	1	&	16.6	$\pm$	2	&	21.7	$\pm$	3	&	17.6	$\pm$	0.3	\\
HE~0035-2853	&	1.638	&	47.309	$\pm$	0.048	&	47.177	$\pm$	0.044	&	1.479	$\pm$	0.17	&	0.447	$\pm$	0.041	&	5141	$\pm$	390	&	5000	$\pm$	370	&	4540	$\pm$	170	&	30.3	$\pm$	1	&	9.4	$\pm$	2	&	27.9	$\pm$	2	&	40.7	$\pm$	3	\\
HE~0048-2804	&	0.847	&	46.032	$\pm$	0.049	&	46.168	$\pm$	0.044	&	0.617	$\pm$	0.114	&	0.102	$\pm$	0.075	&	5484	$\pm$	470	&	4990	$\pm$	270	&	5170	$\pm$	2400	&	40.1	$\pm$	4	&	18.4	$\pm$	1	&	4.9	$\pm$	3	&	22.3	$\pm$	2	\\
HE~0058-3231	&	1.582	&	47.336	$\pm$	0.048	&	47.043	$\pm$	0.055	&	0.589	$\pm$	0.176	&	0.389	$\pm$	0.09	&	5127	$\pm$	160	&	4960	$\pm$	100	&	4910	$\pm$	760	&	55.4	$\pm$	2	&	44.4	$\pm$	8	&	59.6	$\pm$	14	&	27.7	$\pm$	2	\\
HE~0203-4627	&	1.438	&	47.102	$\pm$	0.048	&	46.941	$\pm$	0.045	&	0.759	$\pm$	0.192	&	0.479	$\pm$	0.088	&	5486	$\pm$	810	&	6000	$\pm$	250	&	5960	$\pm$	530	&	26.1	$\pm$	3	&	13.6	$\pm$	3	&	33.2	$\pm$	3	&	14.4	$\pm$	2	\\
HE~0005-2355	&	1.412	&	47.161	$\pm$	0.047	&	46.863	$\pm$	0.045	&	0.933	$\pm$	0.237	&	0.166	$\pm$	0.107	&	4777	$\pm$	710	&	3500	$\pm$	820	&	4600	$\pm$	1070	&	21.8	$\pm$	2	&	7.8	$\pm$	2	&	13.9	$\pm$	8	&	17.2	$\pm$	2	\\
HE~0043-2300	&	1.54	&	47.409	$\pm$	0.05	&	47.257	$\pm$	0.045	&	0.316	$\pm$	0.044	&	0.214	$\pm$	0.025	&	3511	$\pm$	110	&	4000	$\pm$	360	&	4000	$\pm$	150	&	69.3	$\pm$	5	&	24.5	$\pm$	3	&	36.1	$\pm$	3	&	20	$\pm$	1	\\
HE~0349-5249	&	1.541	&	47.567	$\pm$	0.047	&	46.826	$\pm$	0.049	&	1.704	$\pm$	0.102	&	0.165	$\pm$	0.014	&	4000	$\pm$	400	&	3810	$\pm$	220	&	4000	$\pm$	150	&	21.9	$\pm$	2.4	&	13.5	$\pm$	3	&	37.6	$\pm$	8	&	32.1	$\pm$	3.5	\\
HE~0359-3959	&	1.521	&	47.132	$\pm$	0.047	&	46.911	$\pm$	0.049	&	1.173	$\pm$	0.07	&	0.371	$\pm$	0.031	&	4000	$\pm$	400	&	1770	$\pm$	320	&	1560	$\pm$	60	&	40.6	$\pm$	4.4	&	9.2	$\pm$	2.1	&	46.6	$\pm$	9.9	&	43.3	$\pm$	4.7	\\
HE~0436-3709	&	1.445	&	46.949	$\pm$	0.062	&	46.855	$\pm$	0.046	&	1.164	$\pm$	0.07	&	0.335	$\pm$	0.028	&	4491	$\pm$	449	&	4440	$\pm$	250	&	4610	$\pm$	170	&	33.8	$\pm$	3.7	&	13.9	$\pm$	1.9	&	26.8	$\pm$	5.7	&	33.7	$\pm$	3.7	\\
HE~0507-3236	&	1.577	&	47.094	$\pm$	0.044	&	46.939	$\pm$	0.063	&	0.291	$\pm$	0.006	&	0.119	$\pm$	0.034	&	3200	$\pm$	320	&	2830	$\pm$	420	&	3870	$\pm$	800	&	71.3	$\pm$	7.2	&	25.1	$\pm$	6.1	&	23.4	$\pm$	8.2	&	17.8	$\pm$	1.8	\\
HE~0512-3329	&	1.587	&	47.222	$\pm$	0.044	&	47.095	$\pm$	0.064	&	0.81	$\pm$	0.017	&	0.24	$\pm$	0.05	&	3800	$\pm$	380	&	2170	$\pm$	320	&	2320	$\pm$	360	&	75.6	$\pm$	7.6	&	26.7	$\pm$	6.5	&	46.6	$\pm$	13.4	&	53.5	$\pm$	5.4	\\
HE~0926-0201	&	1.682	&	47.572	$\pm$	0.049	&	47.338	$\pm$	0.054	&	1.139	$\pm$	0.082	&	0.374	$\pm$	0.033	&	2900	$\pm$	290	&	4310	$\pm$	380	&	4500	$\pm$	170	&	27.6	$\pm$	3.1	&	28.9	$\pm$	4.3	&	33.1	$\pm$	7	&	27.6	$\pm$	3.1	\\
HE~1039-0724	&	1.458	&	47.027	$\pm$	0.049	&	46.882	$\pm$	0.05	&	0.289	$\pm$	0.021	&	0.033	$\pm$	0.018	&	5141	$\pm$	514	&	3810	$\pm$	890	&	3970	$\pm$	920	&	28.4	$\pm$	3.2	&	11.5	$\pm$	3.4	&	2.7	$\pm$	1.6	&	6.7	$\pm$	0.8	\\
HE~1120+0154	&	1.472	&	47.568	$\pm$	0.047	&	47.363	$\pm$	0.052	&	0.204	$\pm$	0.012	&	0.084	$\pm$	0.011	&	5498	$\pm$	550	&	4030	$\pm$	230	&	4040	$\pm$	220	&	31	$\pm$	3.4	&	13.3	$\pm$	1.9	&	8.1	$\pm$	1.9	&	5.4	$\pm$	0.6	\\
										
\hline\hline\noalign{\vskip 0.1cm}																			
\noalign{\vskip 0.1cm} 																			
\multicolumn{12}{c}{\citet{murilo2016} Sample} \\ 				
\hline \noalign{\vskip 0.1cm}																			
1H~1934-063$\dagger$	&	0.011	&	42.624	$\pm$	0.043	&	42.637	$\pm$	0.044	&	1.404	$\pm$	0.223	&	0.368	$\pm$	0.047	&	1430	$\pm$	100	&	1000	$\pm$	80	&	1205	$\pm$	84	&	35.3	$\pm$	3.5	&	18.8	$\pm$	1.3	&	29.4	$\pm$	1.9	&	53.2	$\pm$	5.1	\\
1H~2107-097$\dagger$	&	0.027	&	43.217	$\pm$	0.043	&	43.465	$\pm$	0.044	&	1.047	$\pm$	0.106	&	0.139	$\pm$	0.019	&	2530	$\pm$	320	&	1720	$\pm$	138	&	1700	$\pm$	136	&	38	$\pm$	3.4	&	14.4	$\pm$	1.5	&	14.2	$\pm$	1.5	&	34.9	$\pm$	3.6	\\
IZw1$\dagger$	&	0.061	&	44.344	$\pm$	0.043	&	44.195	$\pm$	0.044	&	2.286	$\pm$	0.199	&	0.564	$\pm$	0.058	&	1450	$\pm$	110	&	820	$\pm$	57	&	1100	$\pm$	77	&	38.1	$\pm$	2.9	&	33.9	$\pm$	1.8	&	73.4	$\pm$	4.1	&	84.9	$\pm$	4.9	\\
Mrk~1044$\dagger$	&	0.016	&	43.076	$\pm$	0.043	&	43.013	$\pm$	0.046	&	1.181	$\pm$	0.127	&	0.111	$\pm$	0.016	&	1570	$\pm$	145	&	1010	$\pm$	61	&	1200	$\pm$	72	&	57.2	$\pm$	5.1	&	18.9	$\pm$	1.2	&	24.6	$\pm$	2.1	&	65.3	$\pm$	5.7	\\
Mrk~1239$\dagger$	&	0.019	&	43.19	$\pm$	0.043	&	43.366	$\pm$	0.051	&	1.34	$\pm$	0.147	&	0.148	$\pm$	0.016	&	1720	$\pm$	130	&	1220	$\pm$	98	&	1240	$\pm$	74	&	64.3	$\pm$	5.3	&	23.3	$\pm$	1.9	&	20.1	$\pm$	2.2	&	74.8	$\pm$	6.2	\\
Mrk~335$\dagger$	&	0.026	&	43.82	$\pm$	0.043	&	43.721	$\pm$	0.053	&	0.818	$\pm$	0.092	&	0.054	$\pm$	0.007	&	1715	$\pm$	130	&	1140	$\pm$	103	&	1490	$\pm$	119	&	123.8	$\pm$	7.6	&	25.2	$\pm$	1.7	&	11.3	$\pm$	1.1	&	100.5	$\pm$	8.8	\\
Mrk~478$\dagger$	&	0.078	&	44.258	$\pm$	0.043	&	44.45	$\pm$	0.044	&	1.023	$\pm$	0.089	&	0.514	$\pm$	0.056	&	1250	$\pm$	100	&	1300	$\pm$	91	&	1560	$\pm$	94	&	76.9	$\pm$	6.1	&	20.8	$\pm$	1.7	&	19.9	$\pm$	1.6	&	55.2	$\pm$	4.9	\\
Mrk~493$\dagger$	&	0.032	&	43.393	$\pm$	0.043	&	43.515	$\pm$	0.044	&	1.721	$\pm$	0.179	&	0.387	$\pm$	0.046	&	1450	$\pm$	110	&	770	$\pm$	31	&	1065	$\pm$	64	&	62.5	$\pm$	5.9	&	13.3	$\pm$	1.1	&	17.5	$\pm$	1.4	&	33.3	$\pm$	1.8	\\
PG~1448+273$\dagger$	&	0.065	&	44.305	$\pm$	0.043	&	44.064	$\pm$	0.044	&	1.189	$\pm$	0.129	&	0.262	$\pm$	0.034	&	1730	$\pm$	135	&	880	$\pm$	35	&	885	$\pm$	44	&	31.2	$\pm$	3.8	&	15.2	$\pm$	1	&	18.8	$\pm$	1.3	&	32.2	$\pm$	1.8	\\
Tons~180$\dagger$	&	0.062	&	44.283	$\pm$	0.043	&	43.911	$\pm$	0.058	&	0.985	$\pm$	0.11	&	0.145	$\pm$	0.015	&	1470	$\pm$	135	&	930	$\pm$	41	&	990	$\pm$	59	&	32.3	$\pm$	1.9	&	9.6	$\pm$	0.7	&	19.5	$\pm$	1.6	&	31.1	$\pm$	2.2	\\
\hline\hline\noalign{\vskip 0.1cm}																							
\noalign{\vskip 0.1cm} 																													
\multicolumn{12}{c}{\citet{2020arXiv200401811M} Sample} \\ 																
\hline \noalign{\vskip 0.1cm}																						
PHL~1092	&	0.394	&	44.883	$\pm$	0.043	&	44.604	$\pm$	0.047	&	2.576	$\pm$	0.108	&	0.839	$\pm$	0.038	&	1850	$\pm$	100	&	1250	$\pm$	100	&	3750	$\pm$	360	&	34.2	$\pm$	2.2	&	24.4	$\pm$	1.8	&	90.2	$\pm$	7.1	&	88.5	$\pm$	5.3	\\
		
\hline

\end{tabular}}

      \footnotesize
      {\sc Notes.} Columns are as follows: (1) Object name. (2) Redshift. (3) Optical continuum luminosity at 5100\AA. (4) NIR luminosity at 8542\AA. (5) and (6) \rfe\ and { \cat}  values, respectively. (7), (8) and (9) Full-width at half maximum of \hb, \oi\ and { \cat} in units of \kms, respectively. (10), (11), (12) and (13) Equivalent width of \hb, \oi, { \cat}  and \feii\  in units of \AA, respectively.
    
\end{table*}

\newpage

\begin{center}
\begin{longtable}[c]{c c c}
\caption{Black hole parameters}

\label{tab:table2} \\
\hline\hline\noalign{\vskip 0.1cm}         
\multirow{2}{*}{Object} & log \mbh  &  \multirow{2}{*}{log~\eddr} \\

 & [$M_\odot$] & \\

(1) & (2) & (4)  \\

\hline\hline\noalign{\vskip 0.1cm}
\noalign{\vskip 0.1cm} 																
\multicolumn{3}{c}{\citet{persson1988} Sample} \\ 
				
\hline \noalign{\vskip 0.1cm}							

IZw1$\dagger$	&	7.935	$^{+	0.177	}_{-	0.177	}$ &	-0.362	$^{+	0.184	}_{-	0.184	}$ 	\\
Mrk~42	&	6.966	$^{+	0.257	}_{-	0.257	}$ &	-0.501	$^{+	0.262	}_{-	0.262	}$ 	\\
Mrk~478$\dagger$	&	8.035	$^{+	0.158	}_{-	0.158	}$ &	-0.461	$^{+	0.165	}_{-	0.165	}$ 	\\
II~Zw~136	&	8.061	$^{+	0.138	}_{-	0.138	}$ &	-0.66	$^{+	0.147	}_{-	0.147	}$ 	\\
Mrk~231	&	8.46	$^{+	0.121	}_{-	0.121	}$ &	-1.069	$^{+	0.131	}_{-	0.131	}$ 	\\
3C~273	&	9.188	$^{+	0.142	}_{-	0.14	}$ &	-0.485	$^{+	0.15	}_{-	0.148	}$ 	\\
Mrk~6	&	7.724	$^{+	0.141	}_{-	0.141	}$ &	-1.053	$^{+	0.149	}_{-	0.149	}$ 	\\
Mrk~486	&	7.707	$^{+	0.148	}_{-	0.148	}$ &	-0.705	$^{+	0.156	}_{-	0.156	}$ 	\\
Mrk~1239$\dagger$	&	7.037	$^{+	0.201	}_{-	0.201	}$ &	-0.684	$^{+	0.207	}_{-	0.206	}$ 	\\
Mrk~766	&	7.177	$^{+	0.207	}_{-	0.207	}$ &	-0.577	$^{+	0.213	}_{-	0.213	}$ 	\\
Zw~0033+45	&	8.077	$^{+	0.151	}_{-	0.151	}$ &	-0.769	$^{+	0.159	}_{-	0.159	}$ 	\\
Mrk~684	&	7.77	$^{+	0.154	}_{-	0.154	}$ &	-0.615	$^{+	0.161	}_{-	0.161	}$ 	\\
Mrk~335$\dagger$	&	7.709	$^{+	0.148	}_{-	0.148	}$ &	-0.704	$^{+	0.156	}_{-	0.156	}$ 	\\
Mrk~376	&	8.419	$^{+	0.121	}_{-	0.121	}$ &	-1.099	$^{+	0.131	}_{-	0.131	}$ 	\\
Mrk~493$\dagger$	&	7.129	$^{+	0.276	}_{-	0.276	}$ &	-0.361	$^{+	0.28	}_{-	0.28	}$ 	\\
Mrk~841	&	8.232	$^{+	0.12	}_{-	0.12	}$ &	-1.107	$^{+	0.13	}_{-	0.13	}$ 	\\
Ton~1542	&	8.181	$^{+	0.122	}_{-	0.122	}$ &	-0.989	$^{+	0.131	}_{-	0.131	}$ 	\\
VII~Zw~118	&	8.432	$^{+	0.124	}_{-	0.124	}$ &	-0.849	$^{+	0.134	}_{-	0.134	}$ 	\\
Mrk~124	&	7.389	$^{+	0.168	}_{-	0.168	}$ &	-0.69	$^{+	0.175	}_{-	0.175	}$ 	\\
Mrk~9	&	8.118	$^{+	0.123	}_{-	0.123	}$ &	-0.968	$^{+	0.132	}_{-	0.132	}$ 	\\
NGC~7469	&	7.875	$^{+	0.142	}_{-	0.142	}$ &	-0.957	$^{+	0.15	}_{-	0.15	}$ 	\\
Akn~120	&	8.353	$^{+	0.121	}_{-	0.121	}$ &	-1.177	$^{+	0.13	}_{-	0.13	}$ 	\\
Mrk~352	&	7.553	$^{+	0.126	}_{-	0.126	}$ &	-1.297	$^{+	0.136	}_{-	0.135	}$ 	\\
Mrk~304	&	8.289	$^{+	0.135	}_{-	0.135	}$ &	-0.858	$^{+	0.144	}_{-	0.144	}$ 	\\
Mrk~509	&	8.506	$^{+	0.125	}_{-	0.125	}$ &	-1.074	$^{+	0.134	}_{-	0.134	}$ 	\\
														
\hline\hline\noalign{\vskip 0.1cm}														
\noalign{\vskip 0.1cm} 														
\multicolumn{3}{c}{\citet{martinez-aldamaetal15, martinez-aldamaetal15b} Sample} \\  														
\hline \noalign{\vskip 0.1cm}														
														
HE~1349+0007	&	9.834	$^{+	0.183	}_{-	0.179	}$ &	-0.313	$^{+	0.189	}_{-	0.185	}$ 	\\
HE~1409+0101	&	10.023	$^{+	0.178	}_{-	0.173	}$ &	-0.094	$^{+	0.184	}_{-	0.179	}$ 	\\
HE~2349-3800	&	9.746	$^{+	0.166	}_{-	0.162	}$ &	-0.233	$^{+	0.172	}_{-	0.168	}$ 	\\
HE~2147-3212	&	9.817	$^{+	0.219	}_{-	0.216	}$ &	-0.260	$^{+	0.225	}_{-	0.222	}$ 	\\
HE~2202-2557	&	9.981	$^{+	0.181	}_{-	0.177	}$ &	-0.418	$^{+	0.188	}_{-	0.184	}$ 	\\
HE~2340-4443	&	9.479	$^{+	0.157	}_{-	0.154	}$ &	-0.246	$^{+	0.164	}_{-	0.160	}$ 	\\
HE~0248-3628	&	9.917	$^{+	0.174	}_{-	0.169	}$ &	-0.119	$^{+	0.180	}_{-	0.175	}$ 	\\
HE~2352-4010	&	9.960	$^{+	0.180	}_{-	0.175	}$ &	0.048	$^{+	0.185	}_{-	0.180	}$ 	\\
HE~0035-2853	&	9.943	$^{+	0.183	}_{-	0.178	}$ &	-0.270	$^{+	0.189	}_{-	0.185	}$ 	\\
HE~0048-2804	&	9.286	$^{+	0.163	}_{-	0.161	}$ &	-0.634	$^{+	0.171	}_{-	0.169	}$ 	\\
HE~0058-3231	&	9.956	$^{+	0.169	}_{-	0.165	}$ &	-0.262	$^{+	0.176	}_{-	0.172	}$ 	\\
HE~0203-4627	&	9.856	$^{+	0.219	}_{-	0.216	}$ &	-0.348	$^{+	0.224	}_{-	0.222	}$ 	\\
HE~0005-2355	&	9.838	$^{+	0.220	}_{-	0.217	}$ &	-0.283	$^{+	0.226	}_{-	0.223	}$ 	\\
HE~0043-2300	&	9.859	$^{+	0.172	}_{-	0.167	}$ &	-0.106	$^{+	0.179	}_{-	0.175	}$ 	\\
HE~0349-5249	&	9.990	$^{+	0.199	}_{-	0.195	}$ &	-0.110	$^{+	0.205	}_{-	0.201	}$ 	\\
HE~0359-3959	&	9.758	$^{+	0.190	}_{-	0.187	}$ &	-0.227	$^{+	0.196	}_{-	0.193	}$ 	\\
HE~0436-3709	&	9.703	$^{+	0.188	}_{-	0.185	}$ &	-0.317	$^{+	0.198	}_{-	0.195	}$ 	\\
HE~0507-3236	&	9.658	$^{+	0.190	}_{-	0.186	}$ &	-0.156	$^{+	0.195	}_{-	0.192	}$ 	\\
HE~0512-3329	&	9.788	$^{+	0.192	}_{-	0.188	}$ &	-0.184	$^{+	0.197	}_{-	0.193	}$ 	\\
HE~0926-0201	&	9.877	$^{+	0.201	}_{-	0.196	}$ &	0.007	$^{+	0.207	}_{-	0.203	}$ 	\\
HE~1039-0724	&	9.793	$^{+	0.188	}_{-	0.185	}$ &	-0.345	$^{+	0.195	}_{-	0.191	}$ 	\\
HE~1120+0154	&	10.105	$^{+	0.200	}_{-	0.195	}$ &	-0.225	$^{+	0.205	}_{-	0.201	}$ 	\\
														
\hline\hline\noalign{\vskip 0.1cm}														
\noalign{\vskip 0.1cm} 														
\multicolumn{3}{c}{\citet{murilo2016} Sample} \\ 														
\hline \noalign{\vskip 0.1cm}														
														
1H~1934-063$\dagger$	&	6.985	$^{+	0.156	}_{-	0.155	}$ &	-1.059	$^{+	0.162	}_{-	0.161	}$ 	\\
1H~2107-097$\dagger$	&	7.507	$^{+	0.178	}_{-	0.178	}$ &	-1.107	$^{+	0.183	}_{-	0.183	}$ 	\\
IZw1$\dagger$	&	7.907	$^{+	0.151	}_{-	0.151	}$ &	-0.605	$^{+	0.158	}_{-	0.158	}$ 	\\
Mrk~1044$\dagger$	&	7.259	$^{+	0.162	}_{-	0.162	}$ &	-0.973	$^{+	0.168	}_{-	0.167	}$ 	\\
Mrk~1239$\dagger$	&	7.353	$^{+	0.151	}_{-	0.15	}$ &	-0.975	$^{+	0.157	}_{-	0.157	}$ 	\\
Mrk~335$\dagger$	&	7.688	$^{+	0.148	}_{-	0.148	}$ &	-0.805	$^{+	0.155	}_{-	0.155	}$ 	\\
Mrk~478$\dagger$	&	7.807	$^{+	0.156	}_{-	0.156	}$ &	-0.575	$^{+	0.162	}_{-	0.162	}$ 	\\
Mrk~493$\dagger$	&	7.399	$^{+	0.152	}_{-	0.152	}$ &	-0.859	$^{+	0.158	}_{-	0.158	}$ 	\\
PG~1448+273$\dagger$	&	7.949	$^{+	0.15	}_{-	0.149	}$ &	-0.679	$^{+	0.156	}_{-	0.156	}$ 	\\
Tons~180$\dagger$	&	7.879	$^{+	0.16	}_{-	0.16	}$ &	-0.626	$^{+	0.166	}_{-	0.166	}$ 	\\

\hline\hline\noalign{\vskip 0.1cm}														
\noalign{\vskip 0.1cm} 														
\multicolumn{3}{c}{\citet{2020arXiv200401811M} Sample} \\ 														
\hline \noalign{\vskip 0.1cm}														
														
PHL~1092	&	8.281	$^{+	0.140	}_{-	0.140	}$ &	-0.549	$^{+	0.147	}_{-	0.147	}$ 	\\

\hline

\end{longtable}
\end{center}

\footnotesize{{\sc NOTES.} Columns are as follows: (1) Object name. (2) Black hole mass estimated using the classical RL relation (Eq.~\ref{equ:mass_bentz}), in units of M$_\odot$ . (3) Eddington ratio.}

\newpage

\begin{table*}[hbt!]
\caption{Parameters of the correlations for the observed and bootstrap sample}
\resizebox{\columnwidth}{!}{
\begin{tabular}[c]{c c c c c c c c c c c c c c c }
\label{tab:params_corr} \\
\hline\hline\noalign{\vskip 0.1cm}         

 & & \multicolumn{5}{c}{Observational data} & & \multicolumn{6}{c}{Bootstrap results} & \\ \cmidrule{3-7} \cmidrule{9-15}
 
\multicolumn{2}{c}{Relation} &  $\alpha$ & $\beta$ & $\rho$ & $p$-value & $\sigma$ & & $\alpha_{\rm BS}$ & $\beta_{\rm BS}$ & ${\rho_{\rm BS}}$ & $f_{ sig}$ & P. Dist. & $p-$value$_{\rm ran}$ & $f_{\rm ran}$\\ 

\multicolumn{2}{c}{(1)} &  (2) &  (3) & (4) & (5) & (6) & & (7) & (8) & (9) & (10) & (11) & (12) & (13) \\

\hline\hline\noalign{\vskip 0.1cm}         

\multirow{4}{*}{\ewhb}	&	\lopt\		&	-0.064	$\pm$	0.032	&	4.632	$\pm$	1.469	&	-0.412	&	0.001	&	0.200	& &	-0.064	$\pm$	0.031	&	4.618	$\pm$	1.408	&	-0.428$^{+0.21}_{-0.17}$	&	0.5	& c,a,b	& 0.76,0.95, 0.67  & 3.5$\times10^{-3}$	\\
	&	\lnir\		&	-0.064	$\pm$	0.036	&	4.627	$\pm$	1.618	&	-0.352	&	0.007	&	0.204	& &	-0.065	$\pm$	0.034	&	4.656	$\pm$	1.430	&	-0.372$^{+0.22}_{-0.17}$	&	0.2	& c,c,d & 0.76,0.98,0.60 &	1.0$\times10^{-3}$	\\
	&	\mbh\		&	-0.091	$\pm$	0.053	&	2.508	$\pm$	0.455	&	-0.390	&	0.002	&	0.206 &	&	-0.092	$\pm$	0.048	&	2.519	$\pm$	0.409	&	-0.405$^{+0.24}_{-0.19}$	&	0.4	& c,b,d & 0.76,0.96,0.94 &	1.0$\times10^{-3}$	\\
	&	\eddr\		&	{\bf  -0.332	$\pm$	0.149}	&	{\bf  1.53	$\pm$	0.102}	&	{\bf  -0.531}	&	{\bf  1.78$\times10^{-5}$}	&	{\bf  0.195} &	&	{\bf  -0.335	$\pm$	0.148}	&	{\bf  1.529	$\pm$	0.09}	&	{\bf  -0.544$^{+0.19}_{-0.16}$}	&	{\bf  0.85}	& {\bf c,c} &  {\bf 0.76,0.61} &	{\bf 1.5$\times10^{-3}$}	\\
																																
		\hline \noalign{\vskip 0.1cm}																														
																																
\multirow{4}{*}{\ewoi}	&	\lopt\		&	-0.007	$\pm$	0.034	&	1.557	$\pm$	1.548	&	-0.015	&	0.910	&	0.211 &	&	-0.007	$\pm$	0.033	&	1.000	$\pm$	1.510	&	-0.016$^{+0.23}_{-0.27}$	&	0.0	& e,a,b & 0.97,0.95,0.67 &	1.0$\times10^{-3}$	\\
	&	\lnir\		&	-0.003	$\pm$	0.037	&	1.358	$\pm$	1.673	&	0.053	&	0.691	&	0.211 &	&	-0.003	$\pm$	0.034	&	1.358	$\pm$	1.603	&	0.060$^{+0.20}_{-0.28}$	&	0.0	& e,c,d &  0.97,0.98,0.60 &	1.5$\times10^{-3}$	\\
	&	\mbh\		&	-0.006	$\pm$	0.054	&	1.271	$\pm$	0.468	&	-0.008	&	0.951	&	0.211 &	&	-0.005	$\pm$	0.057	&	1.264	$\pm$	0.459	&	0.008$^{+0.23}_{-0.27}$	&	0.0	& e,b,d & 0.97,0.96,0.94 &	1.0$\times10^{-3}$	\\
	&	\eddr\		&	-0.079	$\pm$	0.161	&	1.172	$\pm$	0.110	&	-0.162	&	0.223	&	0.209 &	&	-0.082	$\pm$	0.142	&	1.169	$\pm$	0.107	&	-0.161$^{+0.23}_{-0.26}$	&	0.0	& e,c & 0.97,0.61 &	5.0$\times10^{-4}$ 	\\
																																
		\hline \noalign{\vskip 0.1cm}																														
																																
\multirow{4}{*}{\ewfe}	&	\lopt\		&	-0.091	$\pm$	0.039	&	5.666	$\pm$	1.754	&	-0.409	&	0.001	&	0.239	& &	-0.092	$\pm$	0.039	&	5.678	$\pm$	1.705	&	-0.422$^{+0.22}_{-0.17}$	&	0.5	& a,a,b & 0.94,0.95,0.67 &	5.0$\times10^{-4}$	\\
	&	\lnir\		&	-0.095	$\pm$	0.042	&	5.846	$\pm$	1.917	&	-0.367	&	0.005	&	0.242 &	&	-0.096	$\pm$	0.043	&	5.881	$\pm$	1.809	&	-0.385$^{+0.23}_{-0.18}$	&	0.3	& a,c,d & 0.94,0.95,0.55 & 	5.0$\times10^{-4}$	\\
	&	\mbh\		&	{\bf  -0.151	$\pm$	0.06}	&	{\bf  2.831	$\pm$	0.52}	&	{\bf -0.493}	&	{\bf  8.58$\times10^{-5}$}	&	{\bf  0.234} &	&	{\bf  -0.15	$\pm$	0.059}	&	{\bf  2.824	$\pm$	0.52}	&	{\bf  -0.517$^{+0.21}_{-0.14}$}	&	{\bf  0.75}	& {\bf a,b,d} & {\bf 0.94,0.96,0.94} &	{\bf  1.0$\times10^{-3}$}	\\
	&	\eddr\		&	-0.281	$\pm$	0.203	&	1.373	$\pm$	0.139	&	-0.360	&	0.005	&	0.265	& &	-0.279	$\pm$	0.211	&	1.374	$\pm$	0.142	&	-0.375$^{+0.22}_{-0.19}$	&	0.3	& a,c & 0.94,0.61 &	5.0$\times10^{-4}$	\\
																																
		\hline \noalign{\vskip 0.1cm}																														
																																
\multirow{4}{*}{\ewca}	&	\lopt\		&	0.068	$\pm$	0.053	&	-1.851	$\pm$	2.381	&	0.417	&	0.001	&	0.324 &	&	0.069	$\pm$	0.049	&	-1.896	$\pm$	1.842	&	0.425$^{+0.14}_{-0.25}$	&	0.5	& a,a,b & 0.93,0.95,0.67 &	1.0$\times10^{-3}$ 	\\
	&	\lnir\		&	0.068	$\pm$	0.057	&	-1.827	$\pm$	2.593	&	0.377	&	0.004	&	0.327 &	&	0.069	$\pm$	0.051	&	-1.878	$\pm$	2.451	&	0.380$^{+0.15}_{-0.24}$	&	0.3	& a,c,d & 0.93,0.98, 0.60 &	1.5$\times10^{-3}$	\\
	&	\mbh\		&	0.088	$\pm$	0.085	&	0.470	$\pm$	0.733	&	0.301	&	0.022	&	0.331 &	&	0.089	$\pm$	0.074	&	0.456	$\pm$	0.608	&	0.305$^{+0.17}_{-0.23}$	&	0.1	& a,b,d & 0.93,0.96,0.94 & 1.0$\times10^{-3}$	\\
	&	\eddr\		&	{\bf 0.428	$\pm$	0.237}	&	{\bf  1.477	$\pm$	0.162}	&	{\bf  0.482}	&	{\bf  1.27$\times10^{-4}$}	&	{\bf  0.309} &	&	{\bf  0.431	$\pm$	0.200}	&	{\bf  1.477	$\pm$	0.133}	&	{\bf  0.494$^{+0.14}_{-0.24}$}	&	{\bf  0.71}	& {\bf a,c} & {\bf 0.93,0.61} &	{\bf 1.0$\times10^{-3}$}	\\
																																
		\hline \noalign{\vskip 0.1cm}																														
																																
\multirow{4}{*}{\rfe}	&	\lopt\		&	-0.015	$\pm$	0.049	&	0.520	$\pm$	2.209	&	-0.063	&	0.640	&	0.301 &	&	-0.015	$\pm$	0.045	&	0.536	$\pm$	1.901	&	-0.064$^{+0.22}_{-0.28}$	&	0.0 &  e,a,b &	0.96,0.95,0.67 &	1.5$\times10^{-3}$	\\
	&	\lnir\		&	-0.017	$\pm$	0.053	&	0.631	$\pm$	2.383	&	-0.081	&	0.546	&	0.301 &	&	-0.017	$\pm$	0.047	&	0.602	$\pm$	2.264	&	-0.083$^{+0.22}_{-0.26}$	&	0.0	& e,c,d & 0.96,0.98,0.60 &	5.0$\times10^{-4}$	\\
	&	\mbh\		&	-0.042	$\pm$	0.076	&	0.204	$\pm$	0.662	&	-0.162	&	0.226	&	0.299 &	&	-0.042	$\pm$	0.065	&	0.205	$\pm$	0.566	&	-0.160$^{+0.22}_{-0.25}$	&	0.0	& e,b,d &  0.96,0.96,0.94 &	2.0$\times10^{-3}$	\\
	&	\eddr\		&	0.109	$\pm$	0.230	&	-0.092	$\pm$	0.157	&	0.105	&	0.432	&	0.300 &	&	0.111	$\pm$	0.239	&	-0.088	$\pm$	0.141	&	0.102$^{+0.25}_{-0.29}$	&	0.0	& e,c & 0.96,0.61 &	1.0$\times10^{-3}$	\\
																																
		\hline \noalign{\vskip 0.1cm}																														
																																
\multirow{4}{*}{\rcat}	&	\lopt\		&	0.057	$\pm$	0.063	&	-3.392	$\pm$	2.846	&	0.247	&	0.061	&	0.388 &	&	0.056	$\pm$	0.056	&	-3.350	$\pm$	2.437	&	0.243$^{+0.18}_{-0.23}$	&	0.1	& f,a,b & 0.76,0.95,0.67 &	3.5$\times10^{-3}$	\\
	&	\lnir\		&	0.059	$\pm$	0.068	&	-3.476	$\pm$	3.079	&	0.240	&	0.069	&	0.389 &	&	0.060	$\pm$	0.058	&	-3.502	$\pm$	2.823	&	0.241$^{+0.18}_{-0.24}$	&	0.1	& f,c,d & 0.76,0.98,0.60 &	1.5$\times10^{-3}$	\\
	&	\mbh\		&	0.058	$\pm$	0.101	&	-1.309	$\pm$	0.874	&	0.117	&	0.381	&	0.394 & 	&	0.058	$\pm$	0.091	&	-1.306	$\pm$	0.762	&	0.111$^{+0.20}_{-0.24}$	&	0.0	& f,b,d & 0.76,0.96,0.94 &	1.5$\times10^{-3}$	\\
	&	\eddr\		&	{\bf 0.500	$\pm$	0.275}	&	{\bf  -0.516	$\pm$	0.188}	&	{\bf  0.425}	&	{\bf  8.90$\times10^{-4}$}	&	{\bf  0.359} &	&	{\bf  0.498	$\pm$	0.254}	&	{\bf  -0.516	$\pm$	0.152}	&	{\bf  0.426$^{+0.18}_{-0.25}$}	&	{\bf  0.51}	& {\bf f,c} & {\bf 0.76,0.61} &	{\bf  5.0$\times10^{-4}$}	\\
																																
		\hline \noalign{\vskip 0.1cm}																														
																																
\multirow{4}{*}{\feii/\cat}	&	\lopt\		&	{\bf -0.072	$\pm$	0.039}	&	{\bf  3.912	$\pm$	1.78}	&	{\bf  -0.441}	&	{\bf  5.31$\times10^{-4}$}	&	{\bf  0.242} &	&	{\bf  -0.071	$\pm$	0.039}	&	{\bf  3.877	$\pm$	1.747}	&	{\bf  -0.444$^{+0.22}_{-0.22}$}	&	{\bf  0.54}	& {\bf e,a,b} & {\bf 0.98,0.95,0.67} &	{\bf  5.0$\times10^{-4}$}	\\
	&	\lnir\		&	{\bf  -0.077	$\pm$	0.043}	&	{\bf  4.107	$\pm$	1.929}	&	{\bf  -0.456}	&	{\bf  3.20$\times10^{-4}$} 	&	{\bf  0.244} &	&	{\bf  -0.076	$\pm$	0.042}	&	{\bf  4.086	$\pm$	1.887}	&	{\bf  -0.470$^{+0.23}_{-0.20}$}	&	{\bf 0.61}	& {\bf e,c,d} & {\bf 0.98,0.98,0.60} &	{\bf  5.0$\times10^{-4}$}	\\
	&	\mbh\		&	-0.100	$\pm$	0.064	&	1.513	$\pm$	0.552	&	-0.354	&	0.006	&	0.249 &	&	-0.102	$\pm$	0.061	&	1.532	$\pm$	0.559	&	-0.348$^{+0.23}_{-0.24}$	&	0.3	& e,b,d & 0.98,0.96,0.94 &	1.5$\times10^{-3}$	\\
	&	\eddr\		&	{\bf  -0.391	$\pm$	0.179}	&	{\bf  0.424	$\pm$	0.122}	&	{\bf  -0.554}	&	{\bf  6.44$\times10^{-6}$}	&	{\bf  0.233} &	&	{\bf  -0.394	$\pm$	0.164}	&	{\bf  0.422	$\pm$	0.099}	&	{\bf  -0.560$^{+0.20}_{-0.17}$}	&	{\bf  0.89}	& {\bf e,c} & {\bf 0.98,0.61} &	{\bf  1.0$\times10^{-3}$}	\\
																																
		\hline \noalign{\vskip 0.1cm}																														
																																
{\rfe} 	&	{\rcat}		&	{\bf  0.973	$\pm$	0.239}	&	{\bf  -0.657	$\pm$	0.081}	&	{\bf  0.721}	&	{\bf  1.75$\times10^{-10}$}	&	{\bf  0.270} &	&	{\bf  0.974	$\pm$	0.189}	&	{\bf  -0.658	$\pm$	0.07}	&	{\bf  0.737$^{+0.07}_{-0.19}$}	&	{\bf  1.0}	& {\bf e,f} & {\bf 0.96,0.76} &	{\bf  2.0$\times10^{-3}$}	\\

\hline 		

\end{tabular}}

      \footnotesize
      {\sc NOTES.}  Columns are as follow: (1) Relations. (2) Slope of the observational sample and error at $2\sigma$. (3)  Ordinate of the observational sample and error at $2\sigma$. (4) Spearman rank correlation coefficient for the observational sample. Significant correlations respect to this parameter are bold--faced. (5) $p-$value of the correlation coefficient. (6) Scatter of the observational sample respect to the best fit. (7) Slope of the bootstrap sample and error at $2\sigma$. (8) Ordinate of the bootstrap sample and error at $2\sigma$. (9) Maximum of Spearman rank correlation coefficient distribution for 1000 realizations of the bootstrap sample and the errors at $2\sigma$. (10) Fraction of significant bootstrap realizations respect to the total number. (11) Probability distributions used to model the observational distributions using a random sample. The symbols are as follow: a--skewnorm, b--powernorm, c--powerlaw, d--loglaplace, e--powerlognorm, f--lognorm. The symbols follow the order of col. (1). In correlations involving luminosity and black hole mass, 2 distributions where used in the modelling. (12) $p-$value of the Kolmogorov-Smirnoff test to select the best distribution fitting. Order is the same as col. (11). (13) Fraction of significant correlations assuming a two random samples. 
\end{table*}

\newpage

\begin{center}
\begin{longtable}[c]{cccccccc}
\caption{Correlations  between the first four eigenvectors and the physical parameters}
\label{tab:pca-correlations}\\

\hline\hline\noalign{\vskip 0.1cm}  

\multicolumn{2}{c}{\multirow{2}{*}{Relations}} & \multicolumn{2}{c}{Full} & \multicolumn{2}{c}{Low--$L$} & \multicolumn{2}{c}{High--$L$} \\
\multicolumn{2}{c}{} & $\rho$ & $p$--value & $\rho$ & $p$--value & $\rho$ & $p$--value \\

\multicolumn{2}{c}{(1)} & (2) &  (3) & (4) & (5) & (6) & (7)
\endfirsthead
\endhead

\hline\hline\noalign{\vskip 0.1cm}  

\multirow{14}{*}{PC1} & \fwhmhb\ & \textbf{-0.792} & \textbf{1.32$\times 10^{-13}$} & \textbf{0.669} & \textbf{7.37$\times 10^{-5}$} & \textbf{0.76} & \textbf{1.73$\times 10^{6}$} \\
 & \fwhmoi\ & \textbf{-0.845} & \textbf{7.51$\times 10^{-17}$} & \textbf{0.728} & \textbf{7.65$\times 10^{-6}$} & \textbf{0.884} & \textbf{2.1$\times 10^{-10}$} \\
 & \fwhmca\ & \textbf{-0.844} & \textbf{1.77$\times 10^{-13}$} & 0.659 & 0.00159 & \textbf{0.85} & \textbf{3.94$\times 10^{-8}$} \\
 & \ewhb & \textbf{0.583} & \textbf{1.59$\times 10^{-6}$} & \textbf{0.736} & \textbf{5.46$\times 10^{-6}$} & \textbf{-0.694} & \textbf{2.93$\times 10^{-5}$} \\
 & \ewoi\ & 0.180 & 0.177 & \textbf{0.871} & \textbf{8.11$\times 10^{-10}$} & -0.296 & 0.12 \\
 & \ewfe\ & \textbf{0.703} & \textbf{7.79$\times 10^{-10}$} & 0.476 & 0.00897 & \textbf{-0.681} & \textbf{4.73$\times 10^{-5}$} \\
 & \ewca\ & -0.179 & 0.178 & -0.072 & 0.712 & -0.257 & 0.178 \\
 & $z$ & \textbf{-0.700} & \textbf{1$\times 10^{-9}$} & 0.356 & 0.0578 & 0.366 & 0.0512 \\
 & log \lbol{} & \textbf{-0.748} & \textbf{1.49$\times 10^{-11}$} & 0.495 & 0.00638 & 0.458 & 0.0124 \\
 & log \mbh{} & \textbf{-0.845} & \textbf{6.99$\times 10^{-17}$} & \textbf{0.701} & \textbf{2.24$\times 10^{-5}$} & \textbf{0.662} & \textbf{9.16$\times 10^{-5}$} \\
 & log \eddr{} & \textbf{-0.519} & \textbf{2.94$\times 10^{-5}$} & -0.492 & 0.00672 & 0.174 & 0.367 \\
 & log \rfe{} & 0.164 & 0.218 & -0.316 & 0.0952 & -0.089 & 0.645 \\
 & log \rcat{} & -0.151 & 0.259 & -0.382 & 0.041 & -0.021 & 0.914 \\
 & log \feca\ & \textbf{0.426} & \textbf{8.46$\times 10^{-4}$} & 0.338 & 0.073 & -0.125 & 0.518 \\

\hline \noalign{\vskip 0.1cm}

\multirow{14}{*}{PC2} & \fwhmhb\ & 0.023 & 0.864 & -0.344 & 0.0674 & -0.278 & 0.145 \\
 & \fwhmoi\ & -0.033 & 0.804 & -0.38 & 0.0422 & 0.057 & 0.769 \\
 & \fwhmca\ & -0.098 & 0.518 & 0.456 & 0.0435 & -0.016 & 0.939 \\
 & \ewhb\ & -0.012 & 0.93 & -0.212 & 0.27 & -0.327 & 0.0835 \\
 & \ewoi\ & \textbf{-0.681} & \textbf{3.92$\times 10^{-9}$} & 0.348 & 0.064 & 0.346 & 0.0661 \\
 & \ewfe\ & \textbf{-0.509} & \textbf{4.57$\times 10^{-5}$} & \textbf{0.66} & \textbf{9.88$\times 10^{-5}$} & 0.357 & 0.0572 \\
 & \ewca\ & \textbf{-0.81} & \textbf{1.29$\times 10^{-14}$} & \textbf{0.855} & \textbf{3.57$\times 10^{-9}$} & \textbf{0.86} & \textbf{2.3$\times 10^{-9}$} \\
 & $z$ & -0.291 & 0.0267 & 0.069 & 0.722 & 0.5 & 0.00569 \\
 & log \lbol{} & -0.268 & 0.0416 & 0.065 & 0.738 & 0.492 & 0.00666 \\
 & log \mbh{} & -0.169 & 0.205 & -0.233 & 0.223 & 0.306 & 0.106 \\
 & log \eddr{}  & 0.272 & 0.0392 & 0.354 & 0.0598 & 0.492 & 0.00669 \\
 & log \rfe{} & \textbf{-0.46} & \textbf{2.79$\times 10^{-4}$} & \textbf{0.698} & \textbf{2.6$\times 10^{-5}$} & \textbf{0.669} & \textbf{7.26$\times 10^{-5}$} \\
 & log \rcat{}  & \textbf{-0.506} & \textbf{5.08$\times 10^{-5}$} & \textbf{0.65} & \textbf{1.34$\times 10^{-4}$} & \textbf{0.838} & \textbf{1.46$\times 10^{-8}$} \\
 & log \feca\ & 0.225 & 0.0888 & -0.257 & 0.178 & -0.338 & 0.0725 \\
 
\hline \noalign{\vskip 0.1cm}

\multirow{14}{*}{PC3} & \fwhmhb\ & \textbf{-0.512} & \textbf{3.97$\times 10^{-5}$} & 0.196 & 0.308 & -0.046 & 0.812 \\
 & \fwhmoi\ & \textbf{-0.506} & \textbf{5.15$\times 10^{-5}$} & -0.036 & 0.852 & 0.074 & 0.701 \\
 & \fwhmca\ & \textbf{-0.47} & \textbf{9.7$\times 10^{-4}$} & -0.244 & 0.301 & 0.042 & 0.837 \\
 & \ewhb\ & \textbf{-0.647} & \textbf{4.18$\times 10^{-8}$} & \textbf{-0.743} & \textbf{3.87$\times 10^{-6}$} & 0.478 & 0.00874 \\
 & \ewoi\ & \textbf{-0.562} & \textbf{4.39$\times 10^{-6}$} & -0.342 & 0.0693 & \textbf{0.83} & \textbf{2.57$\times 10^{-8}$} \\
 & \ewfe\ & 0.027 & 0.838 & -0.244 & 0.202 & -0.104 & 0.59 \\
 & \ewca\ & 0.318 & 0.0149 & 0.251 & 0.189 & 0.065 & 0.739 \\
 & $z$ & -0.007 & 0.959 & 0.023 & 0.905 & 0.119 & 0.54 \\
 & log \lbol{} & -0.041 & 0.761 & 0.14 & 0.469 & 0.02 & 0.916 \\
 & log \mbh{} & -0.195 & 0.142 & 0.185 & 0.337 & -0.064 & 0.741 \\
 & log \eddr{} & 0.341 & 0.00884 & -0.124 & 0.521 & 0.097 & 0.617 \\
 & log \rfe{} & \textbf{0.621} & \textbf{2$\times 10^{-7}$} & 0.4 & 0.0313 & -0.501 & 0.00564 \\
 & log \rcat{} & \textbf{0.600} & \textbf{6.52$\times 10^{-7}$} & 0.486 & 0.00759 & -0.173 & 0.368 \\
 & log \feca\ & -0.232 & 0.0794 & -0.351 & 0.0621 & -0.478 & 0.00875 \\
 
\hline \noalign{\vskip 0.1cm}

\multirow{14}{*}{PC4} & \fwhmhb\ & 0.279 & 0.034 & -0.096 & 0.619 & 0.378 & 0.0431 \\
 & \fwhmoi\ & 0.190 & 0.154 & -0.187 & 0.332 & 0.313 & 0.0985 \\
 & \fwhmca\ & 0.118 & 0.435 & \textbf{-0.699} & \textbf{6.02$\times 10^{-4}$} & 0.423 & 0.0313 \\
 & \ewhb\ & -0.381 & 0.00315 & -0.054 & 0.78 & -0.1 & 0.608 \\
 & \ewoi\ & -0.339 & 0.00934 & 0.075 & 0.698 & 0.017 & 0.929 \\
 & \ewfe\ & 0.130 & 0.331 & 0.041 & 0.832 & 0.164 & 0.397 \\
 & \ewca\ & 0.076 & 0.571 & -0.19 & 0.324 & 0.153 & 0.43 \\
 & $z$ & -0.089 & 0.506 & -0.189 & 0.327 & -0.423 & 0.0224 \\
 & log \lbol{} & -0.057 & 0.67 & -0.266 & 0.163 & -0.469 & 0.0104 \\
 & log \mbh{} & 0.016 & 0.907 & -0.197 & 0.307 & -0.281 & 0.14 \\
 & log \eddr{} & -0.239 & 0.0708 & -0.07 & 0.717 & \textbf{-0.693} & \textbf{3.13$\times 10^{-5}$} \\
 & log \rfe{} & 0.42 & 0.00104 & 0.018 & 0.927 & 0.268 & 0.16 \\
 & log \rcat{}  & 0.207 & 0.119 & -0.279 & 0.143 & 0.18 & 0.35 \\
 & log \feca\ & 0.224 & 0.0915 & 0.361 & 0.0547 & 0.244 & 0.201 \\

\hline 	
\end{longtable}
\end{center}

\footnotesize{{\sc NOTES.} Columns are as follows: (1) Relations. (2), (4) and (6) Spearman rank correlation coefficient for the full, low-- and high--$L$ samples, respectively. (3), (5) and (7) $p-$value of the correlation coefficient, for the full, low-- and high--$L$ samples, respectively. Significant correlations are bold-faced.}

\newpage
\section{Random distribution and residuals plots}
\label{appendix:residual_bs}

\begin{figure*}[hbt!]
    \centering
    \includegraphics[width=\columnwidth]{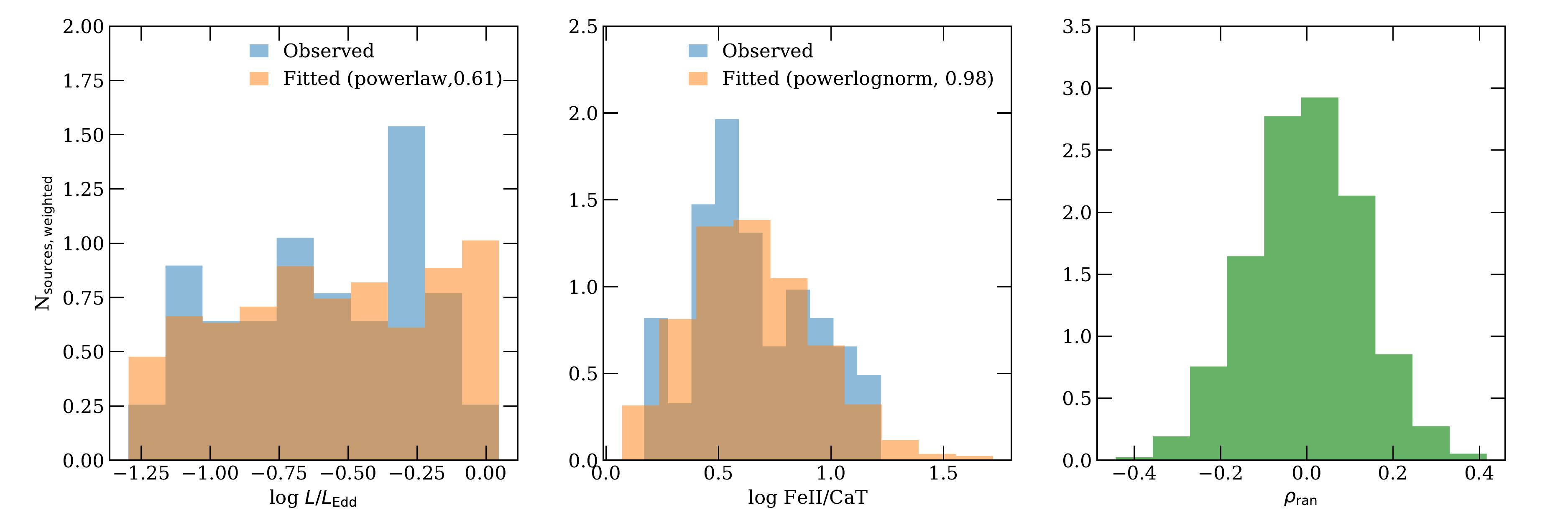}
    \caption{Modeled probability distribution for \eddr\ (left panel) and \feca\ (middle panel) parameters. Blue distribution correspond to the observed one, while the blue one is obtained via a bootstrap analysis, see Sec.\ref{sec:random}. The right panel shows the distribution of the Spearman rank correlation coefficient for 58 randomly selected sources from the distribution in left and middle panel. The number of significant correlations ($\rho_{\rm ran}$ and $p-$val$<0.001$) is below the $3\sigma$ confidence level.}
    \label{fig:dist_ran}
\end{figure*}

{ In this section are included extra figures which are complementary to the results in Sec.~\ref{sec:boots} and \ref{sec:residuals}. Figure~\ref{fig:dist_ran} shows the distributions of \feca\ and \eddr\ parameters compared with the ones from a fitting process, see Sec.~\ref{sec:random}. The  fitted distribution and $p-$value are indicated in left and middle panels. Right panel shows the distribution of a random sample to get a significant correlation \feca-\eddr\ as high as the one from the observational sample. }


\begin{figure*}[hbt!]
    \centering
    \includegraphics[width=\columnwidth]{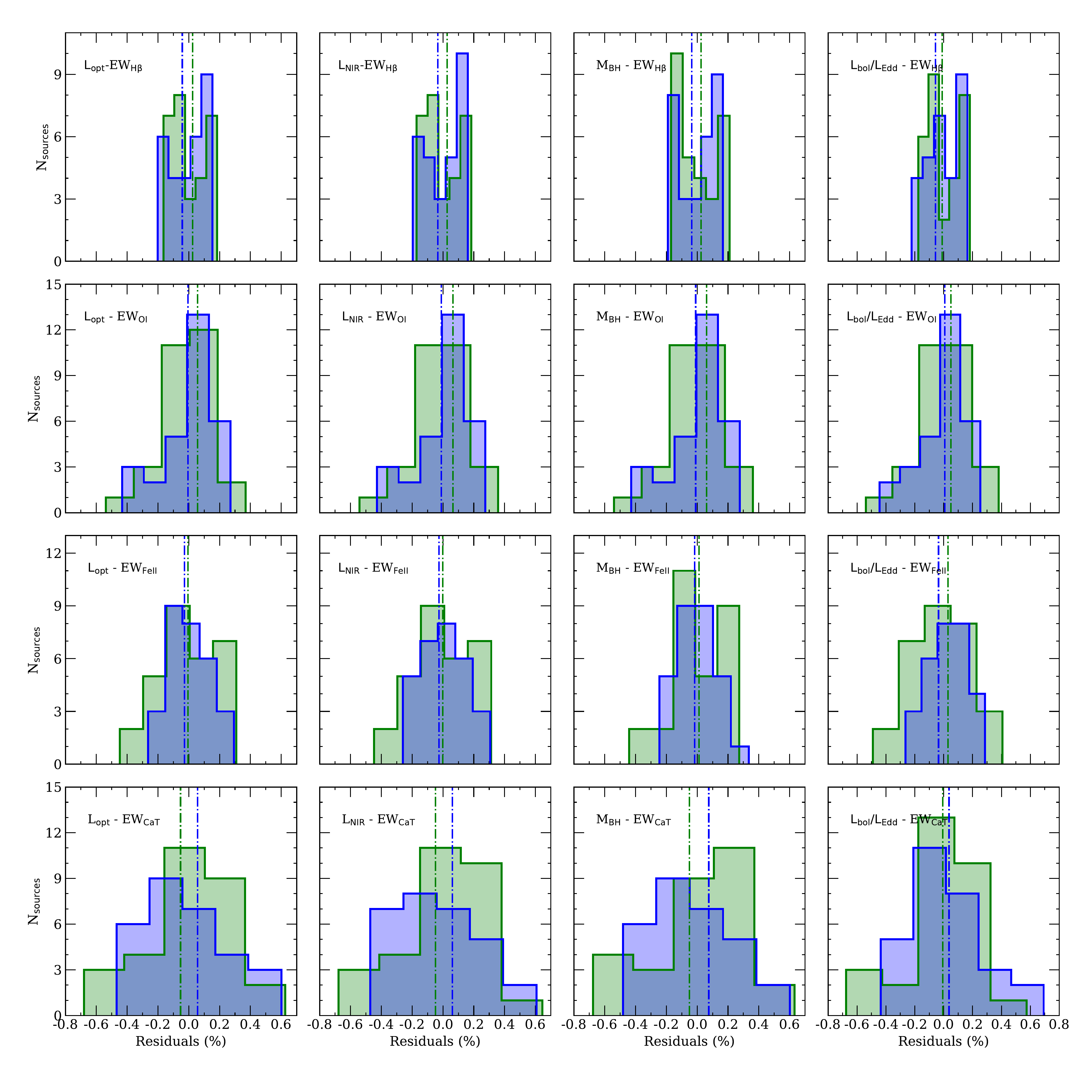}
    \caption{Distribution of the residuals with respect to the the best fits as a function of redshift for the correlation shown in Fig.~\ref{fig:ew_ac}. In each panel is indicated the analyzed relation. Green distribution correspond to the low--$L$ (and low--$z$) subsample, while blue one draws the high--$L$ (and high--$z$) distributions. Vertical lines indicate the median of each distribution.  }
    \label{fig:residual_fig3}
\end{figure*}

\begin{figure*}[hbt!]
    \centering
    \includegraphics[width=\columnwidth]{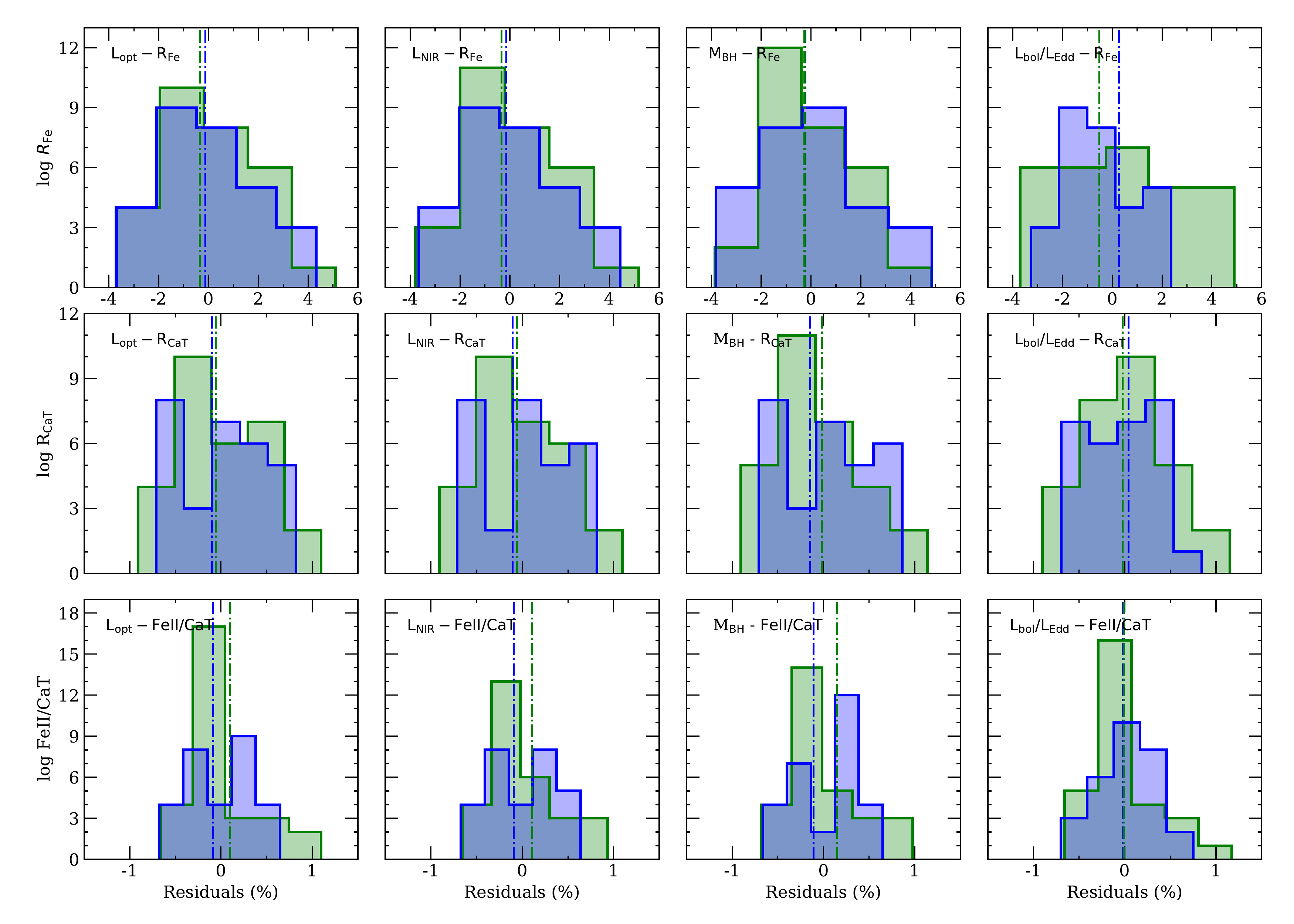}
    \caption{Same as Fig.~\ref{fig:residual_fig3}, but for correlations shown in Fig.~\ref{fig:rfe-rcat_corr}. }
    \label{fig:residual_fig4}
\end{figure*}

{ In order to asses any signatures of redshift effects, Fig.~\ref{fig:residual_fig3} and \ref{fig:residual_fig4} show the distributions of the residuals  as a function of the redshift for low-- and high--$L$ objects. This division is in agreement with low-- and high--redshift sources. The median of each distribution (dashed-vertical lines) does not show a significant difference from the zero residual levels and it is not observed a dependency within  $2\sigma$ confidence level. This suggests that the relation in figures \ref{fig:ew_ac} and \ref{fig:rfe-rcat_corr} are not artificially enhancement by redshift effects.}

\newpage

\section{Principal Component Analysis - effect of redundant parameters}
\label{appendix-pca}

The correlation between a variable and a principal component (PC) is used as the coordinates of the variable on the PC. The representation of variables differs from the plot of the observations: The observations are represented by their projections, but the variables are represented by their correlations \citep{abdi2010}. (a) Positively correlated variables are grouped together; (b) negatively correlated variables are positioned on opposite sides of the plot origin (opposed quadrants); and (c) the distance between variables and the origin measures the quality of the variables on the factor map. Variables that are away from the origin are well represented on the factor map.


\begin{figure*}[hbt!]
    \centering
    \includegraphics[width=\columnwidth]{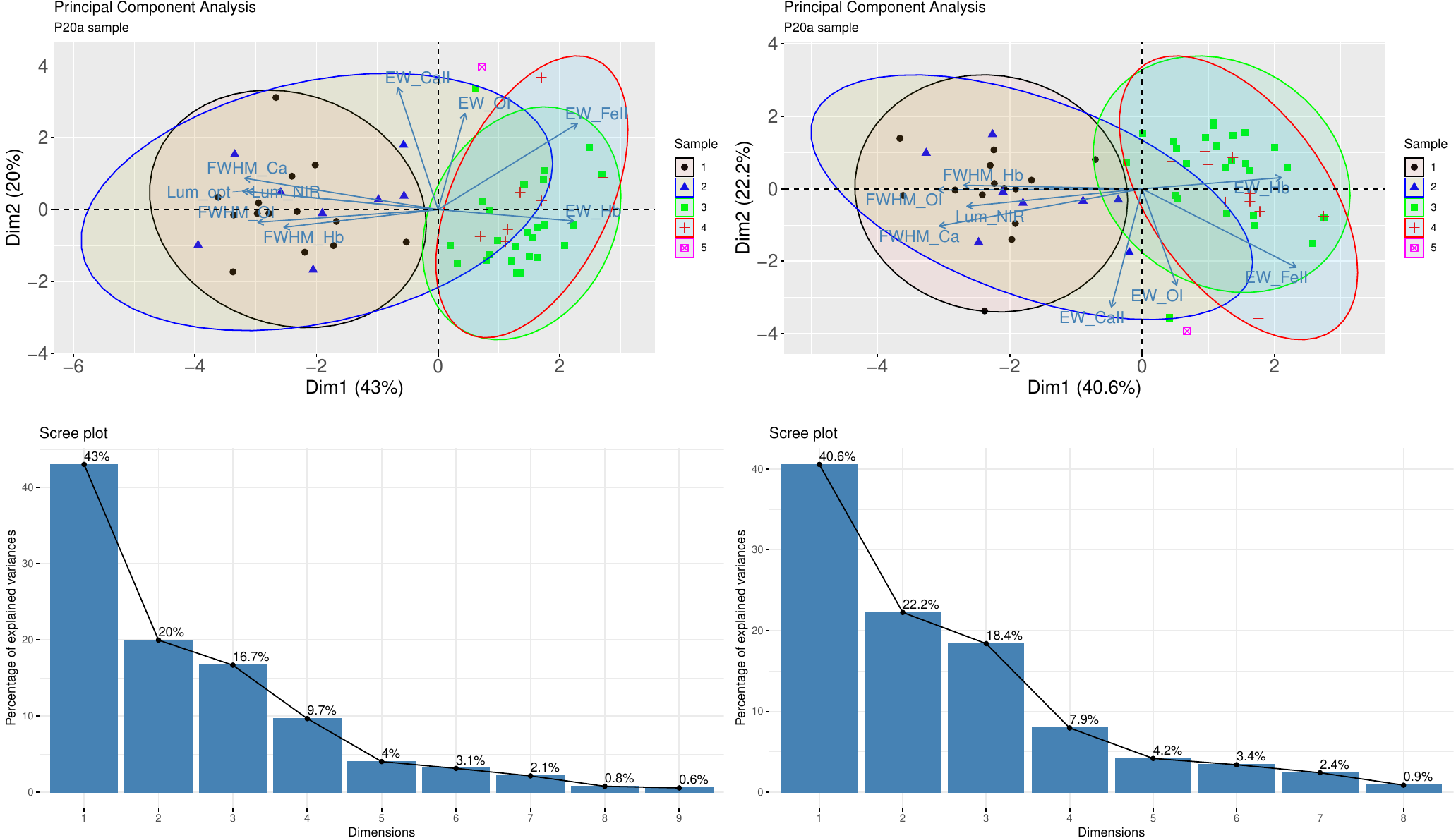}\\
    \caption{Graphical representation of the principal component analysis (PCA) decomposition of our sample (58 sources). The dots represent individual objects on the standardized PC1 - PC2 plane that have variables indicated by the axis labels. The arrows represent the prominence of the variables in the PC1 - PC2 plane. The dashed lines mark the coordinate axes in the PC1 - PC2 plane and the ellipses depict the 95\% occupancy of the sources in their respective subsamples. The sample is categorized based on their original source catalogues \citep[see][for details on the observational sample]{panda_cafe1} - (1) \citet{martinez-aldamaetal15}; (2) \citet{martinez-aldamaetal15b}; (3) \citet{persson1988}; (4) \citet{murilo2016}; and (5) PHL1092 \citep{2020arXiv200401811M}. \textbf{LEFT:} with \lopt{}; \textbf{RIGHT:} without \lopt{}. The lower panels illustrate the precedence of the first 10 principal components in the form of scree plots.}
    \label{fig:pca1}
\end{figure*}

\begin{figure*}[hbt!]
    \centering
    \includegraphics[width=\columnwidth]{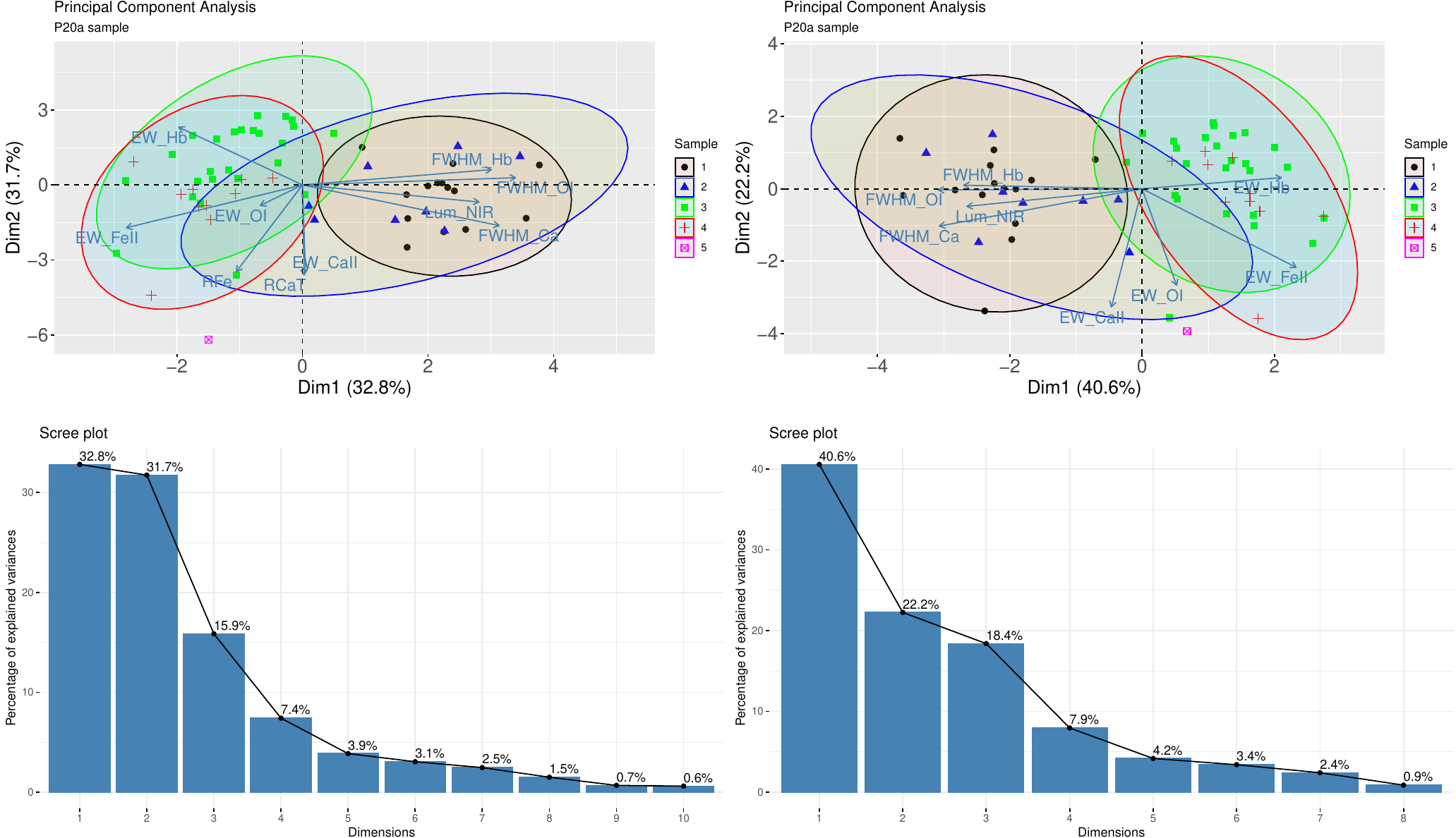}\\
    \caption{Similar to Figure \ref{fig:pca1}. \textbf{LEFT:} with \rfe{} and \rcat{}; \textbf{RIGHT:} without \rfe{} and \rcat{}. The lower panels illustrate the precedence of the first 10 principal components in the form of scree plots.}
    \label{fig:pca3}
\end{figure*}

As a first test for the PCA, we consider the aforementioned 11 parameters. Before drawing any definitive conclusions, we wanted to remove parameters that are redundant/uncorrelated. This allows to eliminates noise in the PCA output due to the presence of these redundant/uncorrelated variables. We perform this testing in two steps. In the first step, we analyze the effect of the presence of both optical and NIR luminosity in the PCA run. The optical (at 5100\AA) and the NIR (8542\AA) luminosities are almost identical for our sample (see bottom-right panel of Figure \ref{fig:spectral_properties}) with a correlation coefficient, \textit{r} = 0.950 ($p-$value = 6.72$\times 10^{-30}$). This result is snychronous to our PCA  (Figure \ref{fig:pca1}). Here, the factor map for the two cases (with and without optical luminosity) are shown adjacent to each other. At a first glance, the differences in the sources constituting our full sample can be clearly seen. The (3) Persson and (4) Marinello samples (i.e. the low luminosity sources) are similarly oriented in the PC1-PC2 diagram (factor-map), while the high-luminosity sources from the two catalogues from Martinez-Aldama (1 and 2) occupy a separate region in the factor-map. We study these subsamples in more detail in the next section. 
The corresponding scree plots are similar and highlight the dominance of the first principal component (43\% for the case with optical luminosity, and 40.6\% for the case without it). The subsequent principal components show similar precedence. We thus make use of only the NIR luminosity henceforth. 

The parameters, \rfe{} and \rcat{} are estimated from the various observations that are tabulated in Table \ref{tab:table1}. These values are estimated from the ratio of the fluxes of the respective emission species (optical \feii{} within the 4434-4684\AA\ and \hb{}; { \cat} and \hb{}, respectively). In our analysis, we use the EWs for the said species which are basically scaled versions of the line fluxes, one that is normalized by the corresponding continuum luminosity (at 5100\AA\ for \feii{} and 8542\AA\ for \cat{}). 
Thus, the \rfe{} and \rcat{} seemingly become redundant in presence of the EWs. We test this effect of redundancy on our sample and the results are presented in Figure \ref{fig:pca3}. The representation is similar to the Figure \ref{fig:pca1}. The factor-map on the left panel shows the case where the \rfe{} and \rcat{} are included in the analysis. Here, the \rcat{} vector is completely aligned with the EW(\cat{}) vector suggesting that the quantity is strongly dependent on this variable itself and is less affected by the EW(\hb{}). On the other hand, the orientation of the \rfe{} vector suggests that the quantity is affected by both EW(\feii{}) and EW(\hb{}). The corresponding scree plots for the two cases highlight the importance of the noise introduced in the PCA due to the presence of \rfe{} and \rcat{}. In the case where these variables were used, the dataset is organised such that the two principal components are almost identical (32.8\% for PC1 and 31.7\% for PC2). This gives a false impression that the dataset is driven by a 2D plane rather than a line. A similar aspect of the optical main sequence being represented as a line or a plane was explored in \citet{wildy2019}. On the other hand, when these two variables are removed and the PCA module is re-run, we see that the dataset is dominated by the variance along the first principal component (40.6\%) and the second principal component becomes less important (22.2\%). Another effect of the removal of redundant variables is the emergence of other quantities, e.g. EW(\oi{}). 

These two tests further confirms that the results of the PCA are dependent on the selection of the sample and the chosen properties \citep{kur09}.

\newpage

\section{Principal Component Analysis. Low- and high-luminosity samples}
\label{appendix:pca_sub}

\begin{figure*}[hbt!]
    \centering
    \includegraphics[width=0.75\columnwidth]{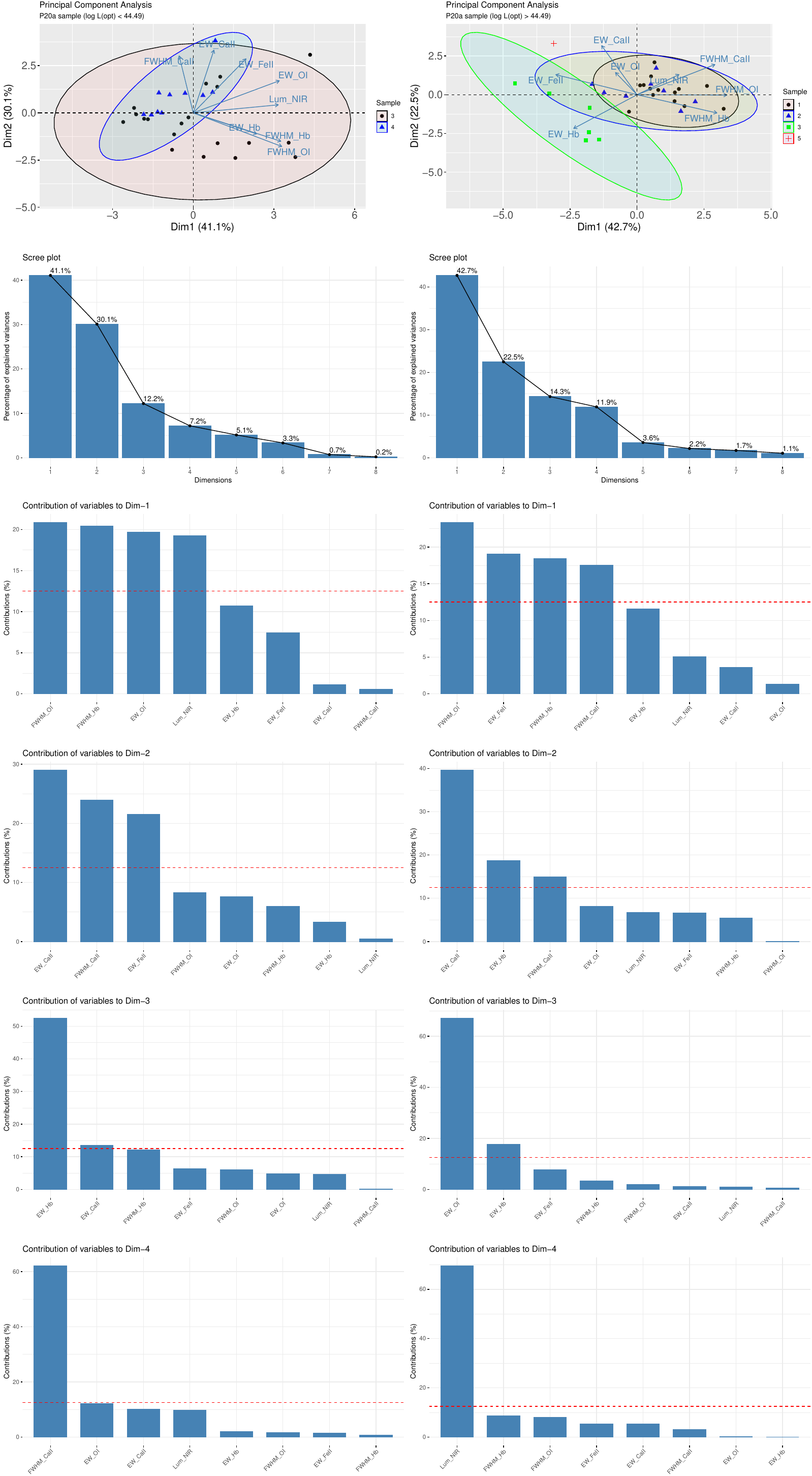}
    \caption{Same as Figure \ref{fig:pca2}.  \textbf{LEFT:} low-luminosity sample (log \lopt{} $\leq$ 44.49 erg s$^{-1}$). \textbf{RIGHT:} high-luminosity sample (log \lopt{} $>$ 44.49 erg s$^{-1}$).}
    \label{fig:pca_subsamples}
\end{figure*}

Taking note from the previous runs and the heterogeneity present in our sample, we now separate the full sample into two subsamples based on the median of the optical luminosity (at 5100~\AA) of the distribution, i.e. log \lopt{} = 44.49 erg s$^{-1}$ (Sec.~\ref{sec:division_sample}), which gives 29 sources in each case. We then perform the PCA on each of the subsamples (low $L_{\rm opt}$ and high $L_{\rm opt}$) and illustrate the results in the left and right columns of Figure \ref{fig:pca_subsamples}, respectively.

\subsection{Low-luminosity subsample}

For the low-luminosity sample, the sources belong to the Persson {(19/25)} and Marinello {(10/10)} samples, and the Marinello sample is {almost} enclosed within the Persson sample in the factor-map. The corresponding scree plot shows the dominance of the primary and secondary principal components ({41.1\% and 30.1\%}), suggesting that the sample seems to be driven majorly by the combination of the two components that can explain {71.2\%} of the variance in the dataset. Accounting for the subsequent two principal components, {90.6\%}  of the total variation in the dataset can be explained in this case.

\textit{First principal component}: Going back to the factor-map, we find that the vectors corresponding to the FWHMs of \hb{} and \oi{}, \ewhb\ are co-aligned, with the FWHM vectors having almost similar magnitudes. These two FWHM vectors are also the major contributors to the variance along the primary principal component (see third panel on the middle column of Figure \ref{fig:pca2}). For the primary principal component, {the EW of \oi{}} follows after these two FWHMs, which then is followed by {the NIR luminosity}.

\textit{Second principal component}: The factor-map highlights the prevalence of the \ewca followed by the \fwhmca and the \ewfe\. This trend is also seen in the contributions to the second principal component and supports our original conclusion that the two species, \feii{} and \cat{} are similar in terms of their excitation and line emissivities (see fourth panel on the middle of Figure \ref{fig:pca2}). We expect that the FWHM$_{\rm FeII}$ would show similar behaviour, likewise of \fwhmca\ \citep[as shown in][where the authors show an almost perfect correlation between the \feii{} emission at 1 $\mu$m and \cat{}]{murilo2016} strengthening the inferences from the photoionization modelling \citep{panda_cafe1, panda_cafe2}. This needs to be tested with a larger, of higher S/N and more-complete sample in the future.


\textit{Third and fourth principal components}: The third and fourth principal components further contribute to {19.4\%} of the {total} variance in the dataset. The third PC is singularly dominated by the \ewhb\ with a minor contribution from {\ewca and \fwhmhb\}. {Similarly, the fourth PC is mainly driven by the \fwhmca} and only a minor contribution from EW of \oi{}.}

\subsection{high-luminosity subsample}

For the high-luminosity sample, the sources belong to the Persson ({6/25}), Marinello (PHL1092) samples, and all of Martinez-Aldama's sources. The {6 sources from Persson (Mrk304, Mrk509, IZw1, Mrk478, VIIZw118 and 3C273) and PHL1092}, outline around the 95\% confidence limit of the Martinez-Aldama's sample shown with the ellipses on the factor-map (top-right panel in Figure \ref{fig:pca2}). This points towards the homogeneity in the subsample as opposed to the earlier scenario when all the sources were bunched together. The corresponding scree plot shows the dominance of the primary and secondary principal components ({42.7\% and 22.5\%}), suggesting that the {high-luminosity} sample {behaves similar to} the low-luminosity case. Likewise to the low-luminosity case, accounting for the subsequent two principal components (PC3 and PC4), {91.4\%} of the total variation in the high-luminosity dataset can be explained.

\textit{First principal component}: Compared to the low-luminosity case, the {FWHM of \oi{} is still the primary dominant driver of the the primary PC, followed by the \ewfe\, \fwhmhb\ and \fwhmca}. The \ewfe\ dominates in the negative space of the PC1.

\textit{Second principal component}: The primary contributor to this PC is still \ewca, but in contrast to the corresponding PC for the low-luminosity case, the \ewfe\ is rather {below the significance threshold} in this case. {Other significant contributors are the \ewhb\ followed by the \fwhmca.}

\textit{Third and fourth principal components}: The third and fourth principal components further contribute to {26.2\%} (earlier this was {19.4\%} for the low-luminosity case) of the {total} variance in the dataset. The third PC is dominated by the {\ewoi\ with contribution from \ewhb.} Whereas, the fourth PC is {singularly driven by the NIR luminosity}.

\subsection{Correlations between the principal eigenvectors and observed/derived parameters for the subsamples}

\begin{figure*}
    \centering
    \includegraphics[width=\columnwidth]{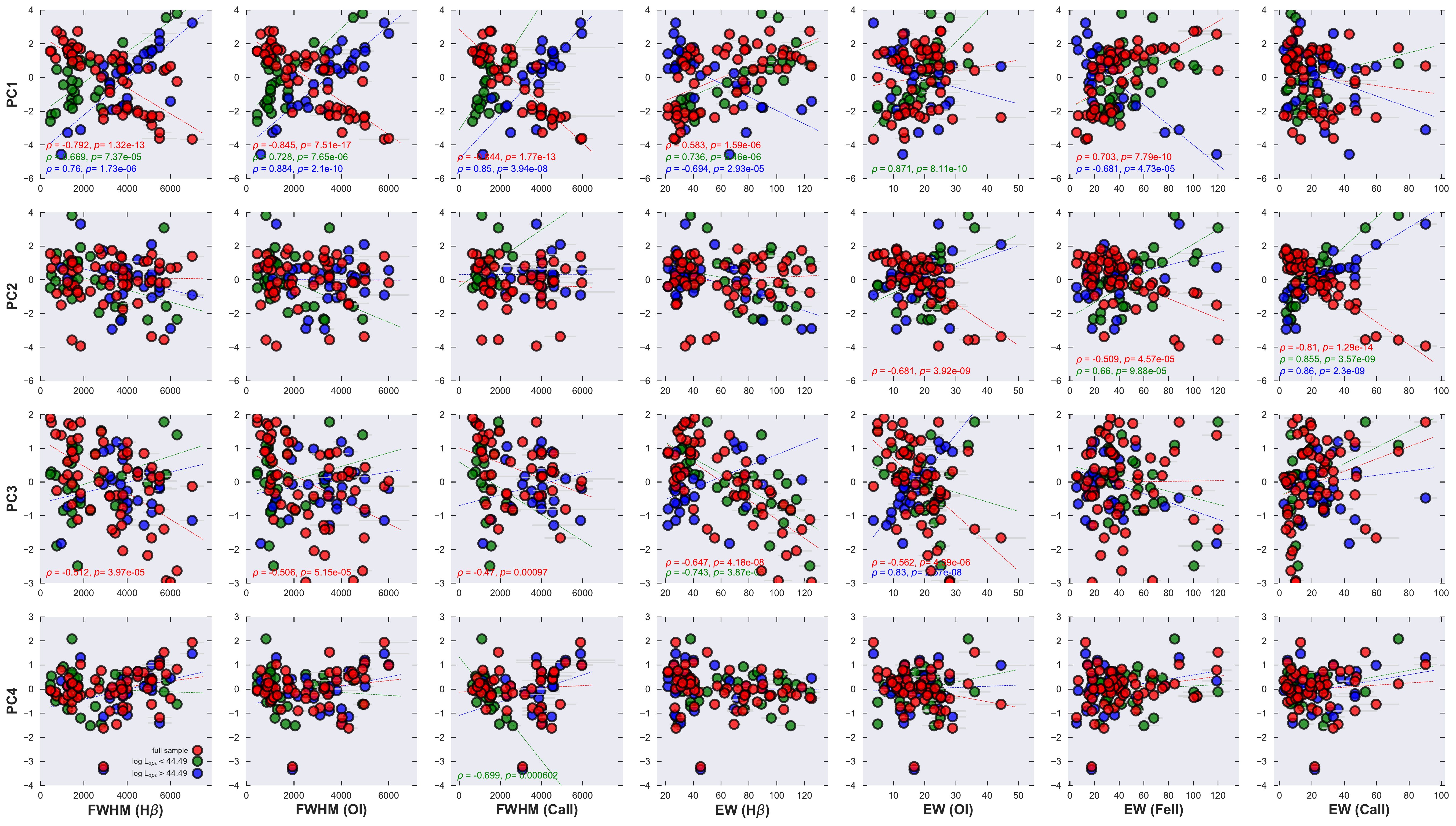}
    \caption{Correlation matrix showing dependence of the first four PCA vectors' loadings versus the physical parameters (\textit{observed}) for our sample. The sample is divided into low luminosity (green dots) and high luminosity (blue dots) subsamples based on the median value of the sample's optical luminosity distribution, i.e. at 44.49 erg s$^{-1}$ (see Appendix \ref{appendix:pca_sub}). The full sample is shown in red dots. The Spearman's rank correlation coefficient ($\rho$) and the $p-$value are reported for the correlations whenever $p$-value $<$ 0.001. The OLS fits for each sample is shown using dashed lines using their respective color.}
    \label{fig:pca_corr2}
\end{figure*}

\begin{figure*}
    \centering
    \includegraphics[width=\columnwidth]{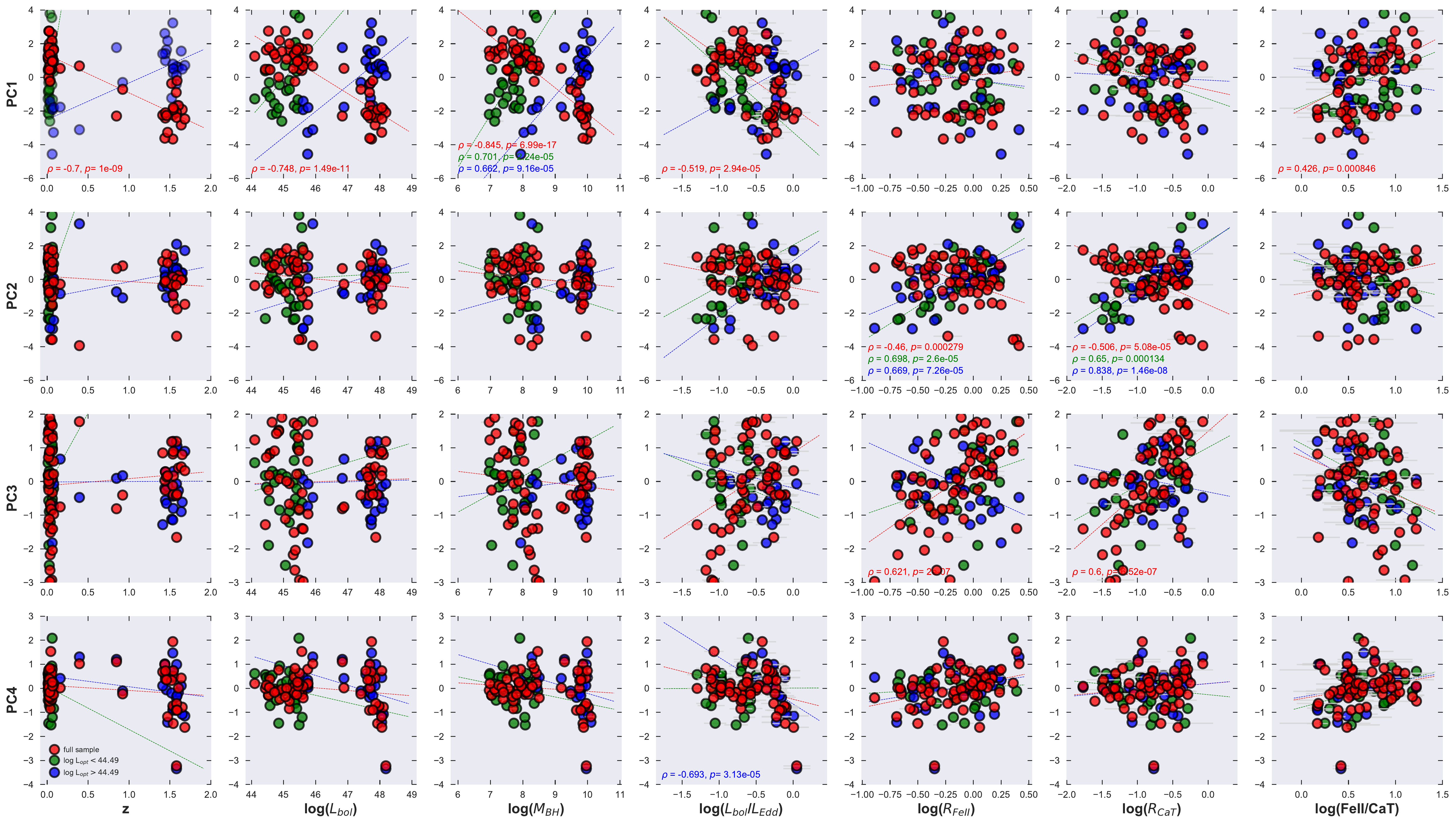}
    \caption{Correlation matrix showing dependence of the first four PCA vectors' loadings versus the physical parameters (\textit{derived}) for our sample. The colors for the data-points are identical to that shown previously in Figure \ref{fig:pca_corr2}. The Spearman's rank correlation coefficients ($\rho$) and the $p-$values are reported for the correlations whenever $p-$value $<$ 0.001. The OLS fits for each sample is shown using dashed lines using their respective color.}
    \label{fig:pca_corr1}
\end{figure*}

Figure \ref{fig:pca_corr2}: For the PC1, there are significant positive correlations for both the subsamples, especially, with respect to \fwhmhb\ (low-luminosity: $\rho$=0.669, $p$=7.37$\times 10^{-5}$; high-luminosity: $\rho$=0.76, $p$=1.73$\times 10^{-6}$) and \fwhmoi\ (low-luminosity: $\rho$=0.728, $p$=7.65$\times 10^{-6}$; high-luminosity: $\rho$=0.884, $p$=2.1$\times 10^{-10}$). This is highlighted by the strong correlation between the PC1 and the black hole mass which is obtained and explained in Figure \ref{fig:pca_corr1} as the \fwhmhb\ is incorporated to estimate the black hole mass. A strong positive correlation is obtained for \fwhmca\
($\rho$=0.85, $p$=3.94$\times 10^{-8}$) for the high luminosity subsample. The \ewhb\ correlation with PC1 behaves differently for the low-luminosity ($\rho$=0.736, $p$=5.46$\times 10^{-6}$) and the high-luminosity ($\rho$=-0.694, $p$=2.93$\times 10^{-5}$) samples. The low-luminosity sample follows the trend of the full sample in this case. For \ewoi, significant correlation is noted only for the low-luminosity case ($\rho$=0.871, $p$=8.11$\times 10^{-10}$). For \ewfe, significant anti-correlation is noted only for the high-luminosity case ($\rho$=-0.681, $p$=4.73$\times 10^{-5}$)  . For PC2, we have two significant correlations for the low-luminosity case - for \ewca\ ($\rho$=0.855, $p$=3.57$\times 10^{-9}$) and \ewfe\ ($\rho$=0.66, $p$=9.88$\times 10^{-5}$). Additionally, there is a correlation obtained for the \ewca\ ($\rho$=0.86, $p$=2.3$\times 10^{-9}$) for the high-luminosity case. For PC3, a significant anti-correlation is observed for the low-luminosity case's \ewhb\ ($\rho$=-0.743, $p$=3.87$\times 10^{-6}$) and a significant correlation for the high luminosity case's \ewoi\ ($\rho$=0.83, $p$=2.57$\times 10^{-8}$). There is a single significant (anti-)correlation observed for PC4, i.e. versus \fwhmca\ ($\rho$=-0.699, $p$=6.02$\times 10^{-4}$).

Figure \ref{fig:pca_corr1}: The strongest and only correlation in the subsamples for the PC1 are with respect to the black hole mass - for the low-luminosity sample, the correlation is relatively stronger ($\rho$=0.701, p=2.24$\times 10^{-5}$) compared to the high-luminosity sample ($\rho$=0.662, $p$=9.16$\times 10^{-5}$). For PC2, significant correlations are obtained only for \rfe{} and \rcat{} cases. For the low-luminosity sample, the correlations of PC2 versus \rfe{} ($\rho$=0.698, $p$=2.6$\times 10^{-5}$) and versus \rcat{} ($\rho$=0.65, $p$=1.34$\times 10^{-4}$), while for the high-luminosity sample, the correlations of PC2 versus \rfe{} ($\rho$=0.669, $p$=7.26$\times 10^{-5}$) and versus \rcat{} ($\rho$=0.838, $p$=1.46$\times 10^{-8}$). This further corroborates the strong connection between these two parameters from our past results obtained in Paper-1 and those obtained from the PCA analysis of the full sample in this paper. There are no significant correlations with respect to PC3. The only significant (anti-)correlation with respect to PC4 is obtained for \ledd{} in high-luminosity sample ($\rho$=-0.693, $p$=3.13$\times 10^{-5}$).

\newpage






\newpage




\end{document}